\documentclass{article} 

\usepackage{amsmath,amssymb,amsthm,graphicx,verbatim,bm,fancyhdr,bbm,enumerate,color,epstopdf,epsfig,latexsym,caption}
\usepackage{verbatim,geometry,subcaption}
\title{Bayesian information-theoretic calibration of patient-specific radiotherapy sensitivity parameters for informing effective scanning protocols in cancer}
\author{Heyrim Cho, Allison L. Lewis,  Kathleen M. Storey}
\date{}

\begin{document}
\maketitle
%%%%%%%%%%%%%%%%%%%%%%%%%%%%%%%%%%%%%%%%%%
\abstract{
With new advancements in technology, it is now possible to collect data for a variety of different metrics describing tumor growth, including tumor volume, composition, and vascularity, among others.  For any proposed model of tumor growth and treatment, we observe large variability among individual patients' parameter values, particularly those relating to treatment response; thus, exploiting the use of these various metrics for model calibration can be helpful to infer such patient-specific parameters both accurately and early, so that treatment protocols can be adjusted mid-course for maximum efficacy.  However, taking measurements can be costly and invasive, limiting clinicians to a sparse collection schedule.  As such, the determination of optimal times and metrics for which to collect data in order to best inform proper treatment protocols could be of great assistance to clinicians.
In this investigation, we employ a Bayesian information-theoretic calibration protocol for experimental design in order to identify the optimal times at which to collect data for informing treatment parameters.  Within this procedure, data collection times are chosen sequentially to maximize the reduction in parameter uncertainty with each added measurement, ensuring that a budget of $n$ high-fidelity experimental measurements results in maximum information gain about the low-fidelity model parameter values. In addition to investigating the optimal temporal pattern for data collection, we also develop a framework for deciding which metrics should be utilized at each data collection point. We illustrate this framework with a variety of toy examples, each utilizing a radiotherapy treatment regimen. For each scenario, we analyze the dependence of the predictive power of the low-fidelity model upon the measurement budget. 
% (A single paragraph of about 200 words maximum.) 
}

\section{Introduction}

Mathematical and computational approaches have long been developed for a better understanding of cancer \cite{Altrock2015,Lavi2012,Swierniak,Byrne2010,Rockne2019}. Complex mechanisms that control carcinogenesis, tumor progression, and dynamics under treatments have been studied through various mathematical models \cite{Altrock2015,Byrne2010}. Such quantitative models are utilized to predict cancer dynamics and therapy response, improve early detection, and design therapeutic protocols \cite{Rockne2019}. Moreover, advances in technology have made it possible to collect considerable amounts of detailed data, including genetic, cellular level, and tissue-scale imaging, describing various aspects of cancer progression in patients \cite{Chambers2019,Bibault}. Incorporating the available data into an appropriate mathematical model can enhance the effectiveness of personalized treatments. However, practically speaking, an abundance of data may not be available in a clinical setting, particularly in the temporal domain. Specifically, data can often only be collected at sparse times, due to physical and financial burdens to patients. Therefore, with a limited number of total scans, it is important to decide when to collect the data to most accurately calibrate the model for maximum predictive power. Moreover, another important purpose of data collection is to aid in making an accurate prediction as early as possible during the treatment so that any necessary adjustments---for example, altering the dosage or discontinuing the treatment---can be made as soon as possible. However, this goal is also prohibited by the limited availability of patient data in the temporal domain.

Many different types of mathematical models have been developed to simulate tumor growth and response to treatment. These range from phenomenological models composed of ordinary differential equations (ODEs)  \cite{Murphy2016,Collis2017,Koziol2020} to complex multi-scale models that combine representations of the tumor microenvironment at the subcellular, cellular, and tissue scales \cite{Liu2007,Ramis2009,Hawkins-Daarud2013, Kannan2019}. Typically the biological accuracy increases as the model complexity increases; however, achieving unique identifiability of model parameter values becomes more challenging. In \cite{Cho}, we present a framework for choosing an appropriate model and calibrating parameter values, given a set of data describing various tumor metrics over time. Here, we extend that work by developing a methodology to determine the optimal data collection times, in order to maximize the utility of the collected samples. 

We focus on applying this methodology to models that simulate tumor response to radiotherapy. Radiotherapy is a common treatment modality applied to many cancer types, and there is a long tradition of mathematical modeling of radiotherapy response. Fractionated radiation dosing is typically modeled using the linear-quadratic (L-Q) model \cite{Thames1982, Fowler1989}. Several recent studies have applied the L-Q model to patient-specific data, in an effort to evaluate and predict individual responses to radiotherapy \cite{Rockne2010, Corwin2013, Sunassee2019, Cho}. Our work extends this body of literature by suggesting optimal sampling times for tumors undergoing radiotherapy, so that the L-Q parameters can be efficiently calibrated and tuned for more accurate post-treatment predictions.

We utilize a Bayesian information-theoretic experimental design framework for choosing optimal design conditions at which to collect data for model calibration.  This methodology is based on the concept of maximizing information gain---and thus reducing uncertainty---about low-fidelity model parameters, while minimizing the number of design conditions at which experimental data must be collected. Originally built around the concept of Shannon entropy \cite{Shannon}, and utilizing the $k$th-Nearest-Neighbor ($k$NN) estimate of mutual information to make the procedure more computationally feasible \cite{Kraskov}, this framework was first introduced in \cite{Terejanu, Bryant, Liepe}.  The methodology was further extended in \cite{Lewis} to incorporate more robust Bayesian methods for supporting highly correlated and nonlinear parameter dependencies. Here, we further amend this framework to be suitable for a temporal data collection setting, emphasizing the minimization of uncertainty in our model parameters while also penalizing our algorithm for choosing points further out in time, which precludes the potential information gain that would result from data collection at intermediate time steps.

In Section \ref{sec:models}, we introduce the mathematical models for tumor growth and radiotherapy that will be used to illustrate our methodology throughout this work.  Section \ref{sec:mi_methodology} outlines the procedure for using mutual information to optimally choose design conditions at which to evaluate tumor data in order to maximize information gain about model parameters for accurate and timely calibration.  This section includes the novel introduction of a score function for temporal data collection, which penalizes the user for skipping too many potential data evaluation times in pursuit of the largest possible information gain.  In Section \ref{sec:simulation}, we illustrate our proposed framework for several examples that showcase a wide variety of scenarios with respect to model complexity and data collection protocols. We summarize our findings and discuss plans for future investigation in Sections \ref{sec:discussion} and \ref{sec:future}. 

\section{Mathematical models used for testing} \label{sec:models}

We begin by presenting two mathematical models that will be used throughout this work as low-fidelity models to be calibrated for clinical predictions, and another model that will be used to generate synthetic high-fidelity data. The first low-fidelity model is a one-compartment ODE model that tracks tumor volume over time---our simplest model for describing tumor growth. The second model is a two-compartment ODE model that incorporates a state variable for tracking the portion of tumor volume that is composed of necrotic tissue, thereby introducing the concept of tumor heterogeneity. 

These two models will be calibrated using high-fidelity data generated by a more complex model, namely, a cellular automaton model. In addition to quantifying both the viable and necrotic cell populations, the cellular automaton model also tracks the cell division cycle, quiescent cells, and oxygen levels, allowing for more accurate reflection of the stochastic and heterogeneous nature of cancer growth in reality \cite{Kannan2019,Cho}. In all three cases, we incorporate treatment via radiation using the linear-quadratic model for radiotherapy \cite{Hall1994, Enderling}, as outlined in Section \ref{sec:radiotherapy}.  

\subsection{The One-Compartment ODE Model}

The one-compartment model describes the time evolution of the total tumor volume, $V(t)$, using a logistic growth model with growth rate $\lambda$ and carrying capacity $K$:

\begin{equation}
    \frac{dV}{dt} = \underbrace{\lambda V \left ( 1 - \frac{V}{K} \right )}_{\mbox{logistic growth}} - 
    \underbrace{\eta V.}_{\mbox{natural cell death}}
    \label{eqn1comp_dimensional}
\end{equation}
We incorporate natural cell death via the term $-\eta V$. In what follows, it will be convenient to re-parameterize Equation (\ref{eqn1comp_dimensional}) to obtain the simpler and parametrically-identifiable form 

\begin{equation}
    \frac{dV}{dt} = AV\left(1-\frac{B}{A}V\right),
    \label{eqn:eqn1comp}
\end{equation}
where $A = \lambda-\eta$ and $B = \frac{\lambda}{K}$. From this point forward, any reference to the one-compartment model is referring to the re-parameterized form, Equation (\ref{eqn:eqn1comp}).

Biologically speaking, as a tumor grows, regions at a distance from oxygen and nutrient sources (e.g., blood vessels for tumors growing \textit{in vivo}) may undergo necrosis in response to sustained oxygen and/or nutrient deprivation.  In this simple one-compartment model, such dead or necrotic cells are assumed to be removed from the tumor instantaneously; that is, we view the tumor as a homogeneous mass of proliferating, viable cells. 

\subsection{The Two-Compartment ODE Model}

In order to account for some aspects of tumor heterogeneity, we next study
a two-compartment model that tracks the time evolution of the viable tumor cell volume, $V(t)$, and the necrotic core volume, $N(t)$, originally developed in \cite{Lewin:PhD} and further analyzed in \cite{Cho}. We consider this model in an effort to better represent reality, as it has been shown that the proportion of necrotic material has a significant impact on one's ability to estimate a tumor's response to radiotherapy \cite{Cho, Lewin2019a, Lewin2019b}. 
We still assume that the population of proliferating (i.e., viable) cells, $V(t)$, grows logistically with growth rate $\lambda$ and carrying capacity $K$, and that viable cells convert to necrotic cells at a constant rate $\eta$. In this second model, we assume that the natural death of viable cells results in the build-up of a necrotic core, denoted by $N(t)$, where the necrotic material then undergoes natural decay at a constant rate $\zeta$. We note that as $\zeta \rightarrow \infty$, the two-compartment model converges to the one-compartment model behaviorally. Combining these processes, we arrive at the following ODE system for $V(t)$ and $N(t)$:
\begin{subequations}
\begin{eqnarray}
\frac{dV}{dt} &=& \lambda V \left(1-\frac{V}{K}\right)-\eta V, \label{eqn:twocompOriginal1}\\
\frac{dN}{dt} &=& \eta V - \zeta N. 
\label{eqn:twocompOriginal2}
\end{eqnarray} \label{eqn:twocomp}
\end{subequations}
Throughout the following investigation, we refer to this system \eqref{eqn:twocomp} as the two-compartment model.

\subsection{The Cellular Automaton Model} \label{sec:CAmodel}

In the absence of experimental data, we generate synthetic total tumor and necrotic volume data using a spatially-explicit, hybrid cellular automaton (CA) model to allow for illustration of our methodology. Throughout this investigation, the synthetic data from the CA model serves as the high-fidelity data in the high-to-low fidelity model calibration framework referred to in \cite{Lewis}. Our cellular automaton model is adapted from that developed in \cite{Kannan2019} and later expanded in \cite{Cho}.  The model incorporates spatially heterogeneous oxygen levels, a stochastic cell life cycle, and a heterogeneous cell population, including proliferating, quiescent, and necrotic cells. 
The cells are arranged on a discrete lattice representing a two-dimensional square cross-section of size 0.36$\times$0.36 cm$^2$ through a three-dimensional cancer spheroid \emph{in vitro}. We identify with each automaton $\textbf{x}=(x,y)$ at time $t$ a dynamical variable with a state and a neighborhood. 

Each automaton can be occupied by a tumor cell in one of three states---proliferating, $\mathcal{P}$, quiescent, $\mathcal{Q}$, or necrotic, $\mathcal{N}$---or can be unoccupied and denoted as empty, $\mathcal{E}$. We note that all lattice cells have an associated oxygen level, regardless of whether they are currently occupied. This oxygen level, $c$, determines the state of the occupying cell using thresholds $c_N$ and $c_Q$: if $c > c_Q$ then the cells proliferate, if $c_N < c < c_Q$ then the cells transition to quiescent cells with a halved oxygen consumption rate, and if $c \leq c_N$ then the cells are considered necrotic. 

We model the single growth-rate-limiting nutrient, oxygen, explicitly via a reaction-diffusion equation. See \cite{Kannan2019,Cho} for a detailed description of the oxygen model. 
If a cell becomes necrotic due to low oxygen concentration or irradiation, which will be discussed in Section \ref{sec:radiotherapy}, the necrotic cells are lysed at rate $p_{NR}$. Lysis involves removing the necrotic cell and then shifting inward a chain of cells starting from the boundary of the spheroid to fill in the removed cell's site. 
Additionally, the CA mimics several other mechanisms observed in tumor spheroids. We incorporate the regulatory process known as contact inhibition of proliferation by reducing the division of cells with a large number of neighbors. We simulate cell-cell adhesion by shifting chains of cells outward after cell division. The details of these model features are described in \cite{Kannan2019,Cho}.

We use the CA model to generate a series of synthetic spheroids that differ in their response to radiosensitivity. 
The parameter values that are used to generate data using the CA model are shown in Table \ref{table:CA_pars} in the Appendix \ref{app:params}. These parameters are estimated using experimental data from the prostate cancer cell line, PC3, in \cite{Kannan2019}.

\subsection{Radiotherapy Treatment}\label{sec:radiotherapy}

Finally, we discuss the incorporation of a radiotherapy (RT) treatment protocol in all three models. We consider a typical tumor treatment regimen in which daily doses of 2 Gy are administered Monday through Friday for six consecutive weeks. We use the linear-quadratic model \cite{Hall1994, Enderling} to account for the effects of RT. This model assumes that the fraction of cells that survive exposure to a single administered dose $d$ of RT is given by

\begin{equation}
\mbox{Survival fraction},\ \  SF = e^{-\alpha d-\beta d^2},
\end{equation}
where $\alpha$ and $\beta$ represent tissue-specific radiosensitivity parameters that model single- and double-strand breaks of the DNA, respectively \cite{Lea1942}. 

For implementation in the one-compartment model, we assume that the effect of RT is instantaneous. That is, the irradiated cell fraction is removed immediately from the tumor volume, akin to the treatment of natural cell death in our model. Under these assumptions, the one-compartment model with RT treatment can be re-formulated as

\begin{align}
   \left\{
\begin{array}{ll}
      \frac{dV}{dt} = A V \left ( 1 - \frac{B}{A}V \right ),  & \mbox{for} \; t_i^+ < t < t_{i+1}^-,\\ \ \\
      V(t_i^{+})=
\exp(-\alpha d-\beta d^2) \: V(t_{i}^{-}), &\\
\end{array} 
\right. 
\end{align}
where $t_i$ (for $i=1, 2, \ldots, n_R$) denote the times at which an RT dose is delivered, and $V(t_i^{\pm})$
denote the tumor volume just before and after radiotherapy is administered.

Because our two-compartment model allows for the existence of a necrotic core, our implementation of RT in the two-compartment model assumes that irradiated cells from the total tumor volume move directly into the necrotic compartment, and later decay naturally. This manifests visually as a delayed reaction to RT in the total tumor volume; it is the heterogeneous composition of the tumor that displays the immediate effects of RT. 

We apply the linear-quadratic model to the cellular automaton model in a similar fashion. At each administration of RT, each living cell converts to a necrotic cell with probability $1-e^{-\alpha d-\beta d^2}$, and then undergoes natural decay during a later iteration with probability $p_{NR}$, as detailed in Section \ref{sec:CAmodel}.

\begin{figure}[!b]
\centerline{  
         \includegraphics[width=1.9in]{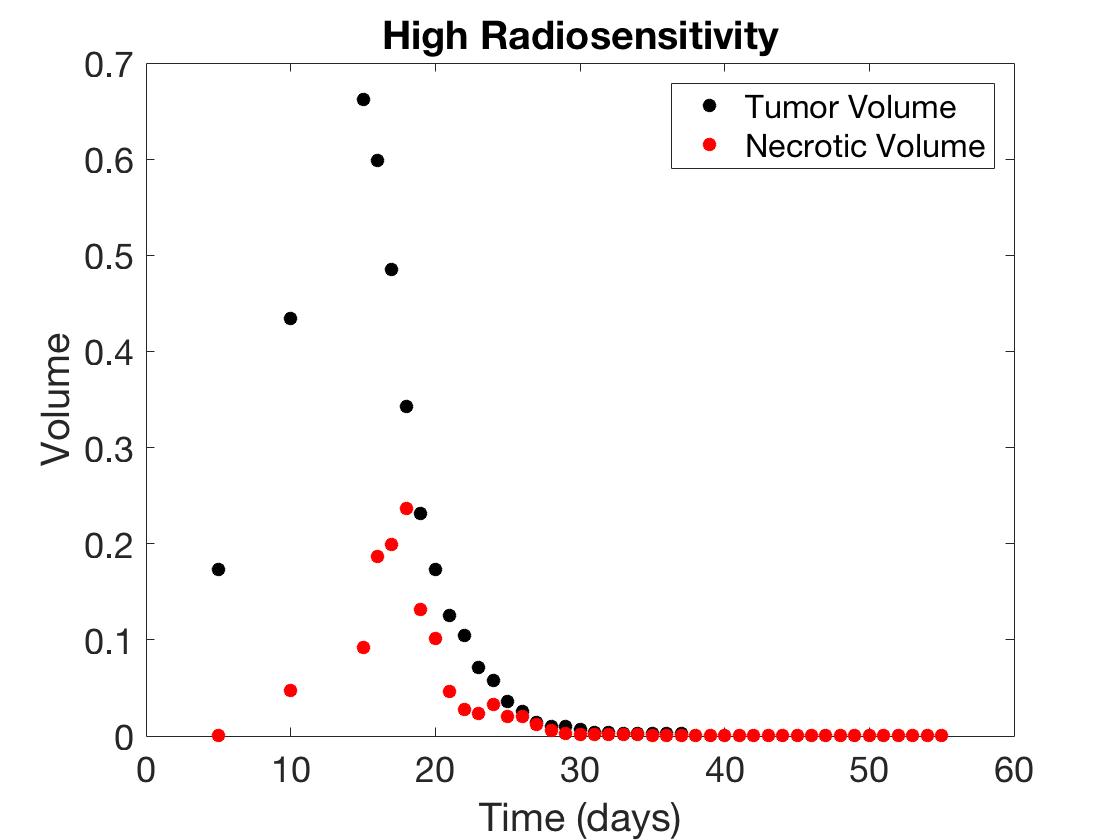}
         \includegraphics[width=1.9in]{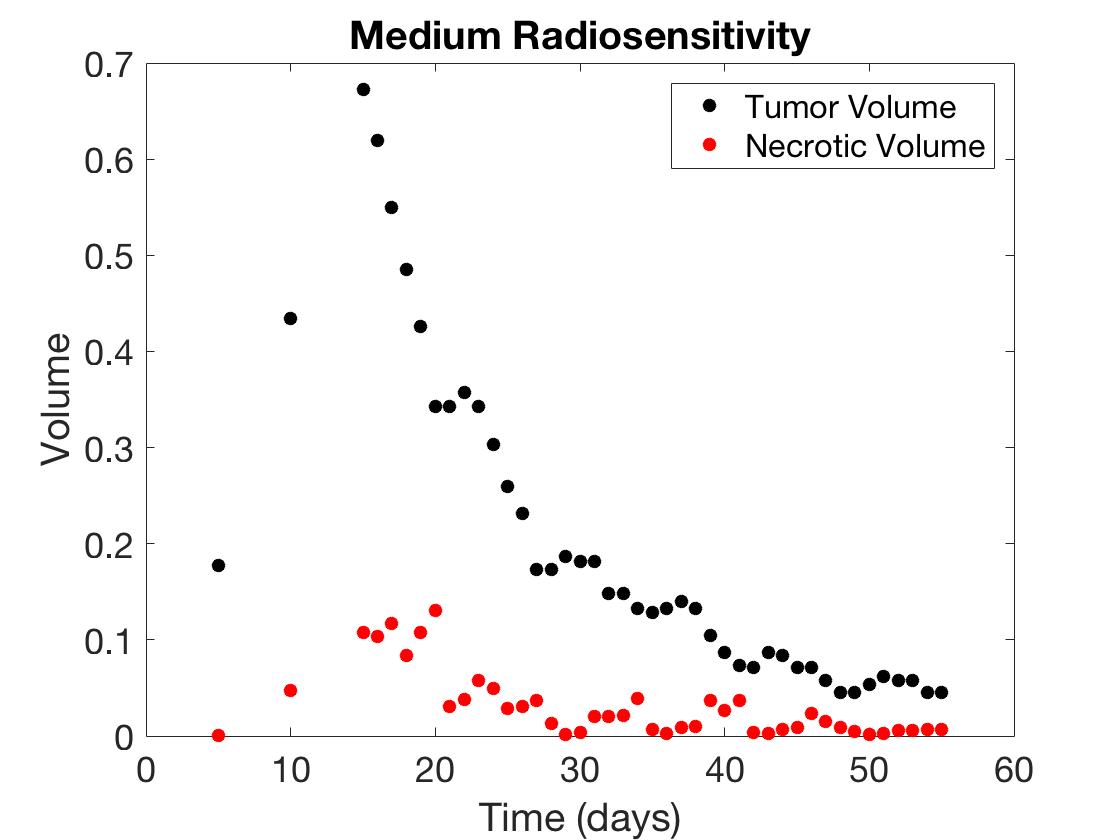}
         \includegraphics[width=1.9in]{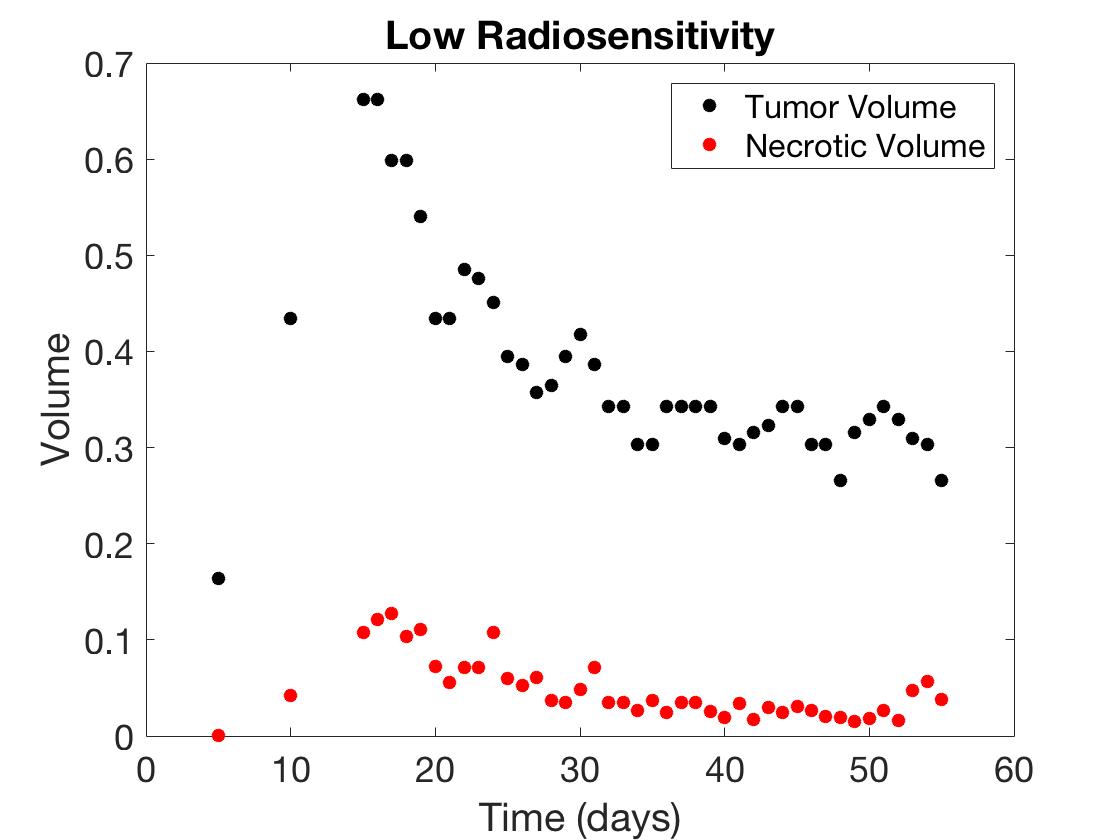}  
         }
    \caption{High-fidelity CA data for each of the three radiosensitivity scenarios: high ($\alpha/\beta = 1$, on left), medium ($\alpha/\beta = 3$, in middle), and low ($\alpha/\beta = 9$, on right).}
    \label{fig:dataplots}
\end{figure}

To illustrate our framework for several scenarios with data displaying a variety of behavior with regard to treatment success, we produce three different sets of test data using the CA model.  These three test scenarios are representative of a patient whose tumor is highly sensitive to radiotherapy (generated in the CA model using $\alpha = 0.14$, $\beta = 0.14$, for a radiosensitivity parameter ratio of 1), a patient whose tumor is somewhat sensitive to RT (using $\alpha = 0.14$ and $\beta = 0.0467$ for a ratio of $\alpha/\beta = 3$), and a patient whose tumor does not respond meaningfully to RT (using $\alpha = 0.14$ and $\beta = 0.0156$ for a ratio of $\alpha/\beta = 9$).  Throughout the investigation, we will refer to these three test scenarios as ``high," ``medium," and ``low" radiosensitivity, respectively. The three sets of CA synthetic data are shown in Figure \ref{fig:dataplots}. In all patients, we consider a typical treatment protocol in which daily doses of 2 Gy are administered Monday through Friday for 6 weeks, as discussed above.

\section{Bayesian Information-Theoretic Methodology} \label{sec:mi_methodology}

Our goal is to calibrate the parameters of a low-fidelity model (Equations \eqref{eqn:eqn1comp} or \eqref{eqn:twocomp}) using as few high-fidelity data collections (or in this case, as few evaluations of the high-fidelity CA model) as possible.  In Section \ref{sec:MI_expdesign}, we outline the mutual information methodology used to choose the most informative high-fidelity data points with respect to low-fidelity parameter uncertainty. Then, in Section \ref{sec:MI_time_series}, we present our adaptation of this method to allow for its use on time series data. 

\subsection{Experimental design framework using mutual information}
\label{sec:MI_expdesign} 

Given a set of initial experimental data (or high-fidelity code evaluations) and a low-fidelity model that we wish to calibrate, we seek to determine the optimal design conditions at which to collect additional data (or evaluate our high-fidelity model) in order to best inform the parameters of the low-fidelity model, $\theta \in \mathbb{R}^p$. We note that throughout this investigation, we will use a high-fidelity code to produce synthetic data to represent experimental data acquisition. We denote the existing high-fidelity data set by $D_{n-1} = \{\tilde d_1, \tilde d_2,\dots, \tilde d_{n-1}\}$, and select the next design condition $\xi_n$ from the set of all possible evaluation strategies or experimental designs, $\Xi$, in such a way as to maximize reduction in the uncertainty in $\theta$ when $\tilde d_n = d_h(\xi_n)$---the data point resulting from evaluation at design $\xi_n$---is appended to the existing data set. 

Because we cannot determine the value of $\tilde d_n$ without first evaluating the high-fidelity code, we must base our decision of $\xi_n$ on predictions from the low-fidelity model. The low-fidelity model prediction for a specified condition is denoted by $d_n = d_{\ell}(\theta,\xi_n)$, where $d_{\ell}(\theta,\xi_n)$ represents the low-fidelity model evaluated at parameter set $\theta$ and design $\xi_n$. Predictions made by this model can be extended into a statistical modeling framework by incorporating a model discrepancy term, $\delta(\xi_n)$, and the possibility of measurement or discretization errors, $\varepsilon_n(\xi_n)$. In this study, we assume no model discrepancy and zero measurement errors, and refer the interested reader to \cite{Lewis} for additional details on the corresponding statistical models for both the low-fidelity and high-fidelity frameworks.

Recall that our objective is to calibrate the low-fidelity model parameters using as little high-fidelity data as possible, so as to minimize unnecessary computational or experimental costs.  Thus, we want to choose design conditions for the high-fidelity code so as to contribute the maximum amount of information about our low-fidelity model parameters.  We do so using a Bayesian framework. We begin with Bayes' Rule, which gives a method by which to update the knowledge about the parameters based on the addition of a new data point:
\begin{eqnarray}
p(\theta | D_n) = \frac{p(D_n | \theta)p(\theta)}{p(D_n)} = \frac{p(\tilde{d}_n, D_{n-1}| \theta)p(\theta)}{p(\tilde{d}_n, D_{n-1})}.\nonumber
\end{eqnarray}
In essence, Bayes' Rule states that the posterior distribution of $\theta$, given data $D_n$, depends both on the likelihood of obtaining data $D_{n}$ given the parameter set $\theta$, represented by $p(D_n|\theta)$, and the prior distribution of $\theta$, $p(\theta)$, which incorporates all known information about $\theta$. 

As a means of quantifying the information gain obtained from including an additional data point, we utilize the mutual information between the low-fidelity model parameters and the high-fidelity data.  We include here the basic layout about the mutual information derivation, and refer the interested reader to \cite{Terejanu, Kraskov, Lewis} for additional details. 

The average level of uncertainty or information inherent in a random variable can be quantified by its Shannon entropy \cite{Shannon}. For a random variable $\Theta$ having a corresponding density $p(\theta)$ for $\theta\in \Omega$, the Shannon entropy is defined as 
\begin{eqnarray}
H(\Theta) = - \int_{\Omega} p(\theta)\log(p(\theta))d\theta \nonumber
\end{eqnarray}
for the prior distribution and
\begin{eqnarray}
H(\Theta | D_{n-1}) = -\int_{\Omega} p(\theta | D_{n-1})\log(p(\theta | D_{n-1}))d\theta \nonumber
\end{eqnarray}
for the posterior distribution, given data $D_{n-1}$.  Since we wish to determine the gain in information that results from incorporating an additional high-fidelity data point $\tilde d_n$, we define our utility function to be the following difference:
\begin{eqnarray}
U(d_n,\xi_n) =
\int_{\Omega} p(\theta | d_n, D_{n-1}) \log p(\theta | d_n, D_{n-1})d\theta  \nonumber \\
- \int_{\Omega} p(\theta | D_{n-1})\log p(\theta | D_{n-1})d\theta.
\label{eq:util_eqn}
\end{eqnarray}
Note that in Equation \eqref{eq:util_eqn}, we use the low-fidelity model prediction $d_n$ as an estimate in place of the high-fidelity output $\tilde d_n$, as we cannot know the value of this high-fidelity output without conducting a potentially expensive experiment.

To compute the average information gain resulting from the contribution of the experimental design $\xi_n$, we marginalize over the full set of all unknown future observations, $\mathcal{D}$, to obtain
\begin{eqnarray}
\mathbb{E}_{d_n}[U(d_n,\xi_n)] = \int_{\cal{D}} U(d_n,\xi_n)p(d_n | D_{n-1},\xi_n)d d_n.
\label{eq:expec}
\end{eqnarray}
By substituting Equation \eqref{eq:util_eqn} into \eqref{eq:expec} and simplifying, we obtain
\begin{eqnarray}
\mathbb{E}_{d_n}[U(d_n,\xi_n)]  
&=& \int_{\mathcal{D}} \int_{\Omega} p(\theta, d_n | D_{n-1},\xi_n) \log \frac{p(\theta, d_n | D_{n-1},\xi_n)}{p(\theta | D_{n-1})p(d_n | D_{n-1},\xi_n)} d\theta d d_n \nonumber \\ \nonumber \\
&=& I(\theta; d_n | D_{n-1}, \xi_n),
\label{eq:mi_eqn}
\end{eqnarray} 
which we define to be the mutual information between the low-fidelity model parameters, $\theta$, and the high-fidelity data collected at design $\xi_n$. Essentially, this is a measure of parameter uncertainty reduction---the larger the mutual information, the more knowledge we expect to gain about the low-fidelity model parameters by collecting data at that experimental design. Thus, we optimize over the set of all available design conditions, choosing the one that maximizes this quantity:
\begin{eqnarray}
\xi_n^* = \arg \max_{\xi_n\in \Xi} I(\theta; d_n | D_{n-1},\xi_n).\nonumber
\end{eqnarray}
Having chosen $\xi_n^*$, we evaluate the high-fidelity model at this point, $\tilde d_n = d_h(\xi_n^*)$, append $\tilde d_n$ to the data set $D_{n-1}$, and re-calibrate the low-fidelity model parameters. For this investigation, we use the Delayed Rejection Adaptive Metropolis (DRAM) algorithm for our model calibration step---see \cite{Haario, Smith, Lewis} for additional details.  Once the low-fidelity model parameters have been re-calibrated, this procedure is repeated in full until either (a) the budget of high-fidelity model evaluations has been exhausted or (b) the uncertainty in the low-fidelity model parameters has been reduced below a user-defined threshold.

We note that Equation \eqref{eq:mi_eqn} often cannot be evaluated directly and may be prohibitively expensive to compute via numerical integration.  As such, we utilize the $k$th-Nearest-Neighbor ($k$NN) estimate of mutual information described fully in \cite{Kraskov, Lewis}. We detail our $k$NN estimate procedure in Appendix \ref{app:kNN_est}.

\subsection{Modified mutual information for time series data}
\label{sec:MI_time_series}

When collecting clinical data to determine a tumor's response to treatment, we consider the set of available design conditions to be a series of time steps at which data can be collected sequentially. We note that it is critical to collect measurements at multiple time points to achieve an accurate calibration of the model. Thus, we adapt the calibration framework described in the previous subsection to apply to time-series data. We present this adaptation using the one-compartment model as the low-fidelity model, and note that it can be extended to the two-compartment model by specifying the type of data to be collected (i.e., total tumor volume versus necrotic volume), in addition to the time that it is to be collected.

Let $t_1,t_2,\hdots, t_{n_T}$ denote the set of ${n_T}$ times at which high-fidelity data can be collected, with $t_i<t_j$ for $i<j$. We use $\tilde d(t_i)$ to denote the value of the high-fidelity data, e.g.~the tumor volume as evaluated by the CA model, collected at time $t_i$; the low-fidelity model prediction at $t_i$ is denoted by $d(t_i)$. Suppose that in the previous step of the algorithm, the data point $\tilde d(t_r)$ was chosen to append to the data set, which we denote by $D_r$. When choosing the next data point to add to this data set, for all $i>r$, we let $I(\theta; d(t_i)|D_r)$ denote the mutual information between the low-fidelity model parameters, $\theta$, and the high-fidelity data collected at time $t_i$, as defined in the general setting in Equation \eqref{eq:mi_eqn}.

In some cases, data collected at later time points may provide more overall mutual information, but choosing such a data point then precludes the subsequent collection of data at any previous time. In order to account for this trade-off, we define a score function based on mutual information, which rewards the user for selecting a point with a large mutual information, while penalizing the user for losing the information that could have been collected from the skipped time points. Before presenting this score function, we first define the mutual information relative to the maximum mutual information at a given step in the algorithm. Let $I^*_r$ denote the maximum possible mutual information in the step after data point $\tilde d(t_r)$ has been chosen, defined as follows:

\begin{equation*}
    I^*_r = \max_{i>r}I(\theta; d(t_i)|D_r).
\end{equation*}

Next we define the corresponding relative mutual information provided by each data prediction $d(t_i)$ by

\begin{equation*}
    R(i,r) = \frac{I(\theta; d(t_i)|D_r)}{I^*_r}.
\end{equation*}

Note that $0\leq R(i,r)\leq 1$ for all $i>r$.
Using this notation, we are now ready to define our score function $S_k(i,r)$, which combines the relative mutual information obtained from choosing data point $d(t_i)$, with a penalty term that captures skipped information. We use the parameter $k$ to vary the weight of the penalty term. We define the score function as follows,

\begin{equation}
    S_k(i,r) = R(i,r) -k\left(\frac{\displaystyle\sum_{j=r+1}^{i-1}{R(j,r)}}{\displaystyle\sum_{l=r+1}^{n_T}{R(l,r)}}\right),
    \label{eq:scorefxn_eqn}
\end{equation}
where the sum in the numerator denotes the total amount of relative mutual information that will be lost by choosing data point $\tilde d(t_i)$, and the sum in the denominator denotes the total relative mutual information from all data points that may collected after $\tilde d(t_r)$.  

In Sections \ref{sec:simulation_one}--\ref{sec:twocomp_both}, we use the score function $S_k(i,r)$ to determine the optimal high-fidelity data points to collect for use in low-fidelity model calibration. At each step of the model calibration algorithm, we choose the data point that maximizes $S_k(i,r)$, that is, $\tilde d(t_{i^*})$ such that $i^* = \arg\max_{r\leq i\leq n_T} S_k(i,r)$. We test this process multiple times, varying the value of the penalty weight parameter, $k$, between 0 and 1, in order to investigate the effect of $k$ on the temporal data collection sequence. Note that $k=0$ is equivalent to using solely the mutual information to determine the next design choice. 

\section{Simulation Results}
\label{sec:simulation} 

In the following section, we outline the results of our algorithm applied to a variety of scenarios. First, we apply our algorithm in a simplified setting where we assume that the clinician has the ability to collect one scan per week, and look for a pattern in the days chosen for scanning.  Then, we extend the use of the algorithm to scenarios in which $n$ scans can be collected at any days during the treatment cycle, and investigate how the algorithm performs for both the one- and two-compartment models for three radiosensitivity levels: high, medium, and low. In all of the following scenarios, we fix the pre-treatment parameters ($A$ and $B$ in the one-compartment model, and $\lambda$, $K$, $\eta$, and $\zeta$ in the two-compartment model) at the values listed in Appendix \ref{app:params}, assuming that these parameters have been identified prior to the start of the treatment regimen.  Additionally, we fix $\alpha = 0.14$ to avoid identifiability issues, as discussed in \cite{Cho}.  Thus, our focus in this investigation is in determining the value of radiosensitivity parameter $\beta$ as quickly and accurately as possible, so as to increase the predictive power of our models and allow for the alteration or discontinuation of a treatment protocol that is predicted to be ineffective.

\subsection{Scenario 1: Collecting one scan per week}
\label{sec:oneScanPerWeek} 

\begin{figure}[!htb]
\centerline{ 
	\includegraphics[width=8cm]{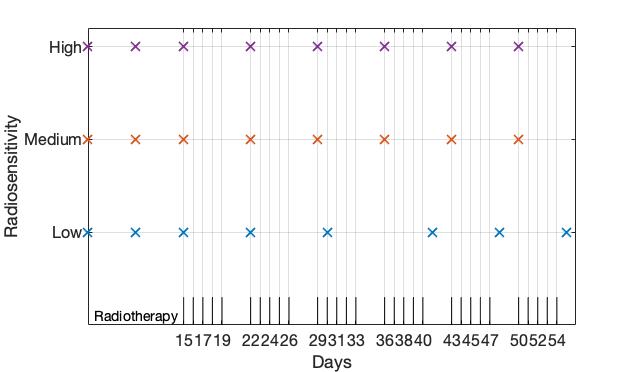}
	}
	\caption{%
	Budget of one scan per week. The scan choices are displayed, with the high radiosensitivity case shown in the upper row in purple, the medium radiosensitivity case shown in the middle in red, and the low radiosensitivity case shown in the lower row in blue. 
	} 
	\label{fig:OneScanPerWeek} 
\end{figure}

Frequently in the clinical setting, tumor data collection is constrained to a strict budget due to limited resources. As an example scenario, we consider the case in which one tumor volume scan can be taken per week during the weeks in which treatment is administered. In addition to assuming that data has been collected prior to the start of treatment at days 5 and 10, we automatically provide the scan for day 15 (day one of treatment week one) to obtain an initial fit for parameter $\beta$, and then enforce a budget of one scan per week in our mutual information framework with the one-compartment ODE as the low-fidelity model, to determine the optimal day to collect weekly data in weeks 2--6. We complete this procedure for three radiosensitivity levels: high, medium, and low. Table \ref{table:OneScanPerWeek} shows the optimal days chosen for each radiosensitivity level, with the day number in the weekly treatment cycle indicated in parentheses (i.e.~1 corresponds to Monday, the first day of treatment each week, and 6 corresponds to Saturday, the day after five sequential days of treatment). Figure \ref{fig:OneScanPerWeek} provides a visualization of the results summarized in the table. 

\begin{figure}[!htb]
\centerline{ \footnotesize \textsf{High Radiosensitivity} \hspace{2cm} \textsf{Medium Radiosensitivity}  \hspace{2cm} \textsf{Low Radiosensitivity} }
\centerline{ 
	\includegraphics[width=1.9in]{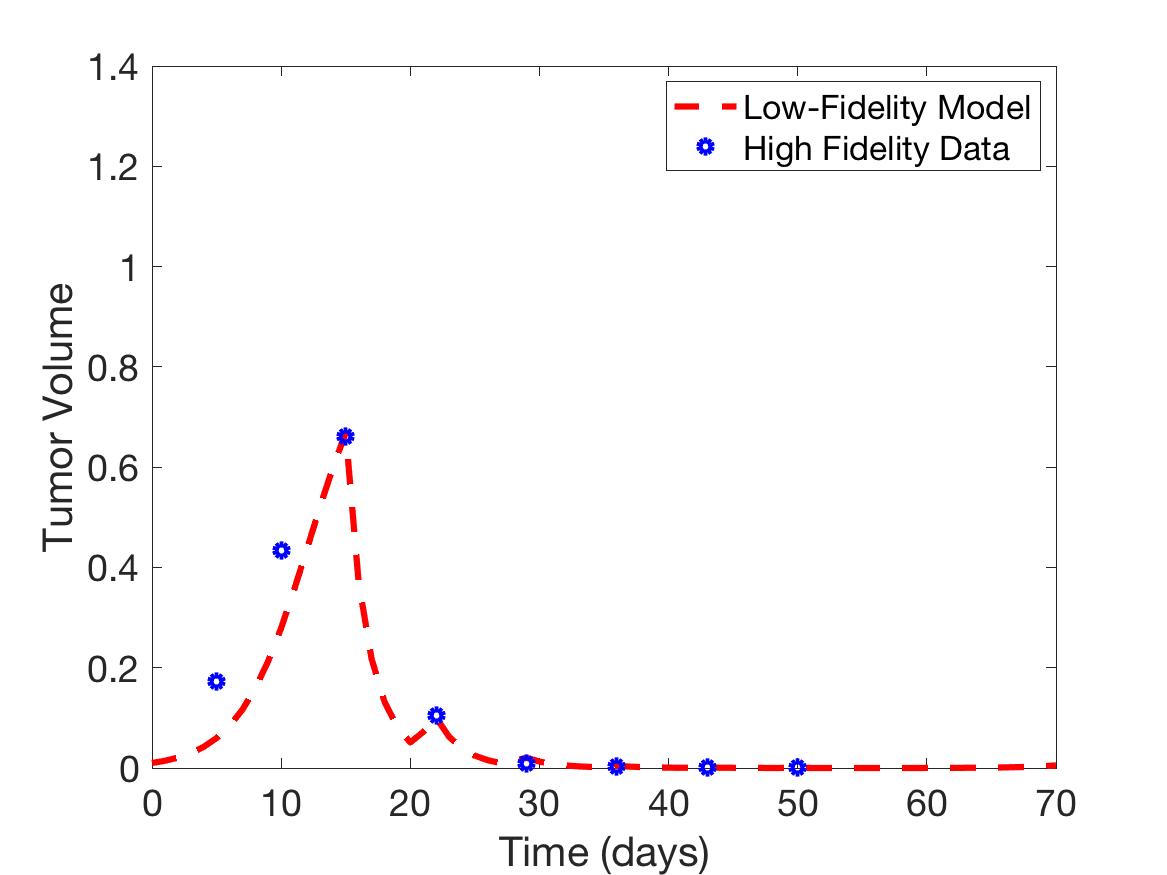}
	\includegraphics[width=1.9in]{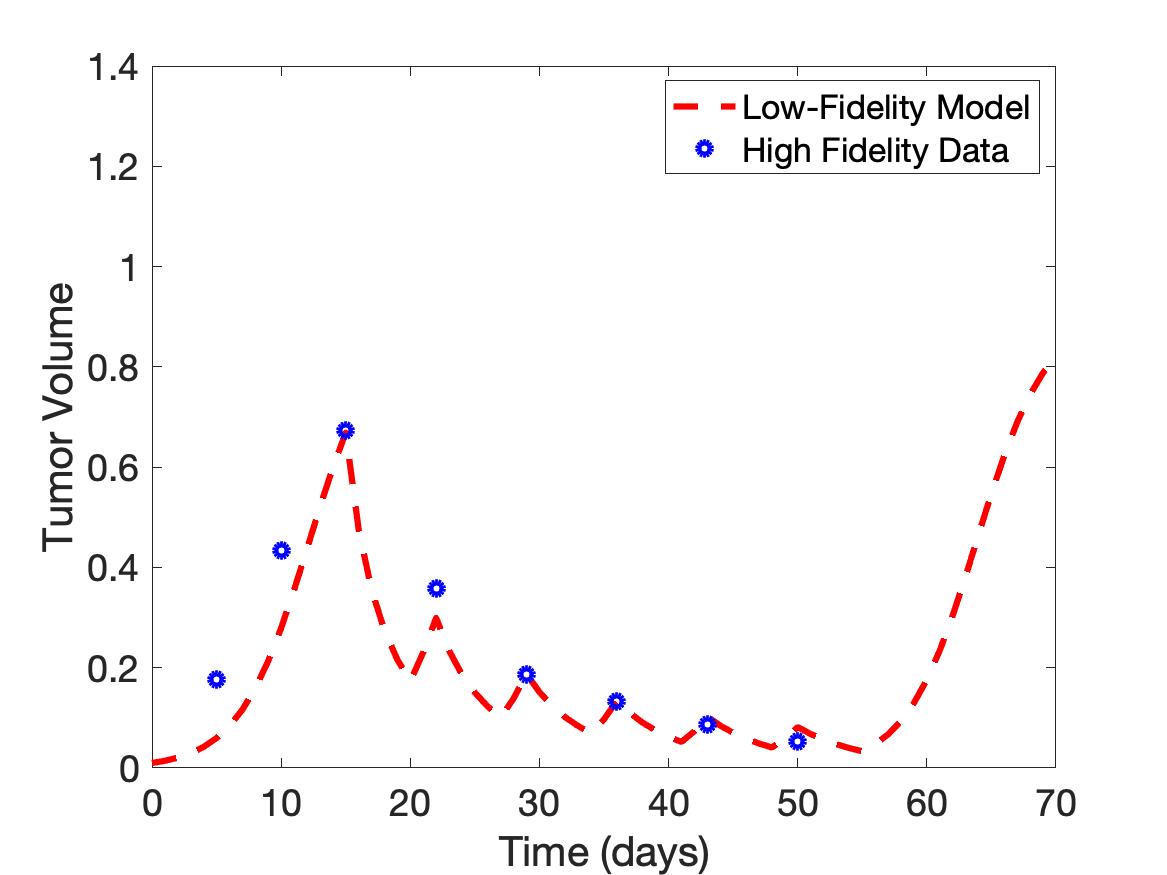}
	\includegraphics[width=1.9in]{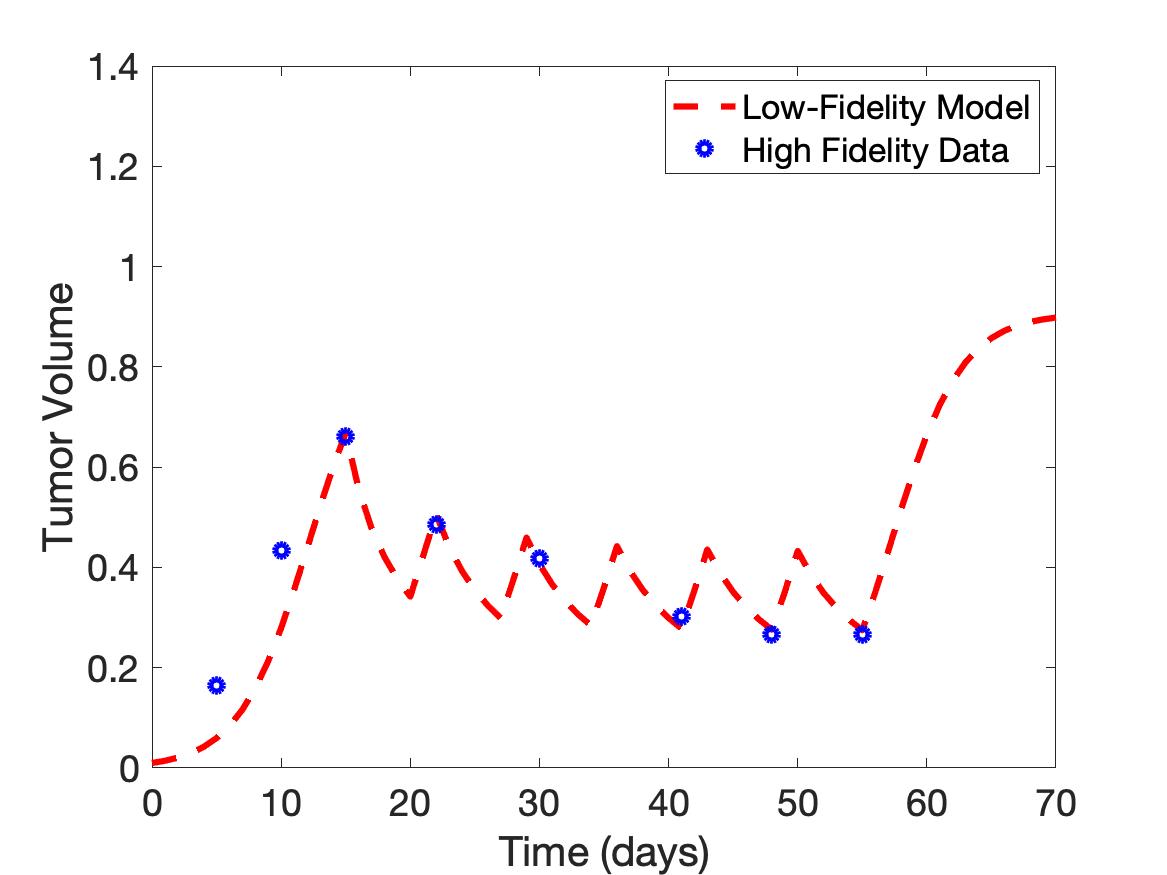}
	}
\caption{Budget of one scan per week. The fitted models are shown after calibrating with one data point per week, for all six treatment weeks. All three radiosensitivity cases are shown, with high radiosensitivity on the left, medium radiosensitivity in the middle, and low radiosensitivity on the right.} 
	\label{fig:OnePerWeek_fits} 
\end{figure}

We observe that in both the high and medium radiosensitivity cases, the first day of each weekly treatment cycle consistently provides the highest level of information for parameter calibration. However, we note that in the high radiosensitivity case, the tumor volume is nearly zero after a the first few treatment weeks, so during the last three cycles, the day 1 scan provides just slightly more mutual information than the other available scans. Figure \ref{fig:OnePerWeek_fits} displays the final model fits for each radiosensitivity case, calibrated using all eight selected scans. In the low radiosensitivity case, the optimal choices are the first day of the treatment cycle in week 2 and the second day of the cycle in week 3---we note that in week 3, days 1 and 2 of the cycle provide nearly identical levels of mutual information. In the subsequent three weeks, day 6 (the Saturday after the treatment cycle ends) provides the highest level of mutual information. Our results suggest that in the second half of the treatment protocol for this low radiosensitivity case, it becomes most important to assess the full extent of the tumor reduction from each week of radiation doses, encoded in the scan collected on day 6 of each week.

 If a clinician has the resources to collect tumor volume data exactly once per week during treatment, then our results suggest that they should begin by collecting this data on the first day of each treatment cycle. If the tumor appears to respond well to the radiotherapy after the first few weeks, then they should continue to collect data on the first treatment day. However, if the tumor is responding slowly after three weeks---for instance, if nearly half of the pre-treatment tumor remains, as in our low radiosensitivity case---then the methodology suggests switching data collection to the end of the week. In particular, it would be optimal in such a scenario to conduct measurements on Saturday, the day following the five-day treatment cycle, for the final three weeks, to provide maximal information for model calibration.

%Results from 1-scan per week run (completed 6/29/2020)
\begin{table}[htbp]
    \centering
\begin{tabular}{| c | c | c | c | c | c | c | c |}
\hline
 & Initial & Week 1 & Week 2 & Week 3 & Week 4 & Week 5 & Week 6 \\
\hline
\textbf{High} & \textcolor{blue}{5 \ 10} & \textcolor{blue}{15}   & 22 (1) & 29 (1) & 36 (1) & 43 (1) & 50 (1)\\
\hline
\textbf{Medium} & \textcolor{blue}{5 \ 10} & \textcolor{blue}{15}  & 22 (1) & 29 (1) & 36 (1) & 43 (1) & 50 (1)\\
\hline
\textbf{Low} & \textcolor{blue}{5 \ 10} & \textcolor{blue}{15}  & 22 (1) & 30 (2) & 41 (6) & 48 (6) & 55 (6)\\
\hline

\end{tabular}
\caption{Budget of one scan per week. This table shows the scan choice for high, medium, and low radiosensitivity levels. The day since the start of the simulation is shown, along with the day of each weekly treatment cycle indicated in parentheses. Note that in all cases, we assume that data at days 5, 10, and 15 is already available, so as to avoid the use of improper priors in our parameter calibration. The other scan days were chosen using the mutual information calibration procedure.}
\label{table:OneScanPerWeek}
\end{table}

 \ \\

\subsection{Scenario 2: One-compartment model with $n$ number of scans}
\label{sec:simulation_one} 

In this section, we explore the possibility of designing a more effective scanning schedule to calibrate the one-compartment model, given by Equation \eqref{eqn:eqn1comp}. We use our proposed score function, \eqref{eq:scorefxn_eqn} in Section \ref{sec:MI_time_series}, to determine the scanning schedule. In particular, we study how the suggested scanning schedule changes for different values of the score function parameter $k$ in Equation \eqref{eq:scorefxn_eqn}. 
Moreover, we aim to find an optimal $k$  for each radiosensitivity cases with respect to the accuracy of the model calibration. To quantify the accuracy of the calibration to data, we define 

\begin{equation} 
    \textrm{error} \doteq \frac{\| \bm d_h - \bm d_l \|_2 }{\| \bm d_h \|_2}, 
    \label{eq:error} 
\end{equation} 
where $\bm d_h = [d_h(t_1), ..., d_h(t_{n_T})]$ is the vector of high-fidelity data at all days during the treatment period, and $\bm d_l = [V(t_1), ..., V(t_{n_T})]$ is the low-fidelity model approximation evaluated at those same days. The error includes all the data points up to the final time $n_T$ so that it can be regarded as a measure of predictive power as well. 

\begin{figure}[!htb]
\centerline{ 
	\includegraphics[width=1.9in]{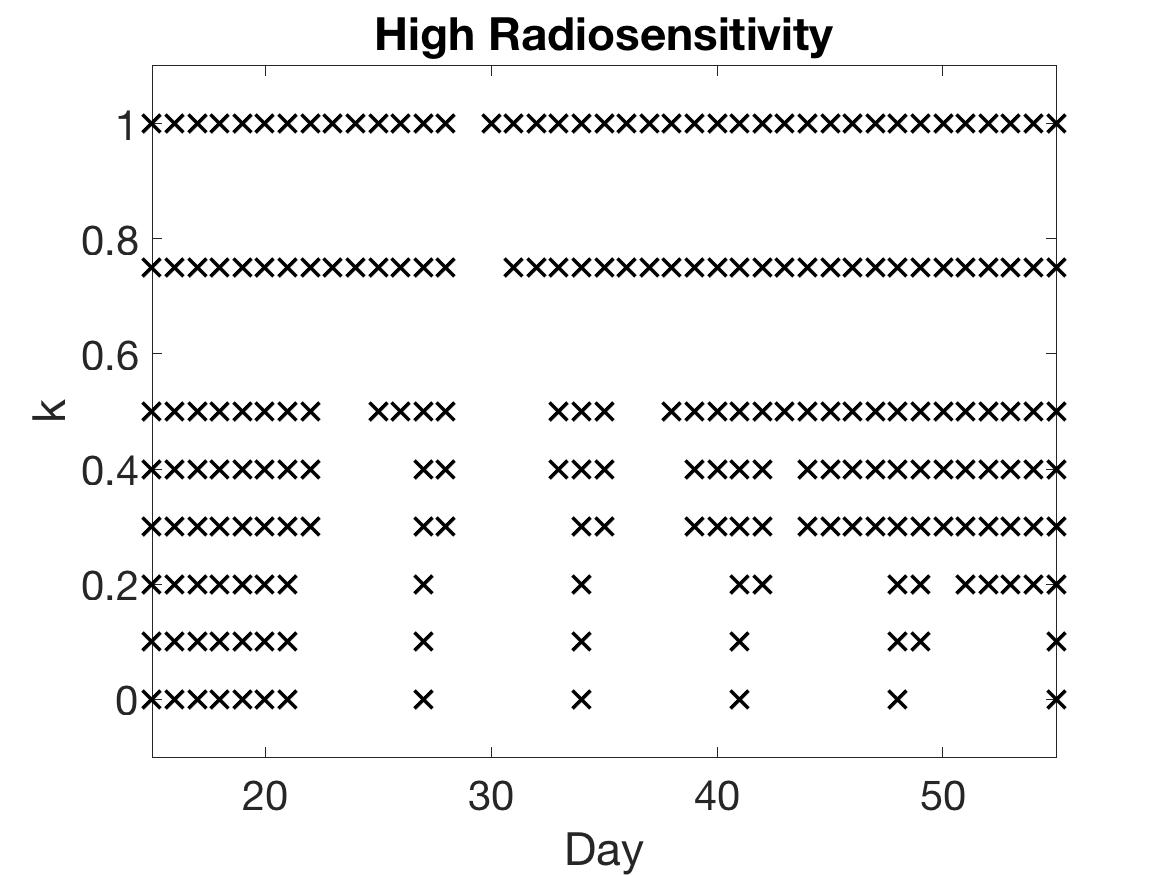}
	\includegraphics[width=1.9in]{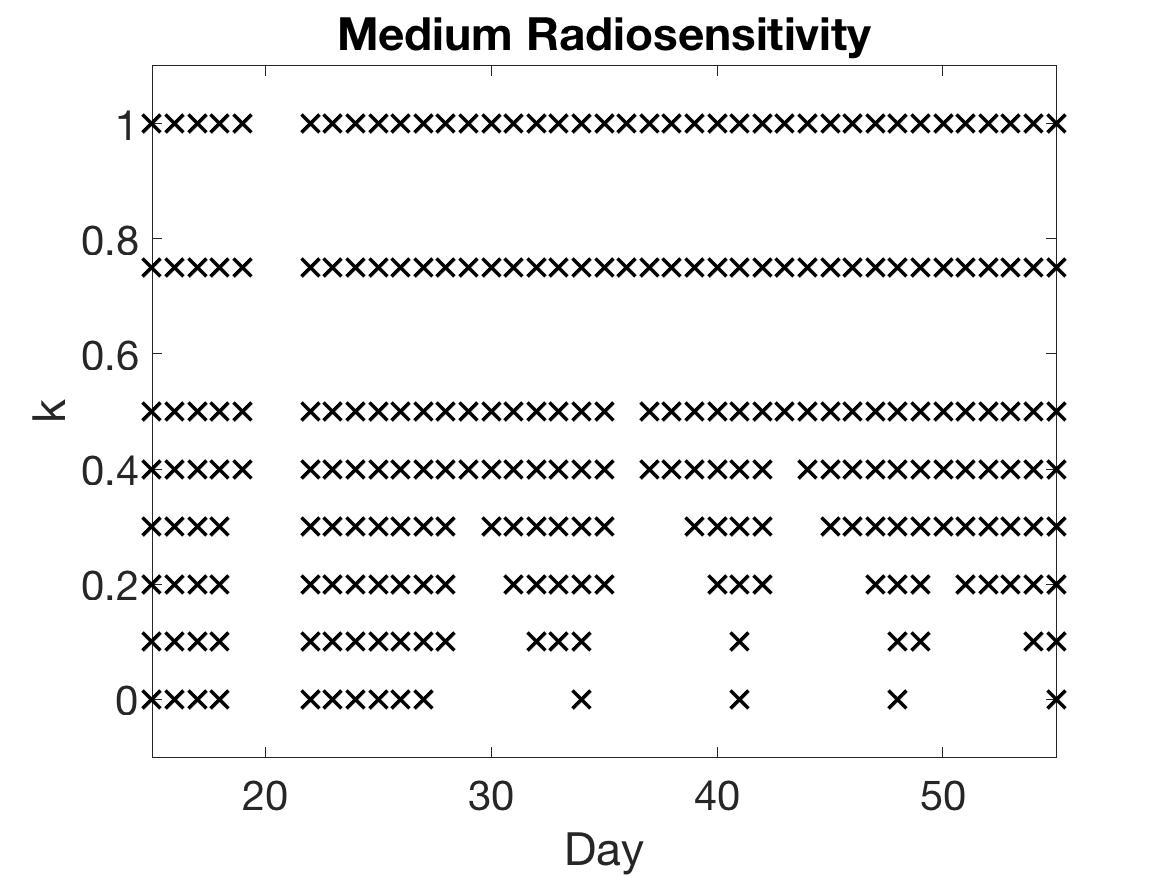}
	\includegraphics[width=1.9in]{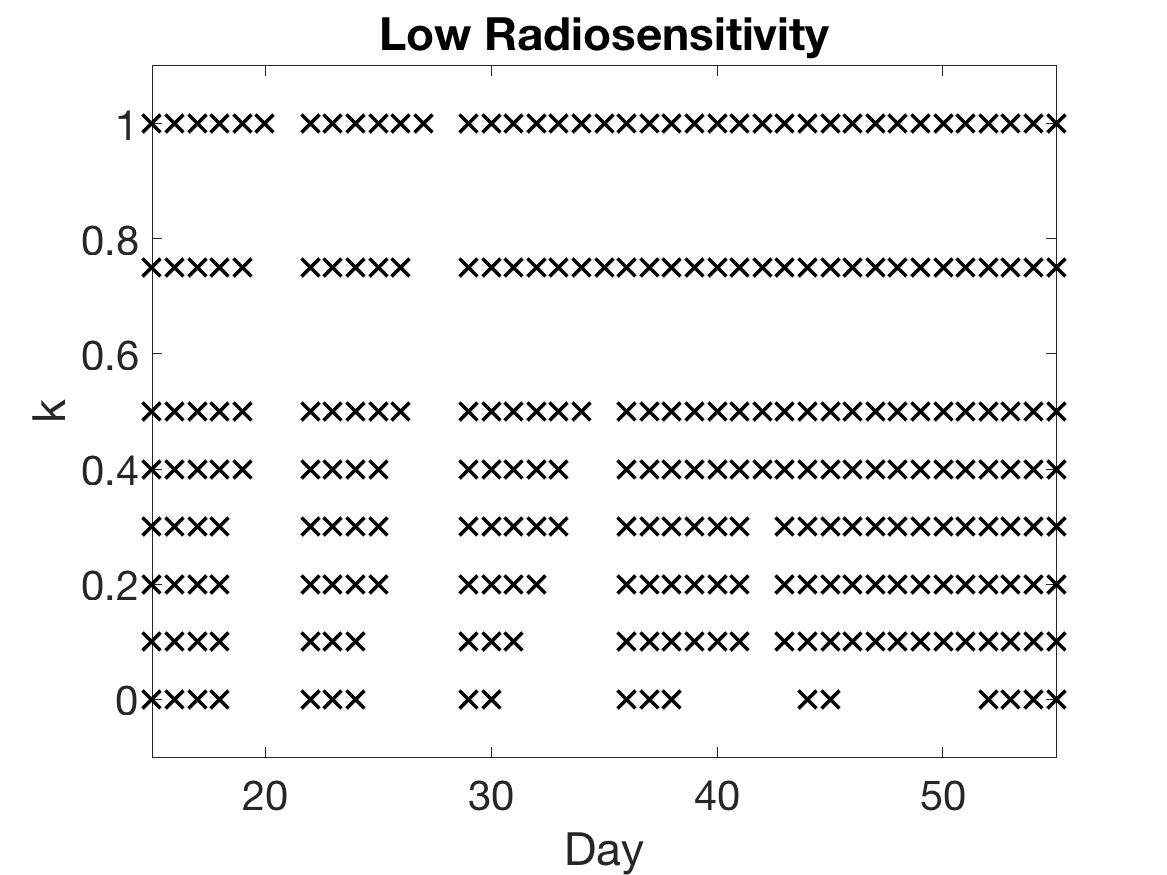}
	}
\caption{One-compartment model. Choice of scan ($\times$) for different values of score function parameter $k = 0, 0.1, ..., 0.5, 0.75, 1$. We observe that using larger values of parameter $k$ tends to favor the choice of earlier time points, since the choice of point is more heavily penalized for skipping over days. On the other hand, smaller values of $k$ allow the algorithm to skip over earlier time points in favor of gathering points with larger mutual information towards the end of the treatment period.  } 
	\label{fig:scanChoice_onecomp} 
\end{figure}
\begin{figure}[!htb]
\centerline{ \rotatebox{90}{\footnotesize \hspace{1.1cm} $S_{0}(i,r)$}
	\includegraphics[width=1.7in]{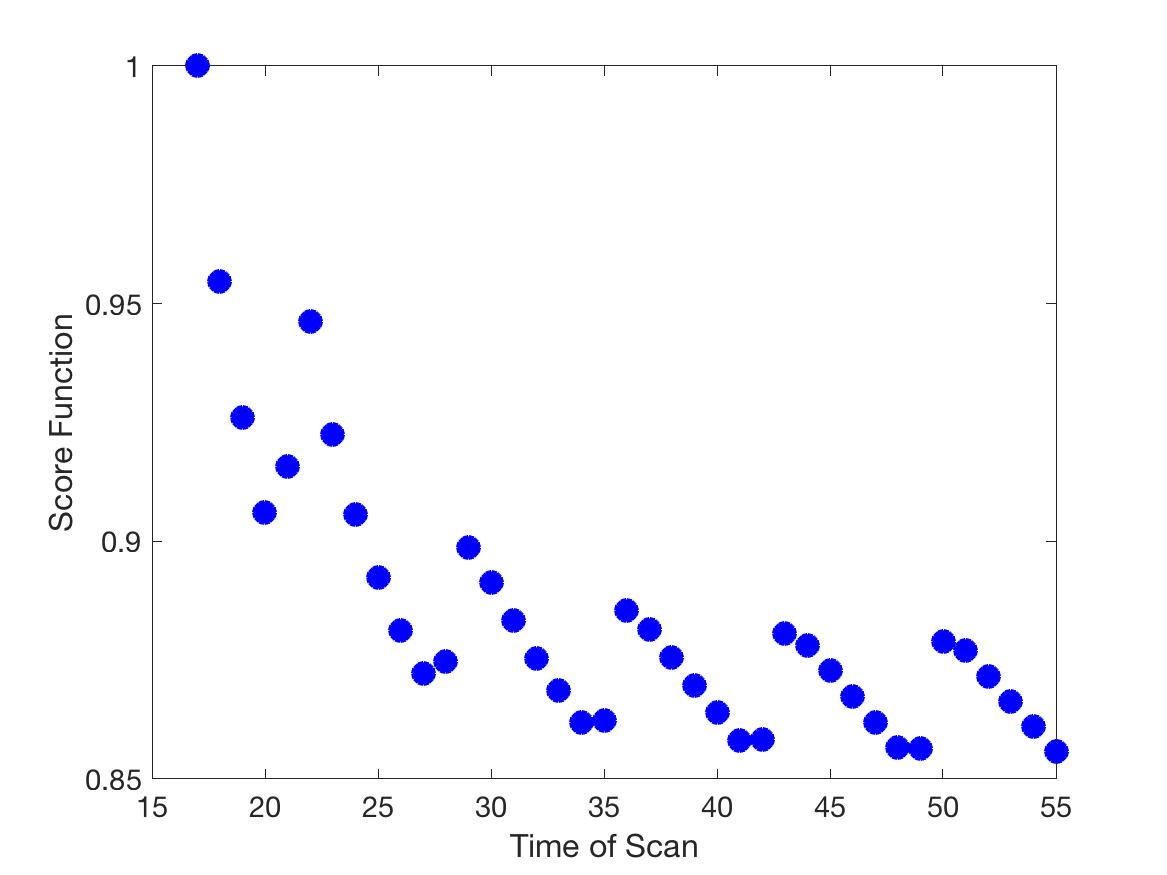}
	\includegraphics[width=1.7in]{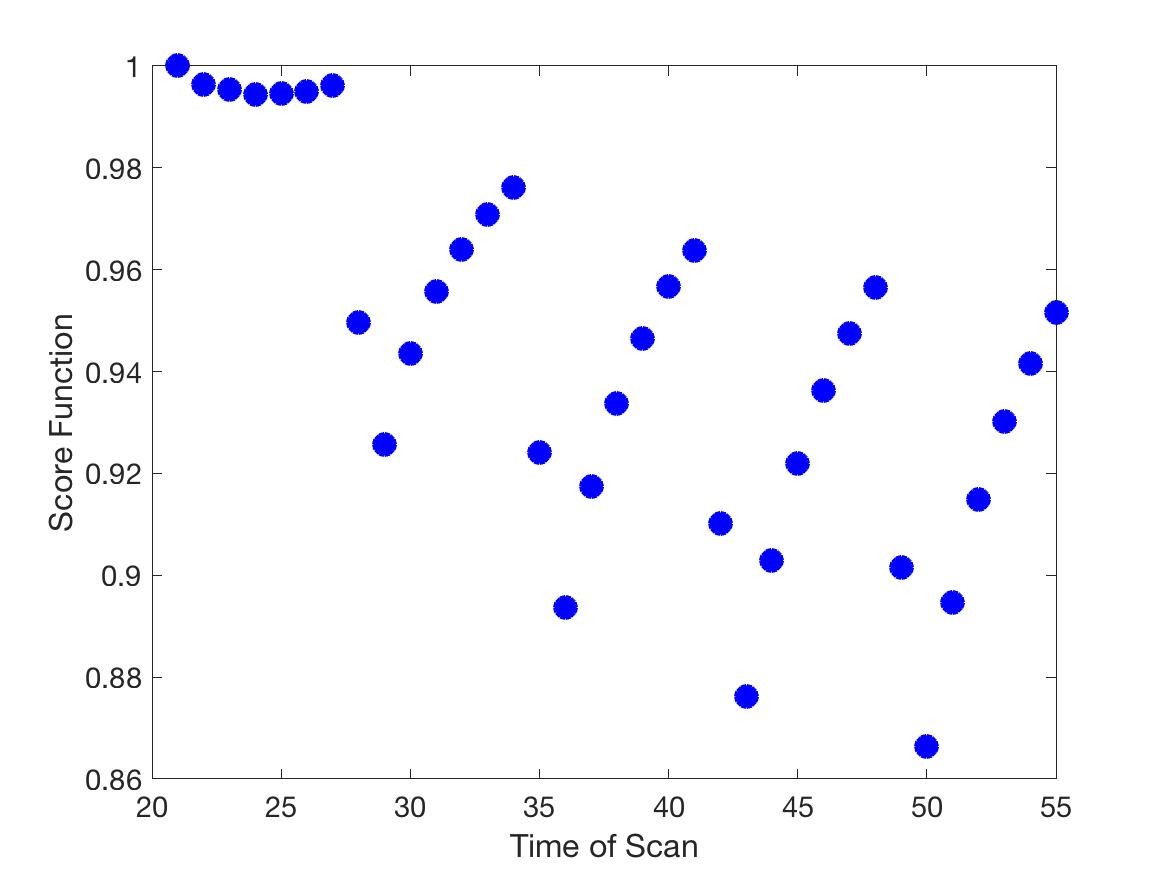}
	\includegraphics[width=1.7in]{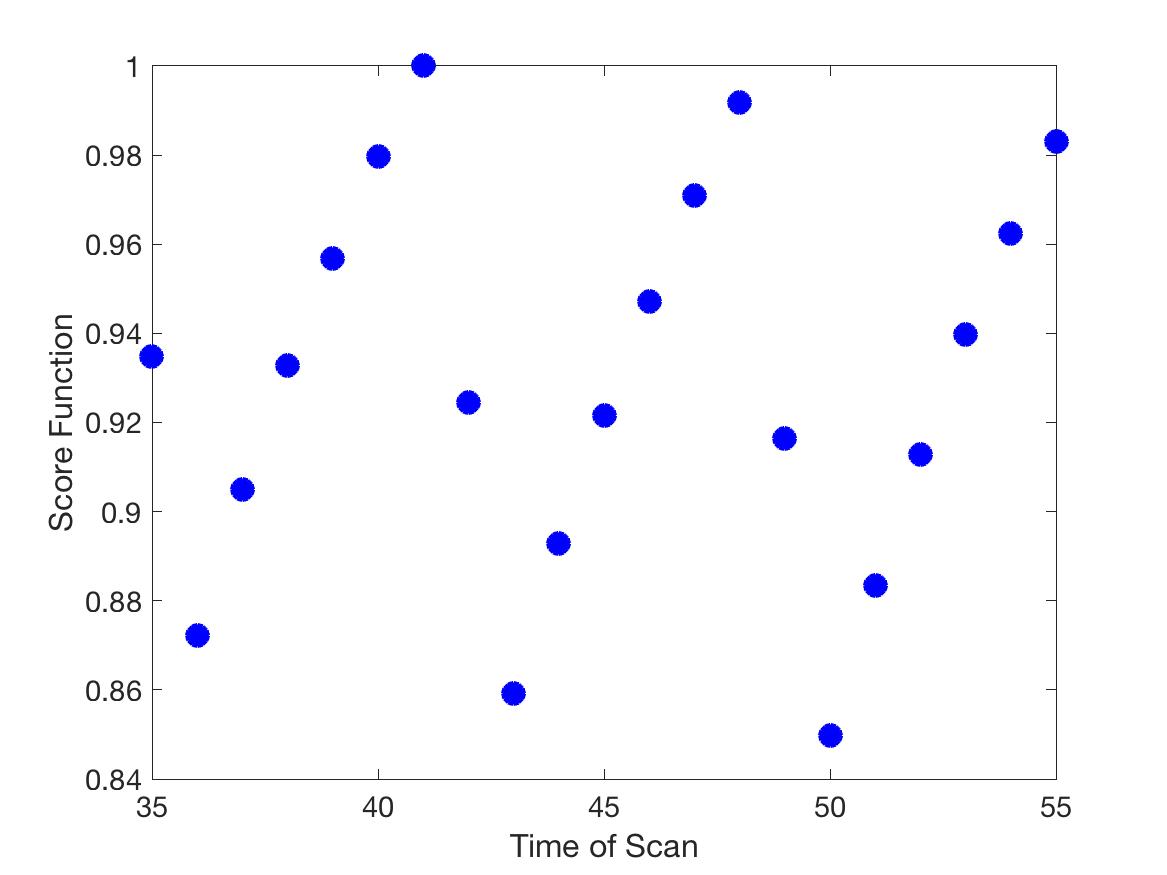}
	}	
\centerline{ \rotatebox{90}{\footnotesize \hspace{1.1cm} $S_{0.5}(i,r)$ }
	\includegraphics[width=1.7in]{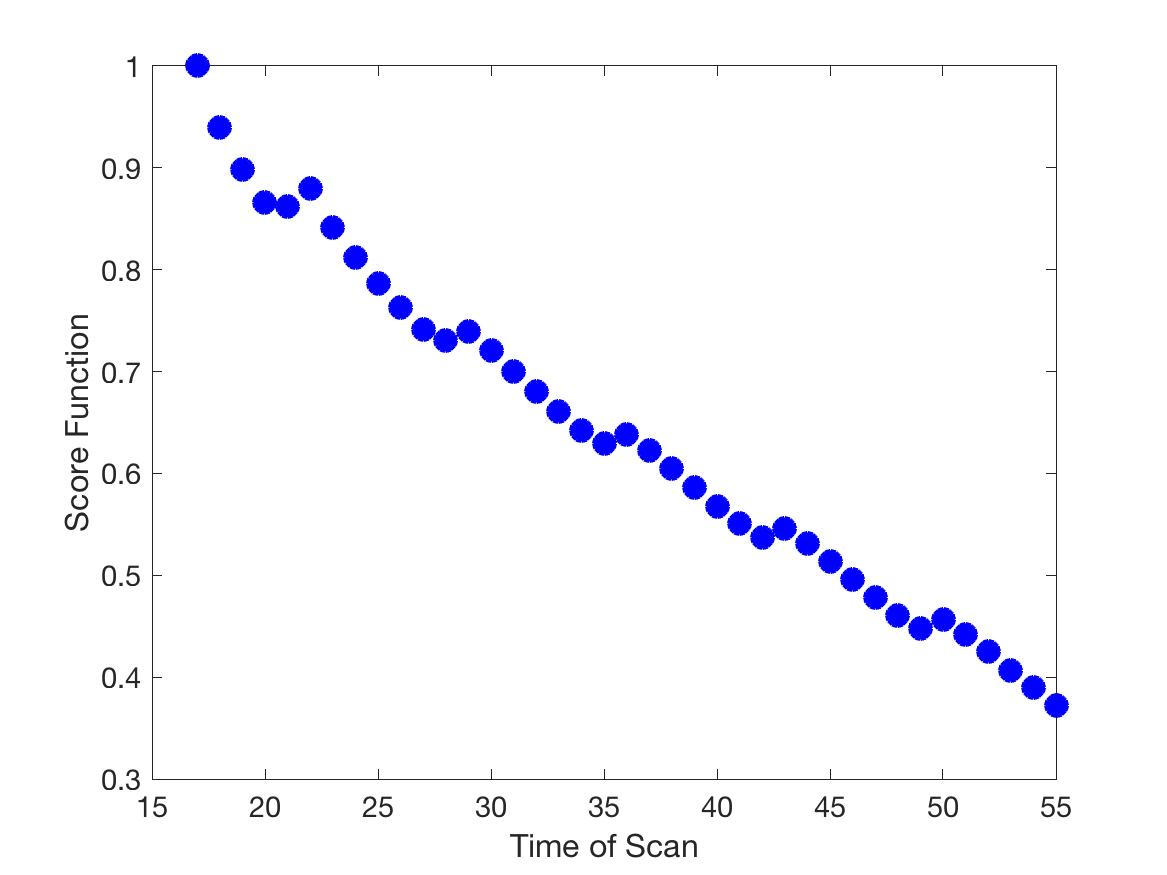}
	\includegraphics[width=1.7in]{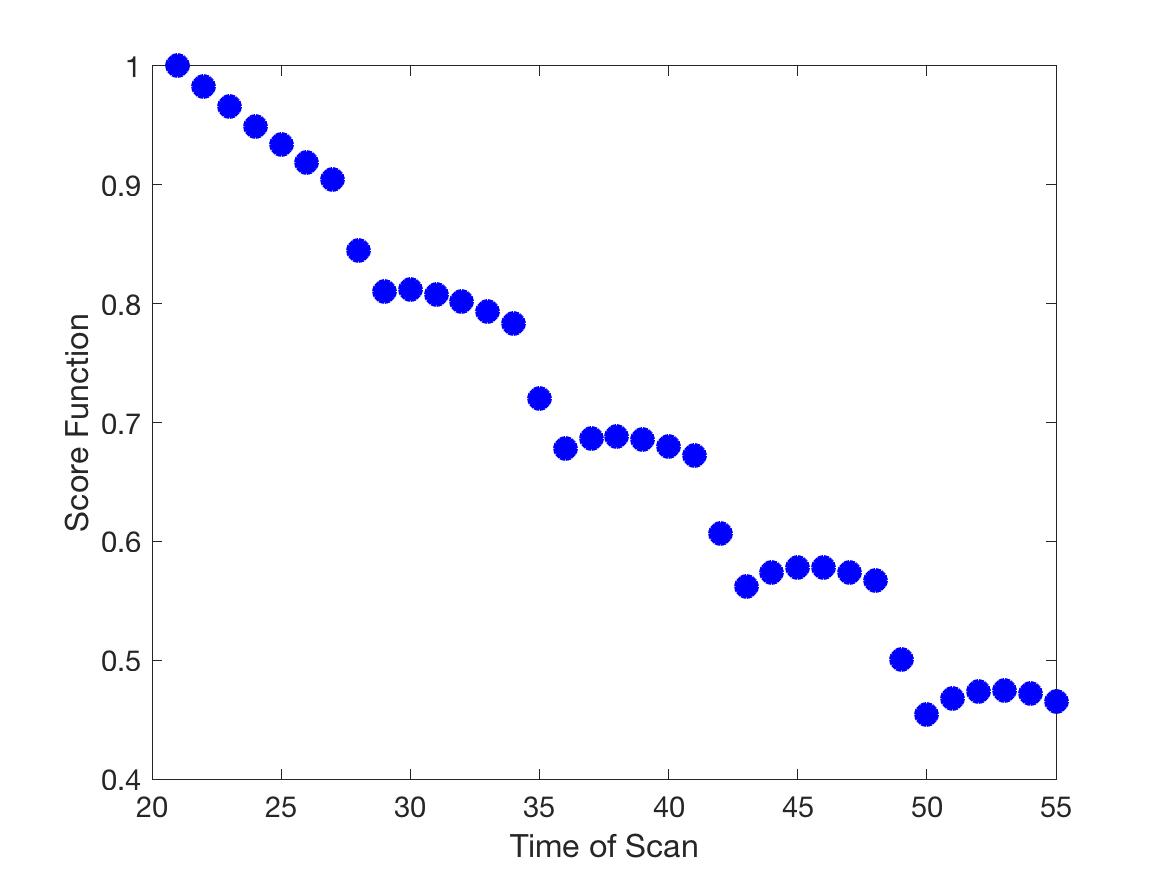}
	\includegraphics[width=1.7in]{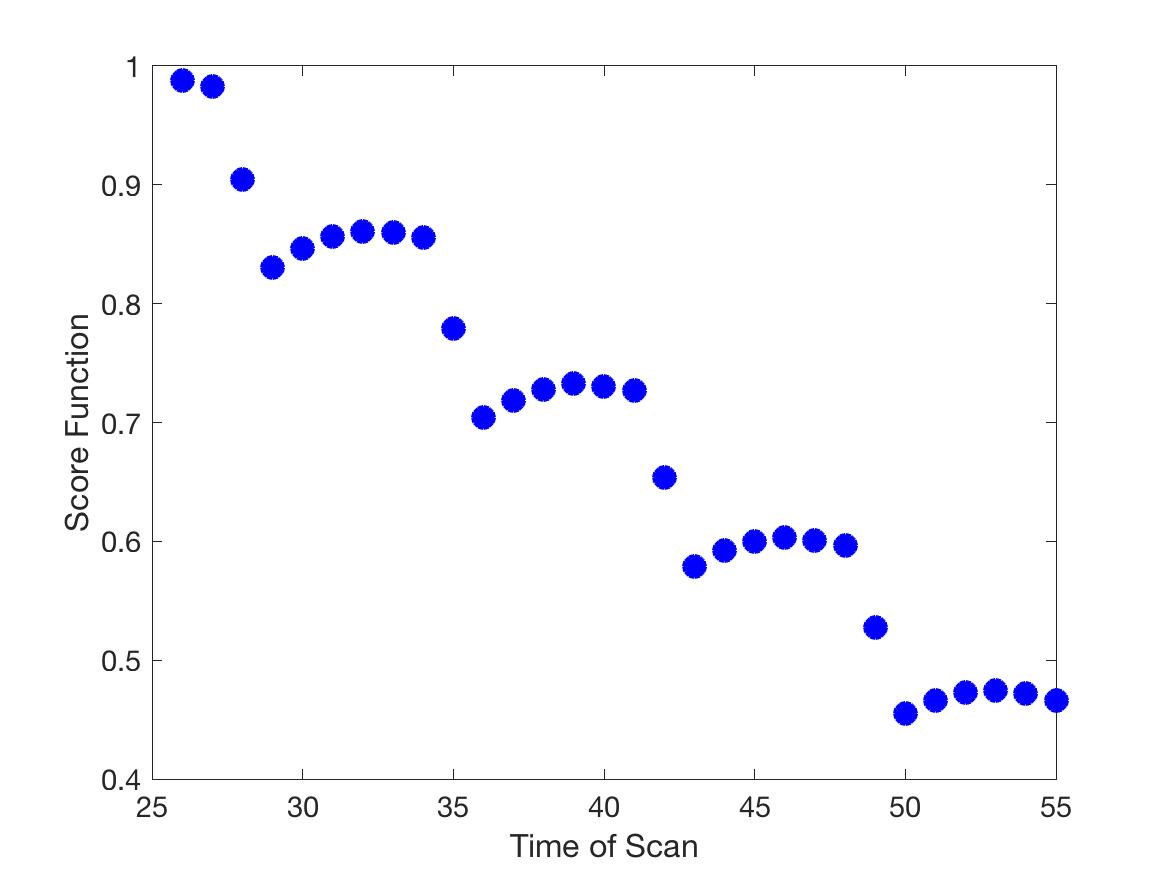}
	}
\caption{One-compartment model for high radiosensitivity case. Score function $S_k(i,r)$ \eqref{eq:scorefxn_eqn} with respect to $i\in [r+1, n_T]$ for different values of score function parameter $k = 0$ and $k=0.5$, when choosing the third (left), seventh (middle), and tenth (right) scan. 
Larger values of $k$, for example, $k=0.5$ give more weight to earlier time points such that the score function values show a decreasing trend, especially shown in the left figure when choosing the third scan. In the case of $k=0$, the optimal scan choice will skip some of the earlier points. 
} 
	\label{fig:ScoreFtn_onecomp} 
\end{figure}
Figure \ref{fig:scanChoice_onecomp} compares the choice of scan for different values of score function parameter $k$. We observe that using larger values of $k$ tends to result in the inclusion of more scans from earlier time points, while small values of $k$ often skip those earlier points as a result of the smaller penalty for choosing later points. 
A sampling of score function values for $k=0$ and $k=0.5$, yielding such scan choices, are shown in Figure \ref{fig:ScoreFtn_onecomp}, for the case of high radiosensitivity. The displayed score function values correspond to choosing the third, seventh, and tenth scan. For $k=0$, which coincides with the standard mutual information of Equation \eqref{eq:mi_eqn}, we observe that the score function values vary widely within a single week. In particular, either the first or the last day of the week have large score function values, and thus will be selected. Therefore, by using this score function, the intermediate time points of scans will be skipped. However, by using our score function with $k=0.5$, more weight is given to earlier time points. This makes it possible to have more frequent earlier scan choices, as shown in Figure \ref{fig:scanChoice_onecomp}.

\begin{figure}[!htb]
\centerline{ 
	\includegraphics[width=1.9in]{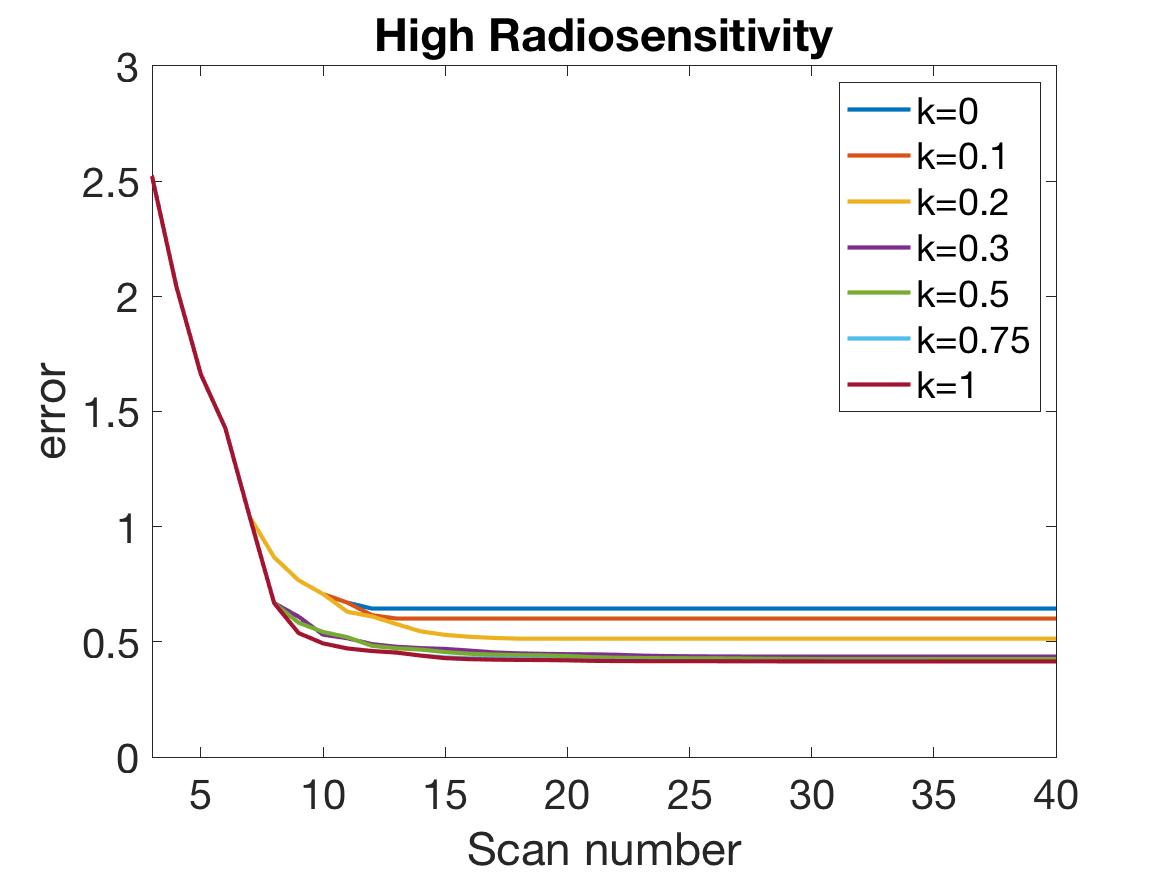}
	\includegraphics[width=1.9in]{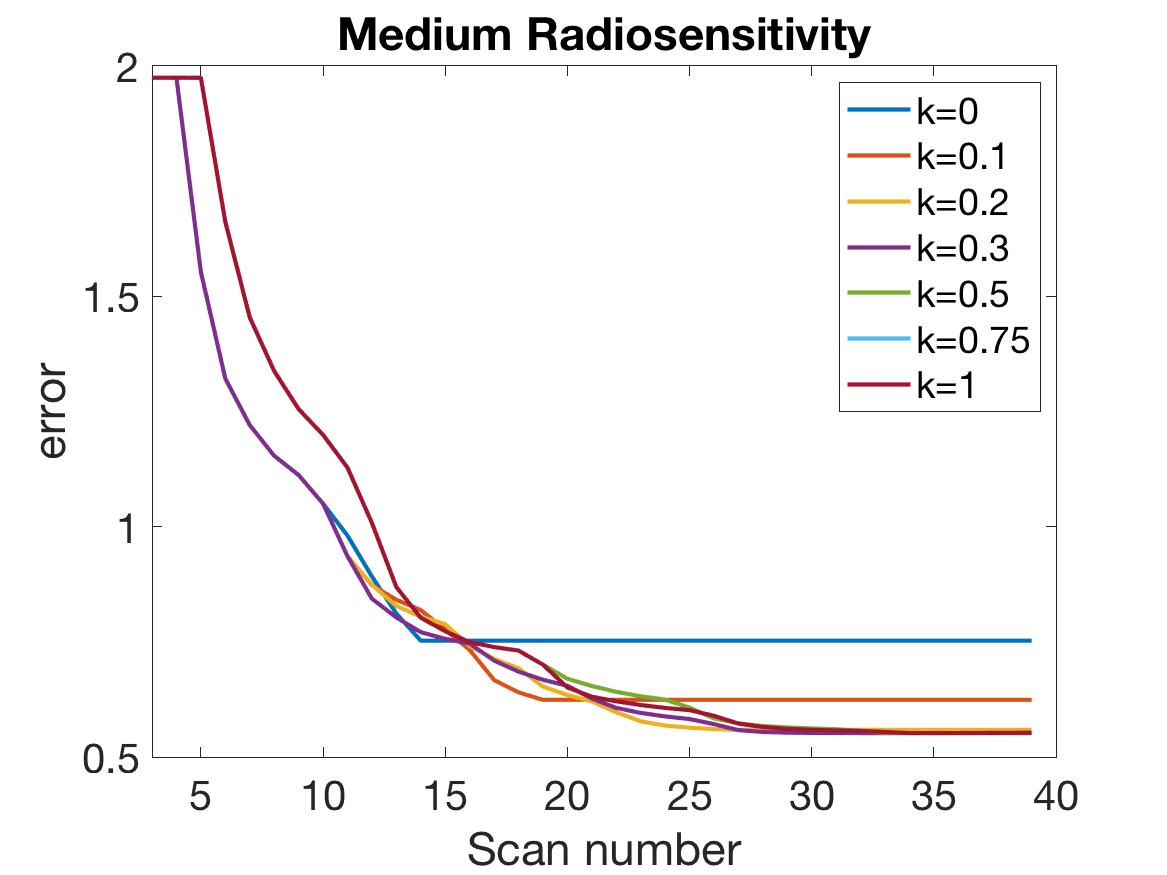}
	\includegraphics[width=1.9in]{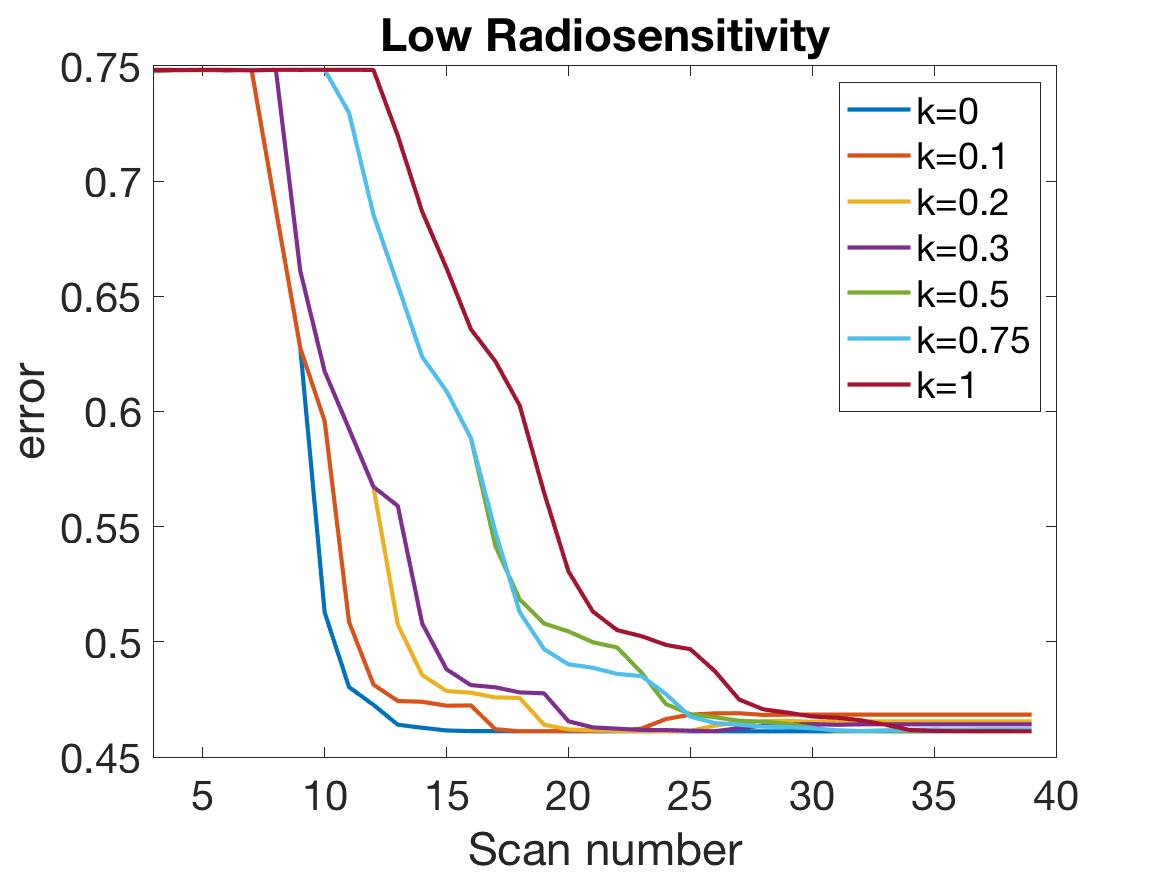}
	}
	\centerline{ 
	\includegraphics[width=1.9in]{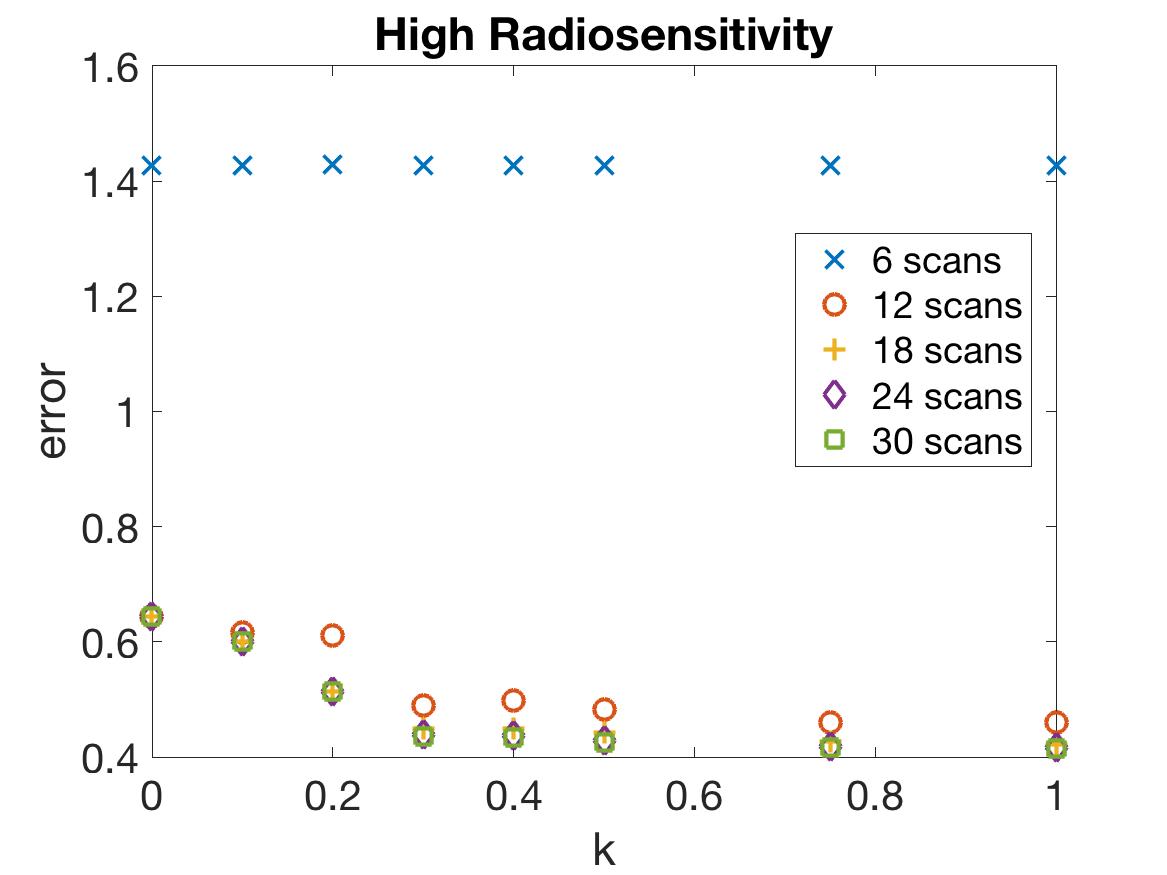}
	\includegraphics[width=1.9in]{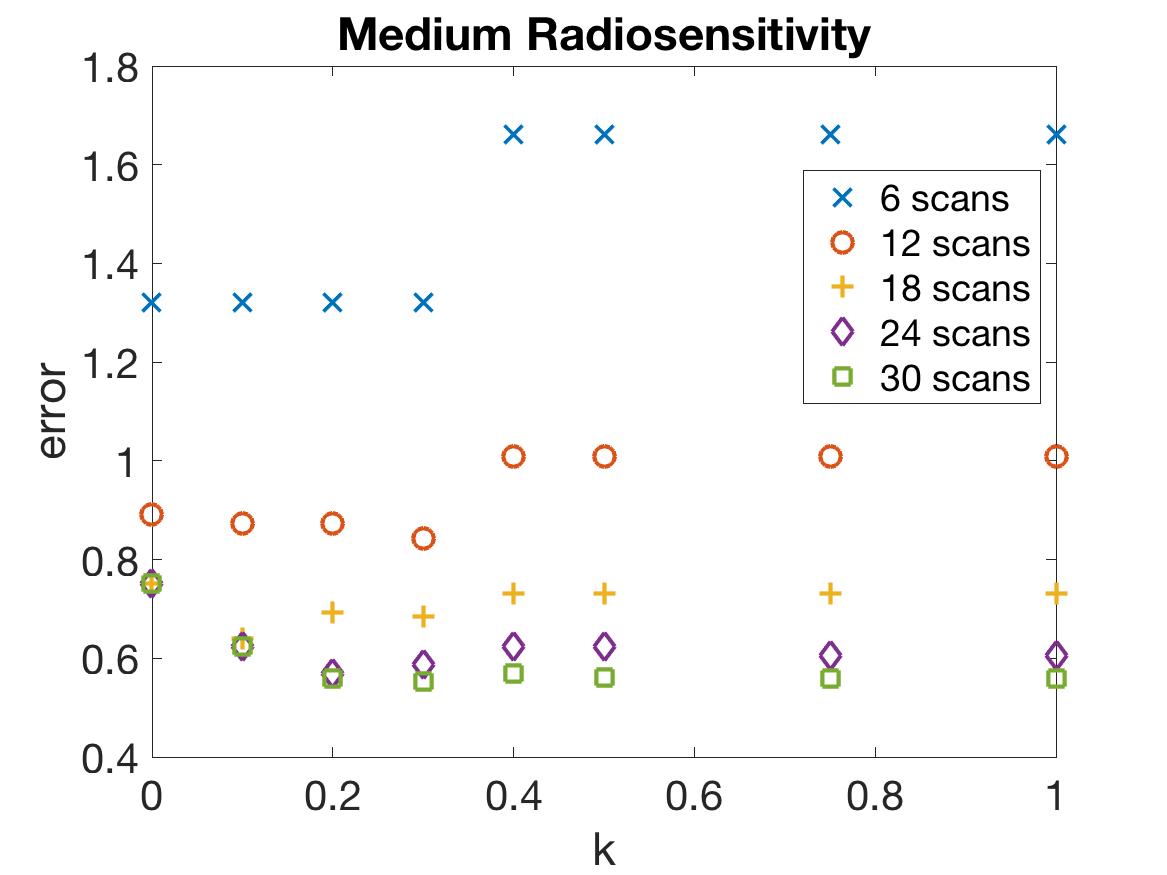}
	\includegraphics[width=1.9in]{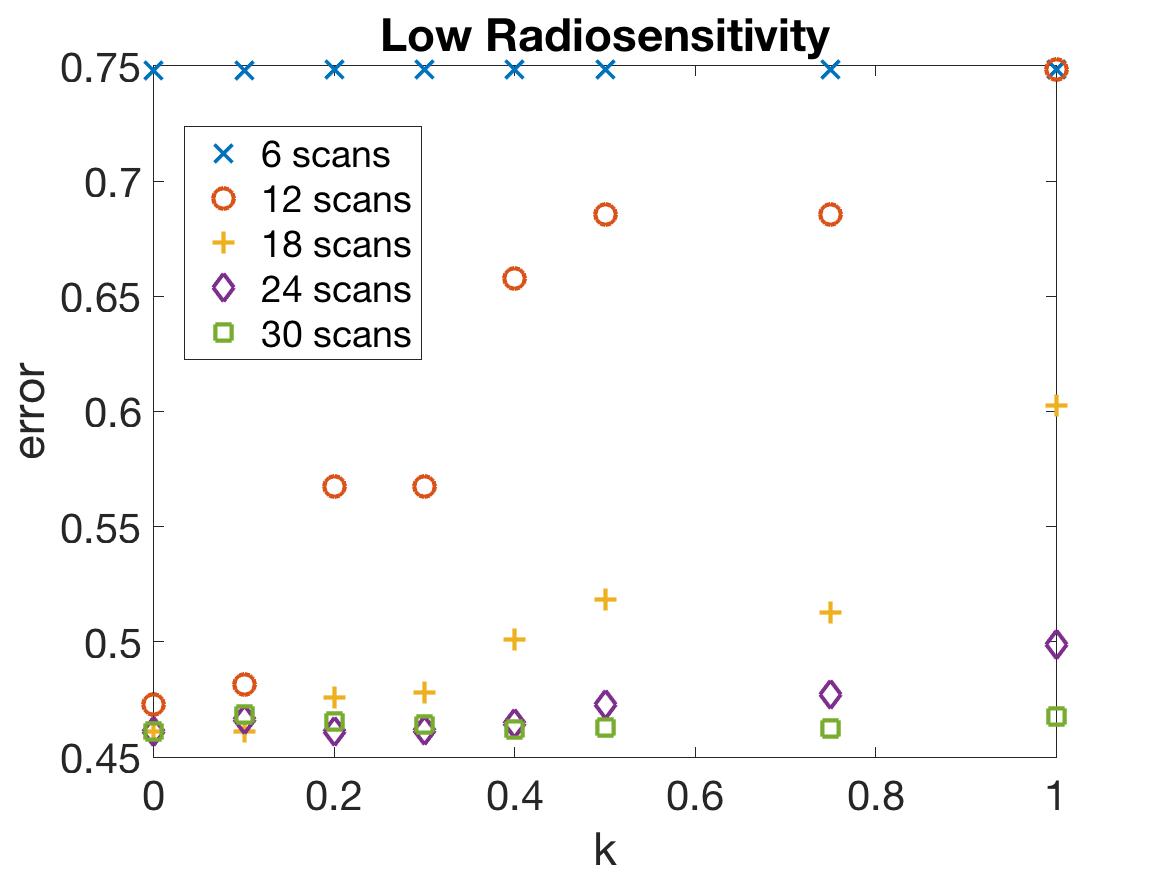}
	}
	\caption{One-compartment model. Relative error of model fitness to the data as defined in Equation  \eqref{eq:error} with respect to the number of scans for different score function parameter $k$ values (top), and error with respect to $k$ using a fixed number of 6, 12, ..., 30 scans (bottom). In general, when the scan number is limited to a small number, (6 and 12 scans), $k=0$ gives accurate results, particularly when the tumor is less responsive to radiotherapy. However, if a large number of scans is available, $k>0$ gives more accurate results, especially when the tumor is highly responsive to radiotherapy. } 
	\label{fig:errorVSk_onecomp} 
\end{figure}

In Figure \ref{fig:errorVSk_onecomp}, we compare the relative error of model fitness to the data as defined in Equation  \eqref{eq:error} with respect to the number of scans for different values of score function parameter $k$ in three radiosensitivity levels: high, medium, and low. 
We observe that when using $k=0$, the error is reduced most rapidly in the small scan number range. 
In the case of medium radiosensitivity, any $k$ value less than or equal to $0.4$ gives similar accuracy when using less than 12 scans. 
In the low radiosensitivity level, with a fixed budget of 12 scans, $k=0$ gives the most accurate result. 
However, larger $k$ values give more accurate results as the available scan number become larger. Specifically, $k$ values around $0.3$ give more accurate outcomes in the medium and low radiosensitivity cases when more than 18 scans are included in the budget. For the high radiosensitivity level, where the tumor responds quickly and drastically to radiation, using larger values of $k$ close to $k=1$ gives the most accurate result with nearly any scan budget. 
We conclude that using $k = 0$ or a relatively small value of $k$ is preferable when the available scan number is small, especially when the tumor is not responsive to treatment, as these small $k$ values allow one to skip forward and obtain at least one measurement towards the end of the treatment cycle so as to capture the overall trend in the data. However, when the available scan number increases, $k>0$ gives more accurate results,  especially when the tumor is highly responsive to treatment.

\begin{figure}[!htb]
\ \\
\centerline{ \footnotesize \textsf{High Radiosensitivity} \hspace{2.5cm}  \textsf{Low Radiosensitivity} }	
\centerline{ \rotatebox{90}{ \hspace{1.35cm} $k=0.0$}
	\includegraphics[width=2in]{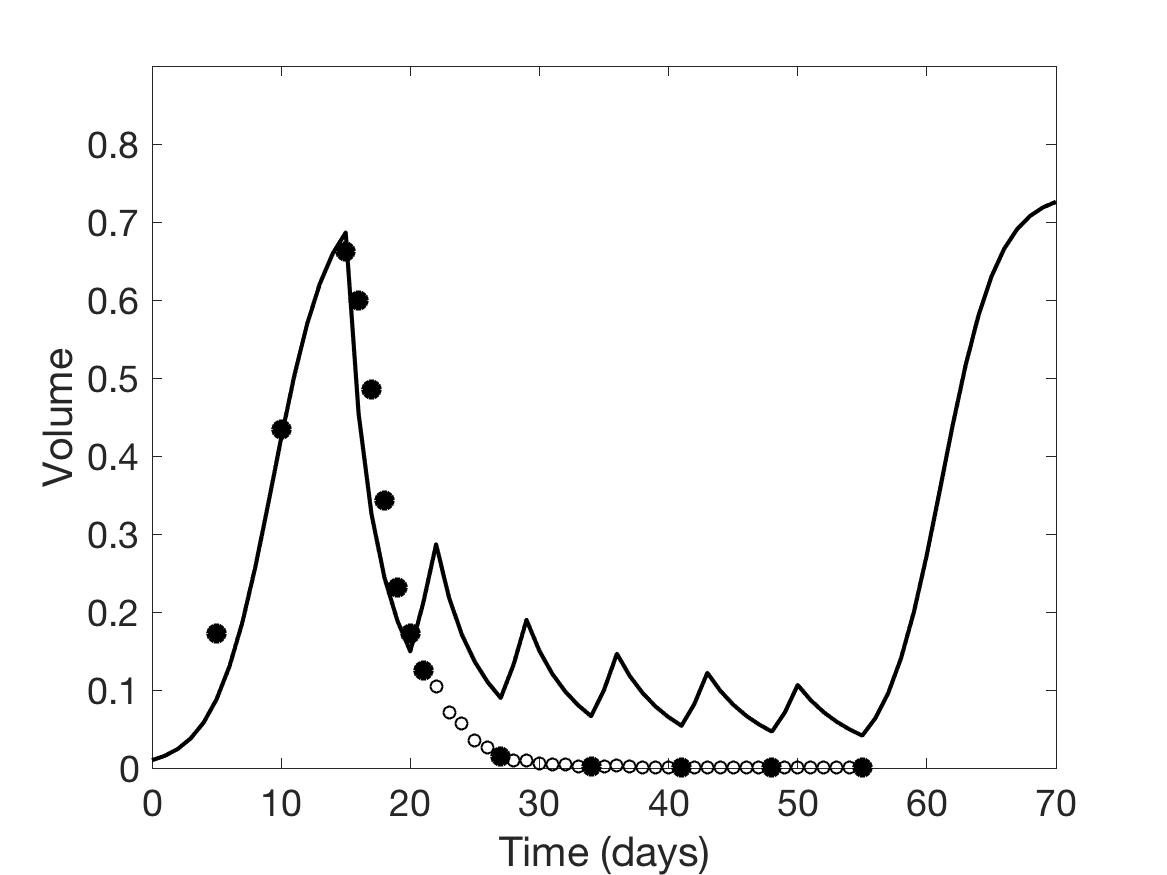}
	\includegraphics[width=2in]{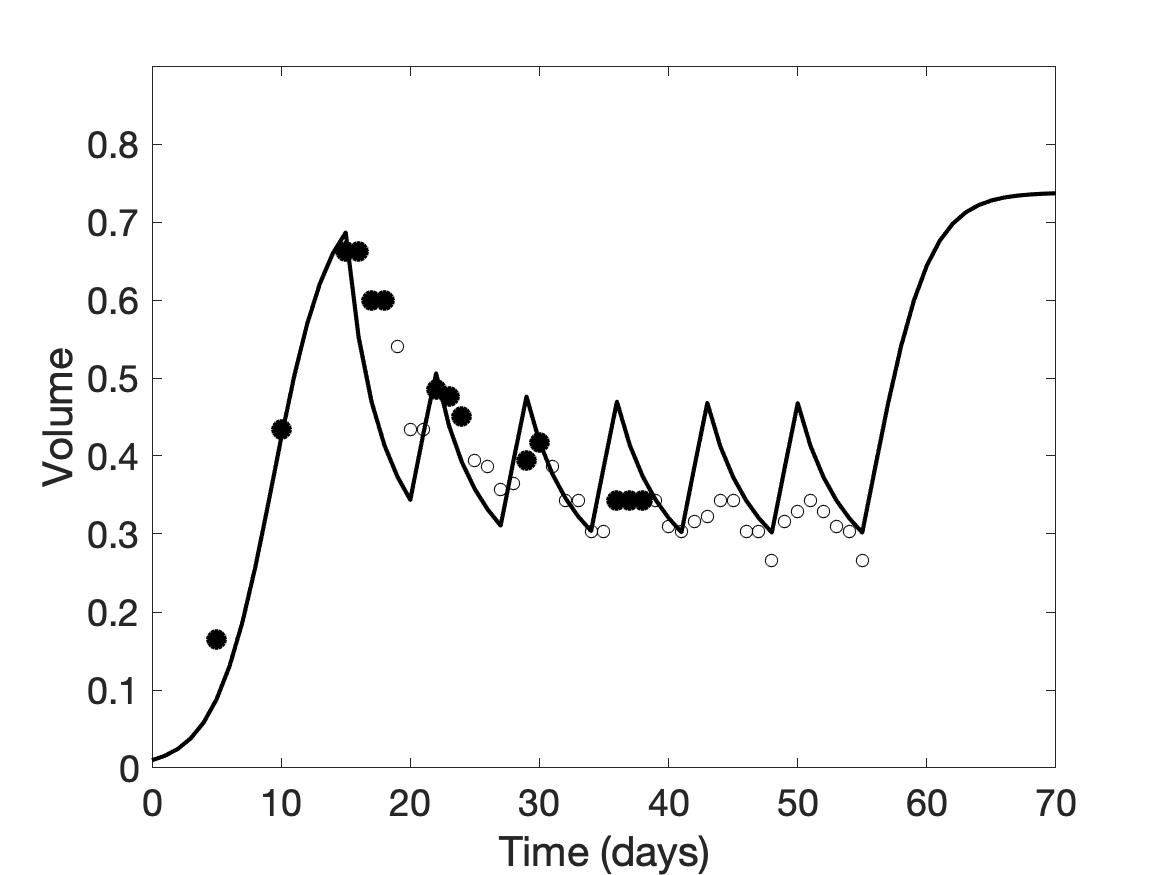}
    }	
\centerline{  \rotatebox{90}{\hspace{1.35cm} $k=0.5$}
	\includegraphics[width=2in]{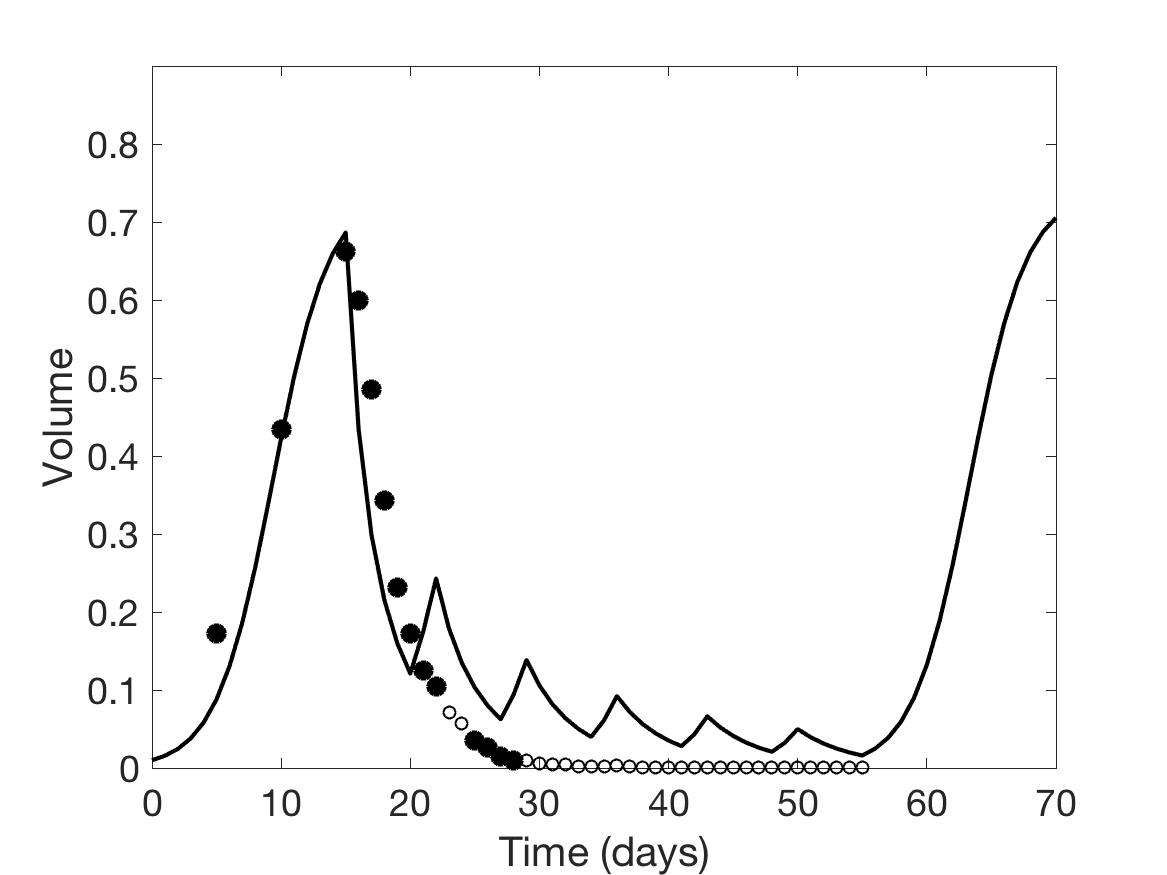}  
	\includegraphics[width=2in]{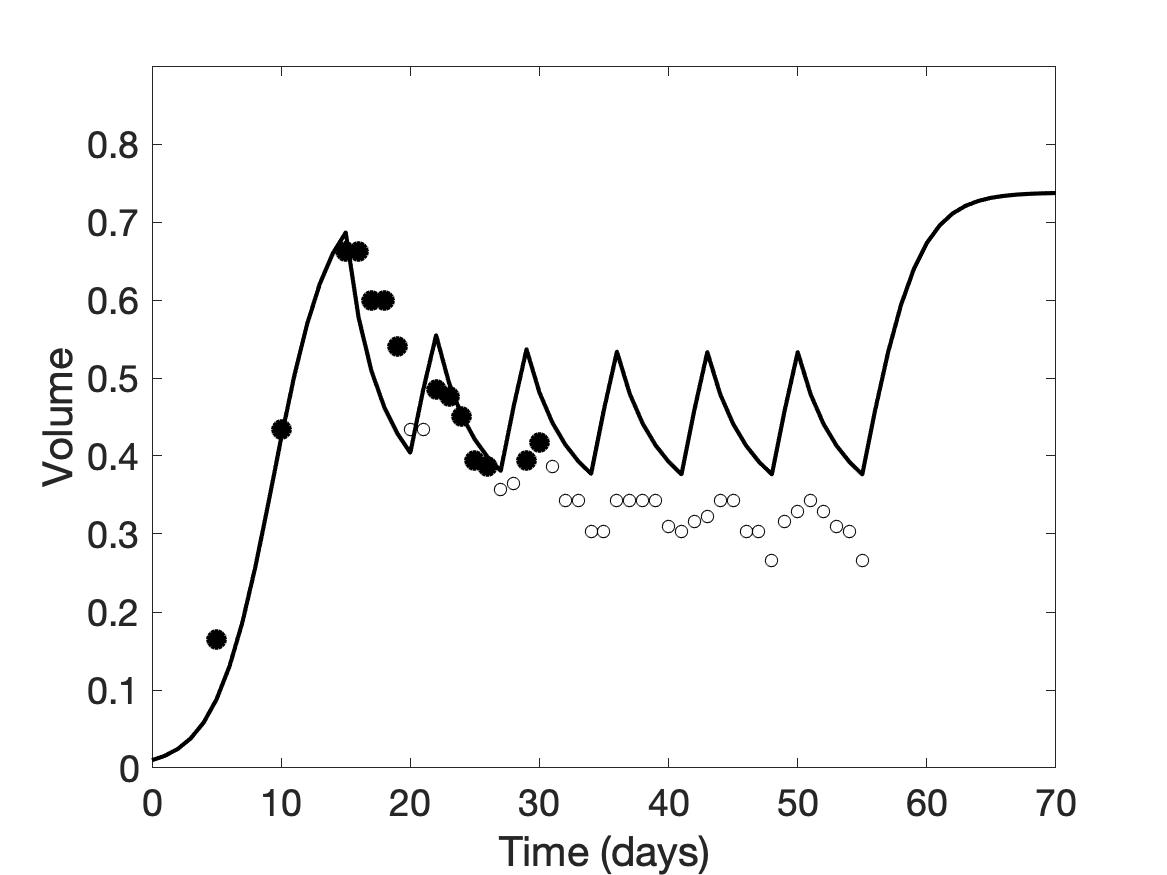}
    }
\caption{One-compartment model. Choice of high fidelity data ($\bullet$) among the potential data ($\circ$) using a 12 scan budget, and the fitted model prediction (--) of tumor volume using Equation \eqref{eqn:eqn1comp} for parameter values $k = 0$ and $0.5$. We observe a better fit using $k=0.5$ in the case of high radiosensitivity and $k=0$ in the case of low radiosensitivity. } 
	\label{fig:Modelfit_onecomp} 
\end{figure}

\begin{figure}
\centerline{  
         \includegraphics[width=5.5in]{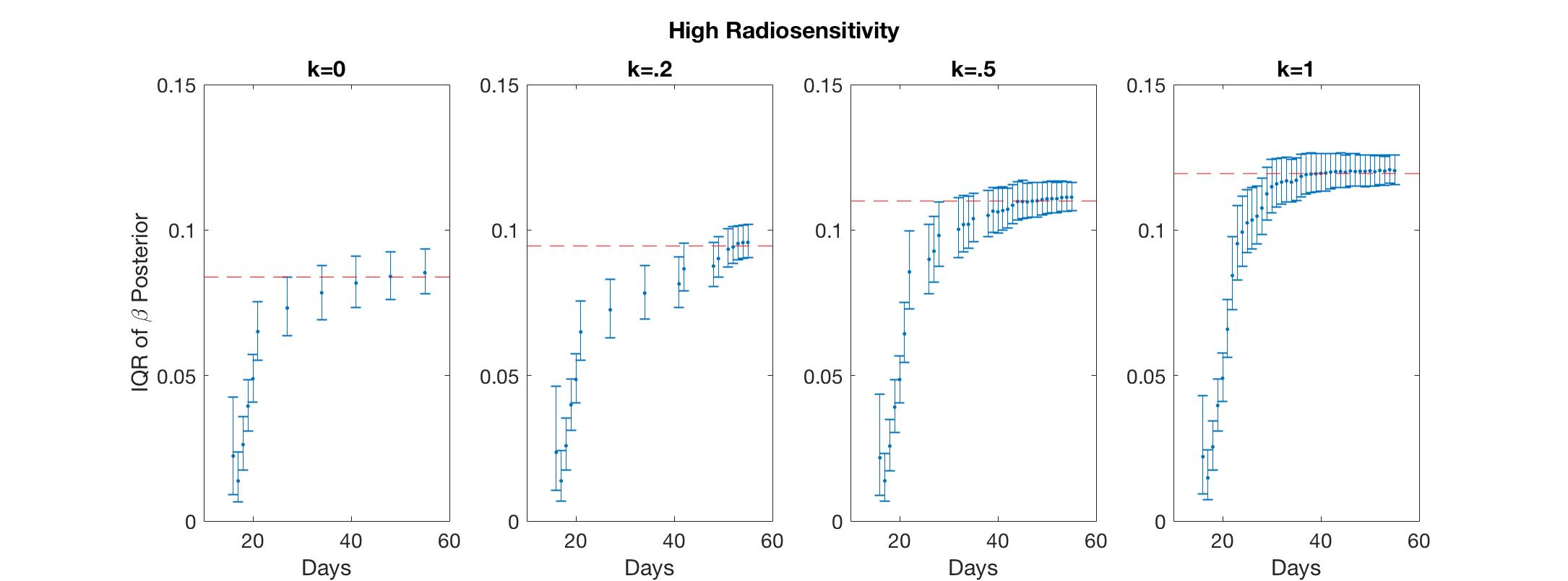}}
         \centerline{
         \includegraphics[width=5.5in]{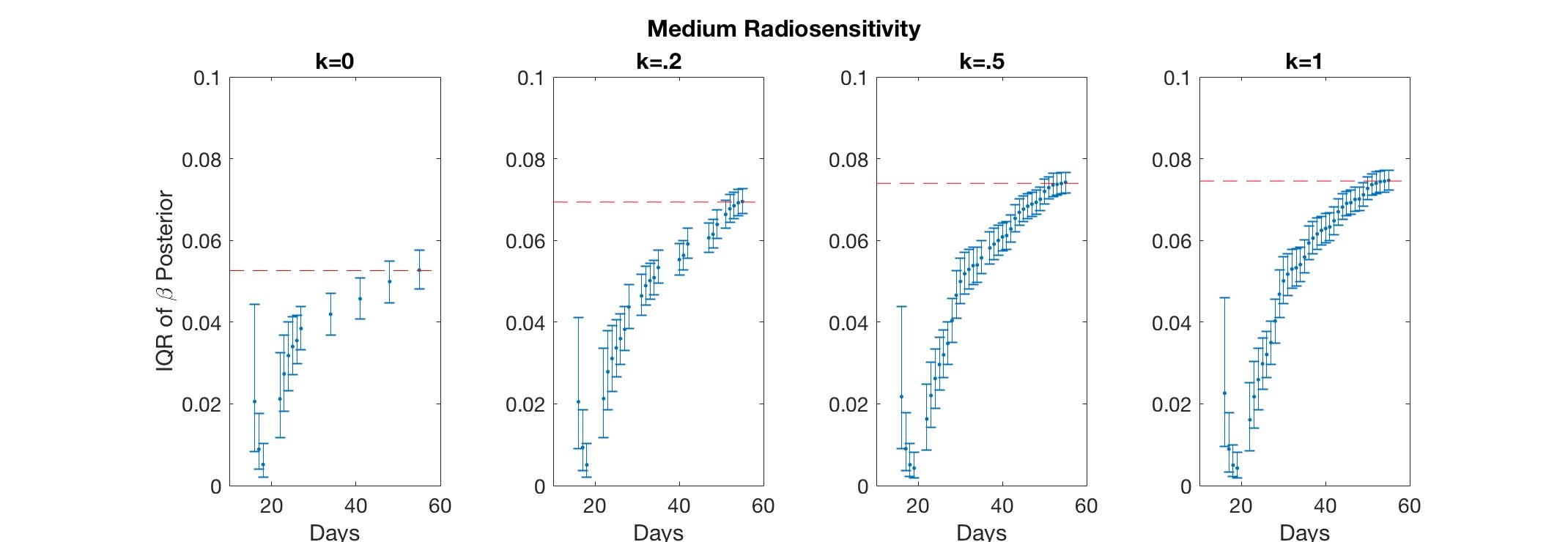}}
         \centerline{
         \includegraphics[width=5.5in]{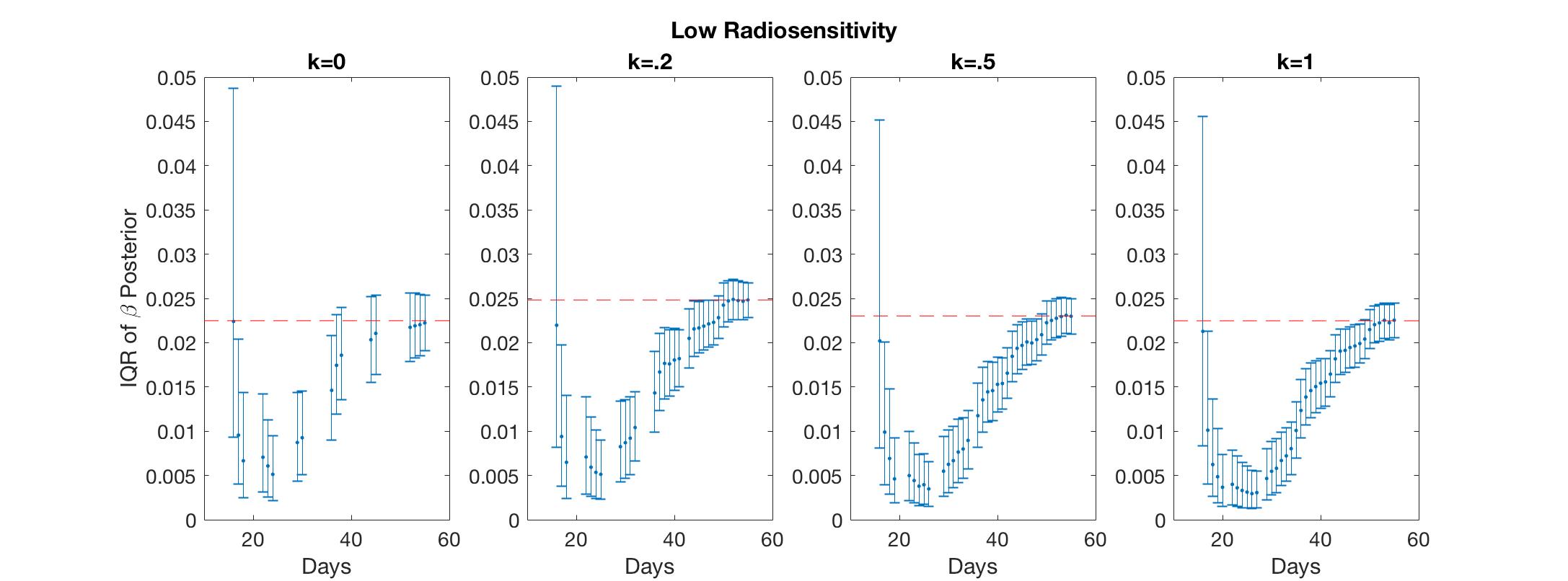}  
         }
    \caption{One-compartment model. Interquartile ranges of the posterior distribution for $\beta$ after the addition of each high-fidelity data point. As more information is gathered, we observe shrinkage in the IQR demonstrating increased precision, and convergence to the final parameter estimate demonstrating increased accuracy. The red dashed lines indicate the final value of the $\beta$ estimate. For the high and medium radiosensitivity cases, it can be seen that using $k=0$ does not allow the parameter estimate to fully converge to its actual value, leading to larger errors. In these cases, where the gradient of the data is rapidly changing, we are better off using more scans to achieve our desired accuracy in the final fit.\\ }
    \label{fig:iqrplots_onecomp}
\end{figure}

\begin{table}[!htb]
\centering
\begin{tabular}{|c||c|c|c|c|}
\hline
Radiosensitivity & $k=0$ & $k=0.2$ & $k=0.5$ & $k=1$\\
\hline
High & .0838 & .0945& .1100& .1194\\
Medium & .0526 & .0694 & .0740 & .0745\\
Low & .0225 & .0248 & .0239 & .0225\\
\hline
\end{tabular}
\caption{One-compartment model. Final calibrated $\beta$ values for the scenarios shown in Figure \ref{fig:iqrplots_onecomp}. We observe that while $k=0$ uses the fewest number of scans---a fact that makes it the more appealing option in terms of expense---using more scans as in the larger $k$ values for the high and medium radiosensitivity cases results in a significantly different final value, suggesting that more scans may be needed for $\beta$ to converge to a best estimate.}
\label{table:beta_values}
\end{table}

Figure \ref{fig:Modelfit_onecomp} shows the data selected by using a 12 scan budget, with parameter values $k = 0$ and $k = 0.5$, and the fitted model prediction using the one-compartment model \eqref{eqn:eqn1comp}. When the tumor is highly responsive to radiotherapy, we observe that for larger values of $k$---for example, $k=0.5$---the fitted model prediction is more accurate. This is unsurprising; since the tumor decay is rapid in these scenarios, it is more informative to learn about the rapidly changing gradient of the data by frequently measuring the tumor data at the earlier time points. On the other hand, when the tumor is less responsive to radiotherapy, learning the overall trend with sparsely placed scans is sufficient, so that using smaller values of $k$, including $k=0$, gives fairly accurate results. However, the accuracy can be improved with a larger number of scans, using larger $k$ values. This is supported by Figure \ref{fig:iqrplots_onecomp}, where it can be seen that $k=0$ results in the fewest total scans, but including more scans (as with the use of larger $k$ values) may lead to a more refined final parameter estimate, particularly for the low and medium radiosensitivity cases; the final calibrated $\beta$ parameter values are listed in Table \ref{table:beta_values} for each of the scenarios illustrated in Figure \ref{fig:iqrplots_onecomp}. We see this theme recurring throughout our investigation; small $k$ values require fewer scans and are thus less expensive to implement, but in many scenarios---particularly those in which the overall gradient of the data is steeper---choosing a small $k$ results in larger prediction errors.

\subsection{Scenario 3: Two-compartment model with $n$ number of scans } \label{sec:twocompOAT}

In this section, we present results of the selected scanning schedule and accuracy of the calibration in the two-compartment model \eqref{eqn:twocomp}, by selecting either the tumor volume data or the necrotic volume data at each step in the algorithm. Though only one metric can be chosen at a time, we do allow for the next step in the algorithm to pick the second metric at the same time point; i.e., if the algorithm first chooses to measure the tumor volume at day 16, the next point chosen could be the necrotic volume at day 16. The results are similar to the one-compartment model in Section \ref{sec:simulation_one}.

\begin{figure}[!htb]
\centerline{  
         \includegraphics[width=1.9in]{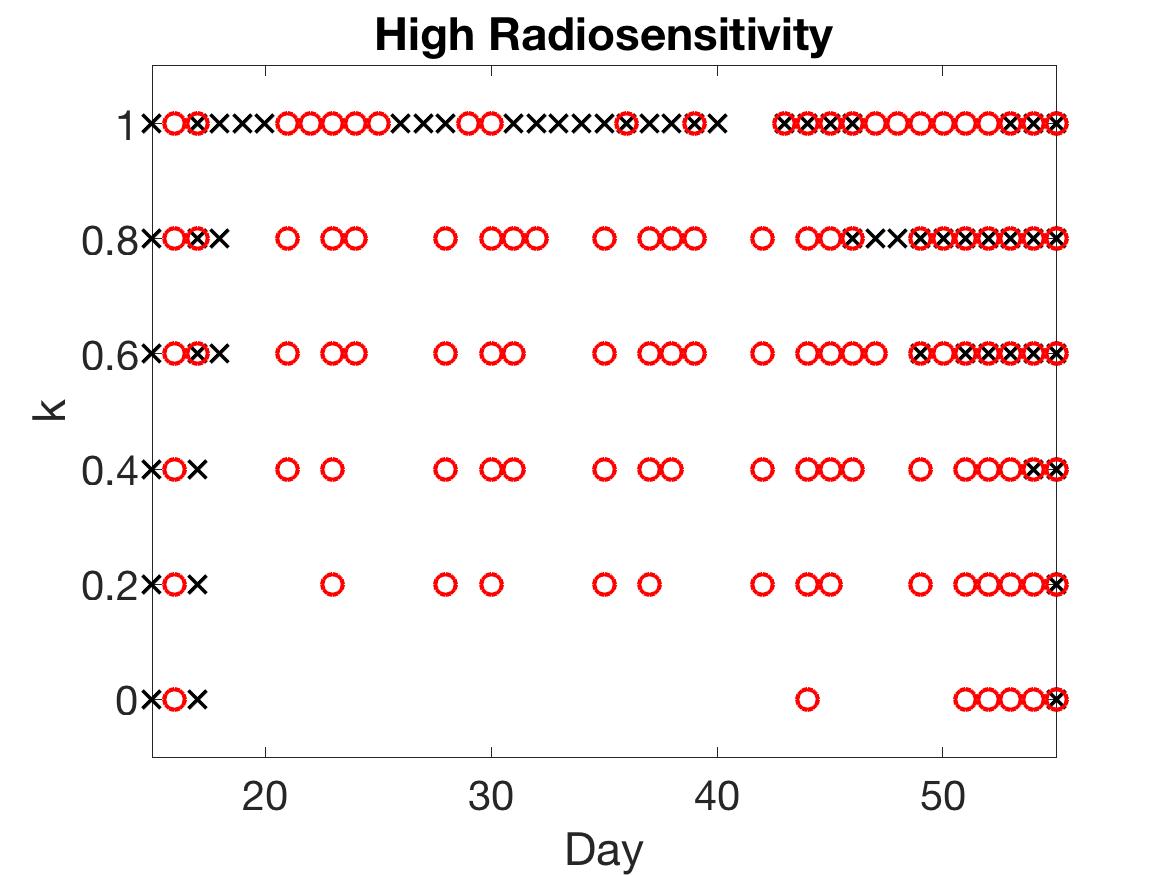}
         \includegraphics[width=1.9in]{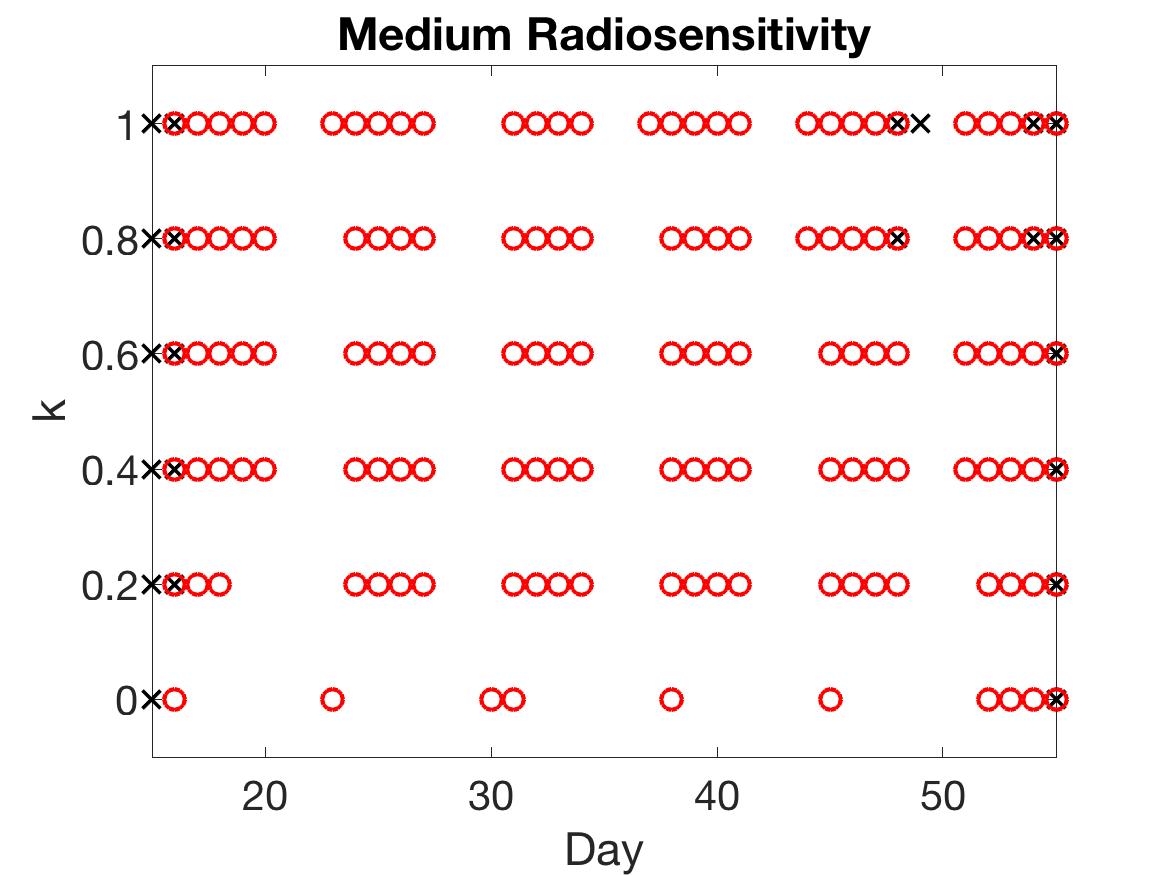} 
         \includegraphics[width=1.9in]{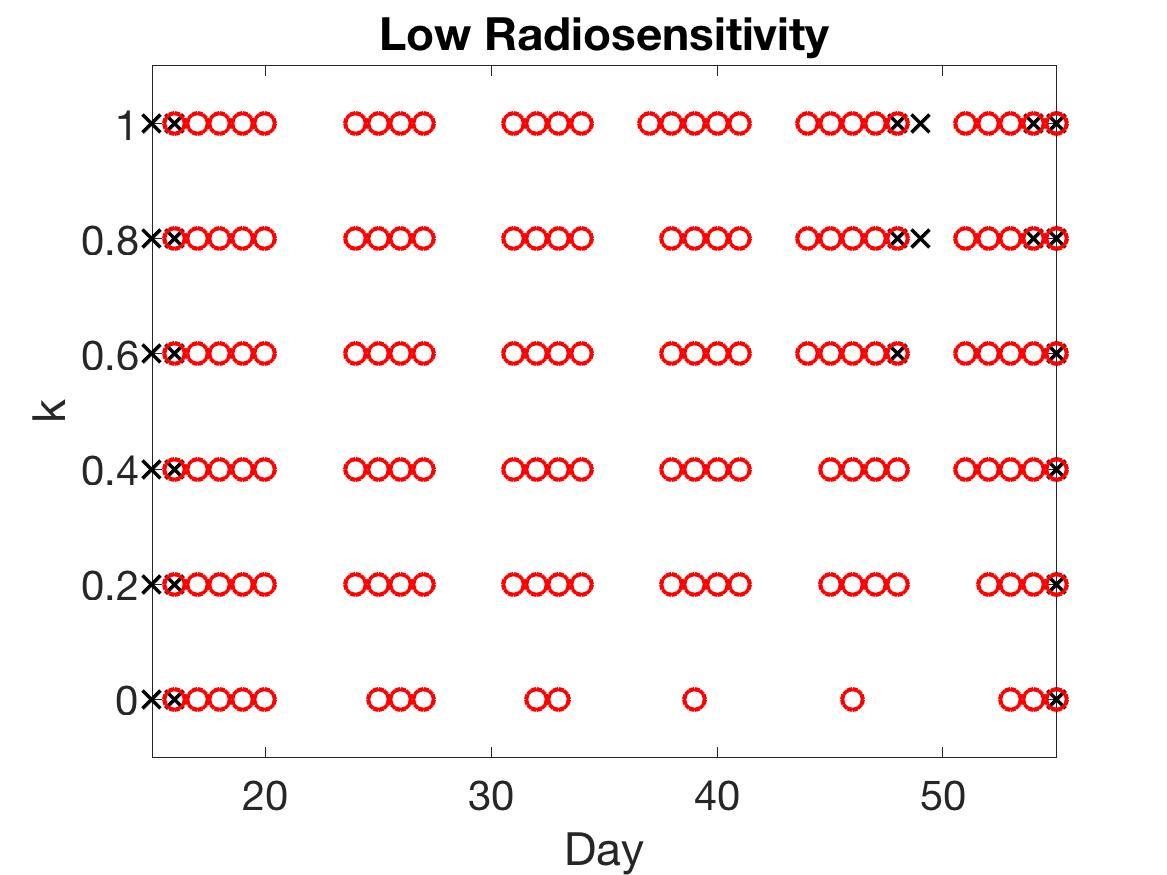}  
         }
    \caption{Two-compartment model with one type of point chosen at each step of the algorithm. 
    Choice of scan of either tumor volume ($\times$) or necrotic volume (\textcolor{red}{$\circ$}) for different values of score function parameter $k = 0, 0.2, ..., 1$. Larger values of $k$ result in choosing earlier time points in all three radiosensitivity levels.}
    \label{fig:ptschosen_twocomp}
\end{figure}

In Figure \ref{fig:ptschosen_twocomp}, we observe that the selected scanning schedule becomes more refined in the earlier times for larger values of score function parameter $k$. 
We observe that our mutual information criterion often chooses the necrotic volume data over the tumor volume data. This is reasonable, considering that the necrotic volume is more informative for estimating the effect of radiotherapy parameters accurately, since the administration of RT has an immediate effect on the composition of the tumor through the conversion of viable cells to necrotic cells, but only a delayed effect on the total tumor volume. 

\begin{figure}[]
\centerline{  
        \includegraphics[width=1.9in]{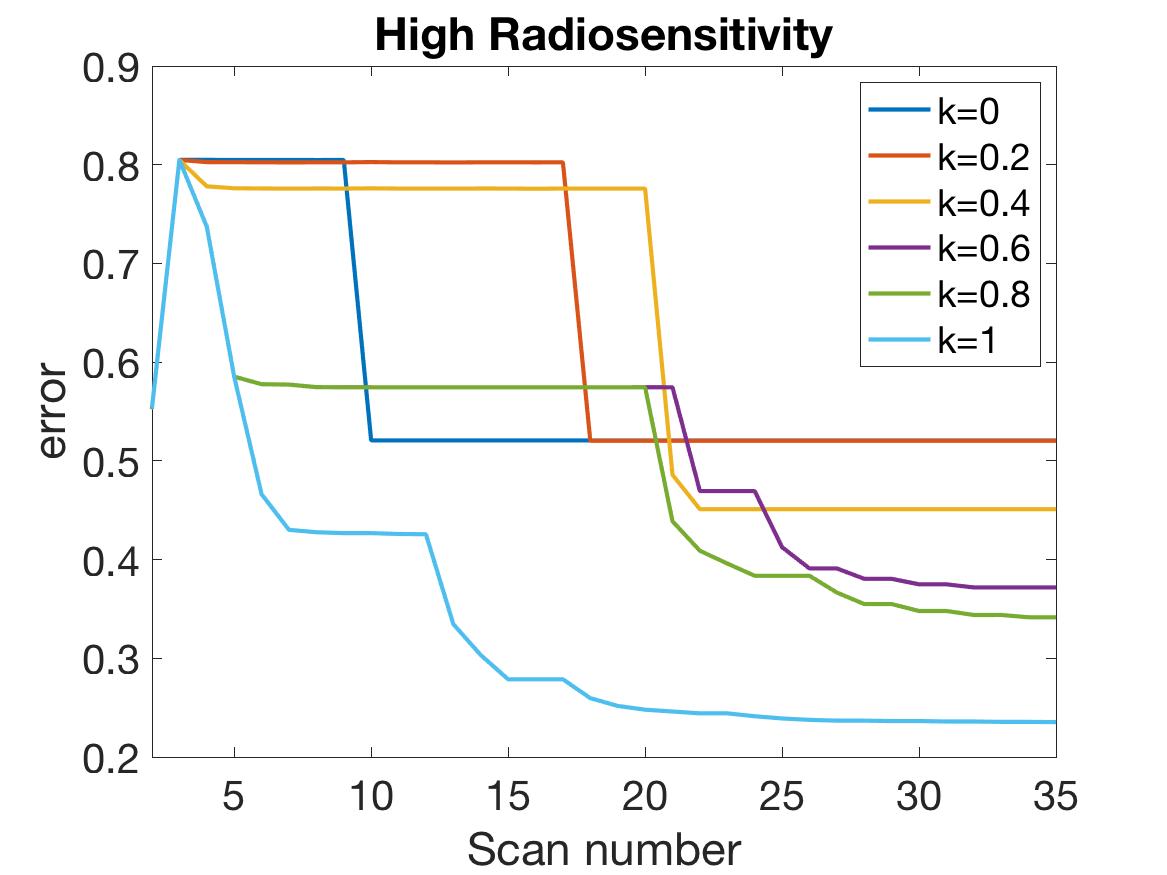}
         \includegraphics[width=1.9in]{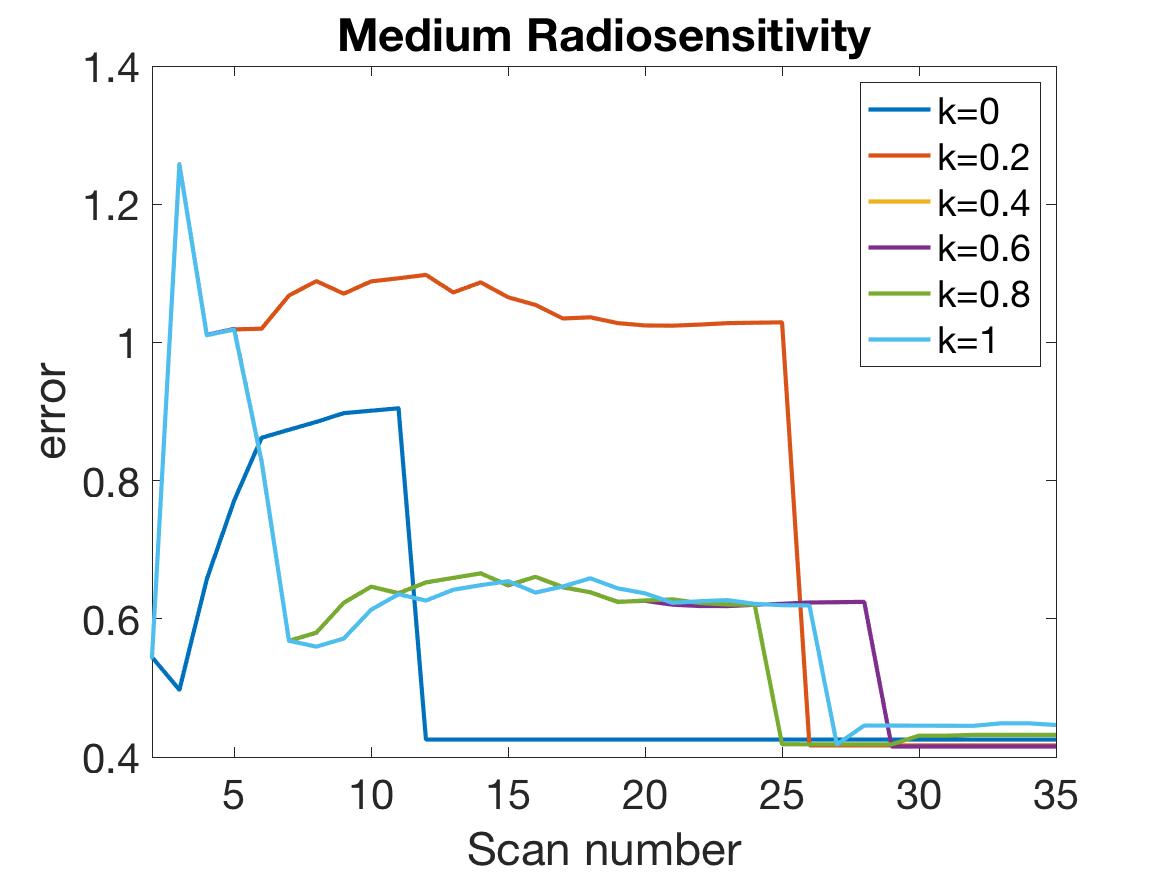} \includegraphics[width=1.9in]{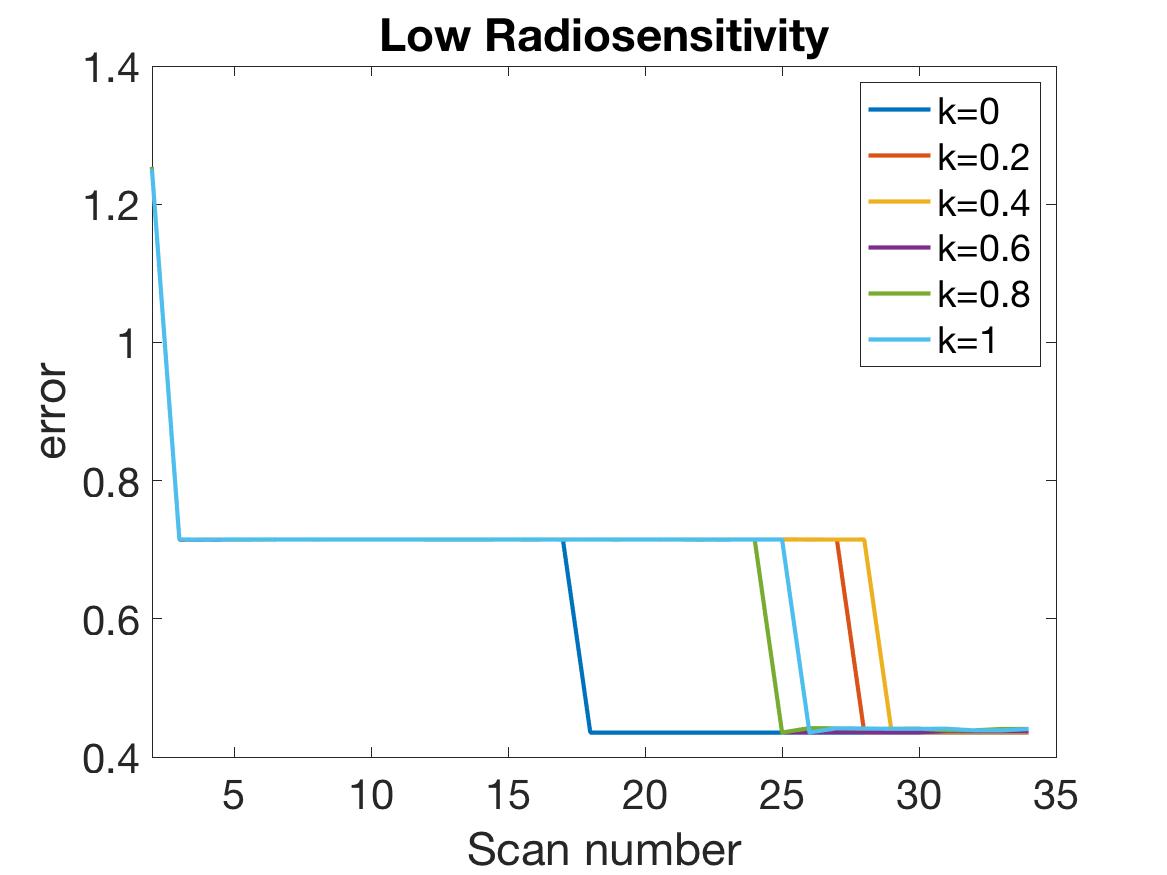}  
    }  
    \caption{Two-compartment model with one type of point chosen at each step of the algorithm. Relative error of model fitness to the data as defined in Equation \eqref{eq:error} with respect to number of scans for each of three radiosensitivity levels. Using $k=1$ gives the most accurate result in high radiosensitivity level, while $k=0$ is most beneficial for a low radiosensitivity response. }
    \label{fig:errorvsscan_twocomp}
\end{figure}

We remark that choosing necrotic volume data may result in non-monotonic decay in error due to the fact that only providing necrotic volume without tumor volume can deteriorate the accuracy in predicting total tumor volume, as shown in Figure \ref{fig:errorvsscan_twocomp}. 
However, the optimal $k$ value with respect to accuracy displays similar results to the one-compartment model. In the case of high radiosensitivity, larger $k$ values, especially $k=1$, provide the most accurate result. For the medium to low radiosensitivity cases, $k=0$ is more accurate when the scan budget is small (for example, less than 25 scans), as it is able to capture some of the final points in this limited scan budget and is thus cognizant of the overall shape of the data across the full treatment period.  However, larger values of $k$ (i.e., $k\geq 0.8$) provide a more accurate final fit when large scan budgets are allowed---the inclusion of additional scans in the beginning of the treatment regimen allow for a more refined fit. 

% \begin{figure}[]
% \centerline{  
%          \includegraphics[width=1.5in]{Fig/TotalMIvsDays_TwoComp_Rad1.jpg}\includegraphics[width=1.5in]{Fig/TotalMIvsDays_TwoComp_Rad3.jpg}   \includegraphics[width=1.5in]{Fig/TotalMIvsDays_TwoComp_Rad9.jpg}}
%          \centerline{
%          \includegraphics[width=1.5in]{Fig/TotalMIvsPts_TwoComp_Rad1.jpg}\includegraphics[width=1.5in]{Fig/TotalMIvsPts_TwoComp_Rad3.jpg}   \includegraphics[width=1.5in]{Fig/TotalMIvsPts_TwoComp_Rad9.jpg}\\
%     }
%     \caption{THESE ARE UP TO DATE. Two-compartment model testing with one point chosen at a time. Total MI versus day and versus number of points.}
%     \label{fig:miplots_twocomp}
% \end{figure}

\subsection{Scenario 4: Two-compartment model in practical setting}
\label{sec:twocomp_both}

While the two-compartment model analysis described in Section \ref{sec:twocompOAT} gives an interesting look at the performance of the algorithm when presented with two alternate metric choices at each time, from a practical standpoint, a clinician who chooses to measure a necrotic proportion at time $t$ would essentially get a tumor volume estimate at time $t$ for ``free" from the imaging scan.  Thus, in this section we repeat the analysis from Section \ref{sec:twocompOAT}, but this time incorporate a tumor volume measurement automatically whenever a necrotic proportion metric is chosen, to better represent the clinical setting.

\begin{figure}[!b]
\centerline{ 
         \includegraphics[width=1.9in]{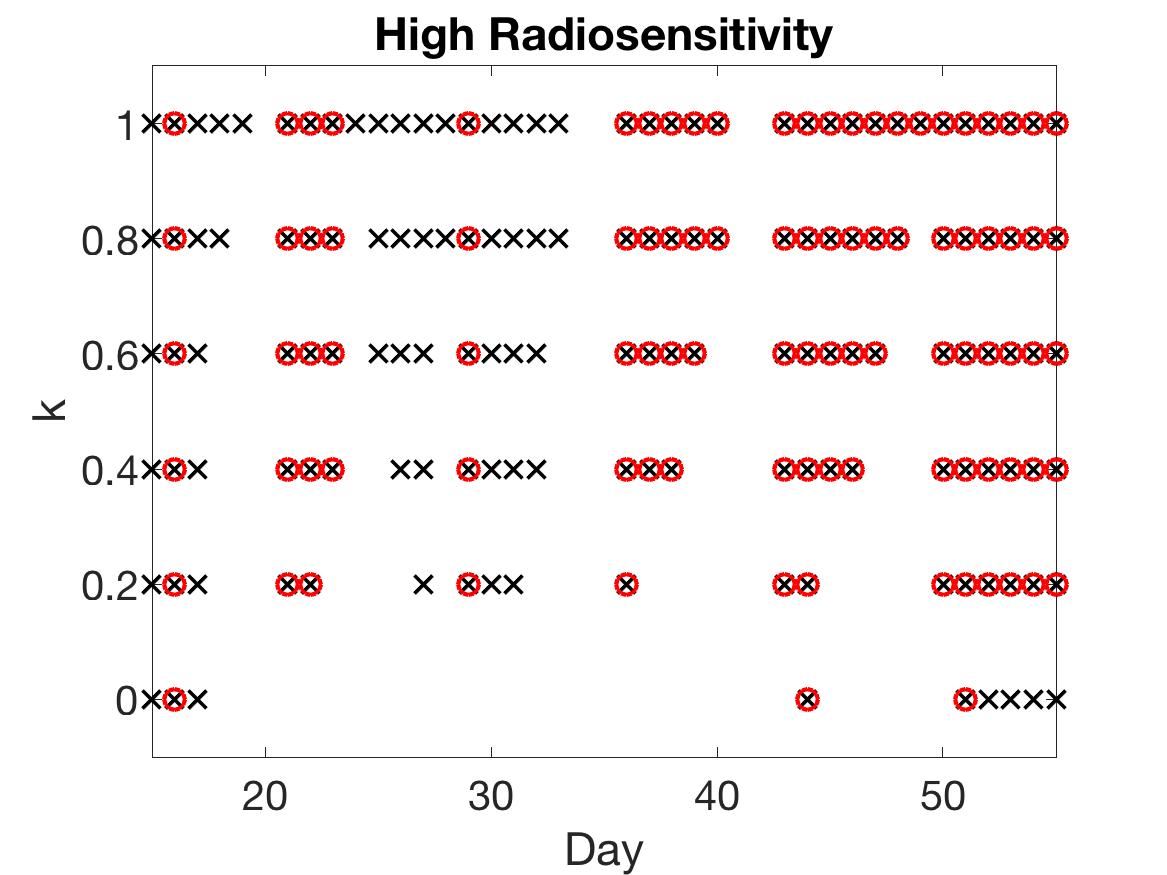}
         \includegraphics[width=1.9in]{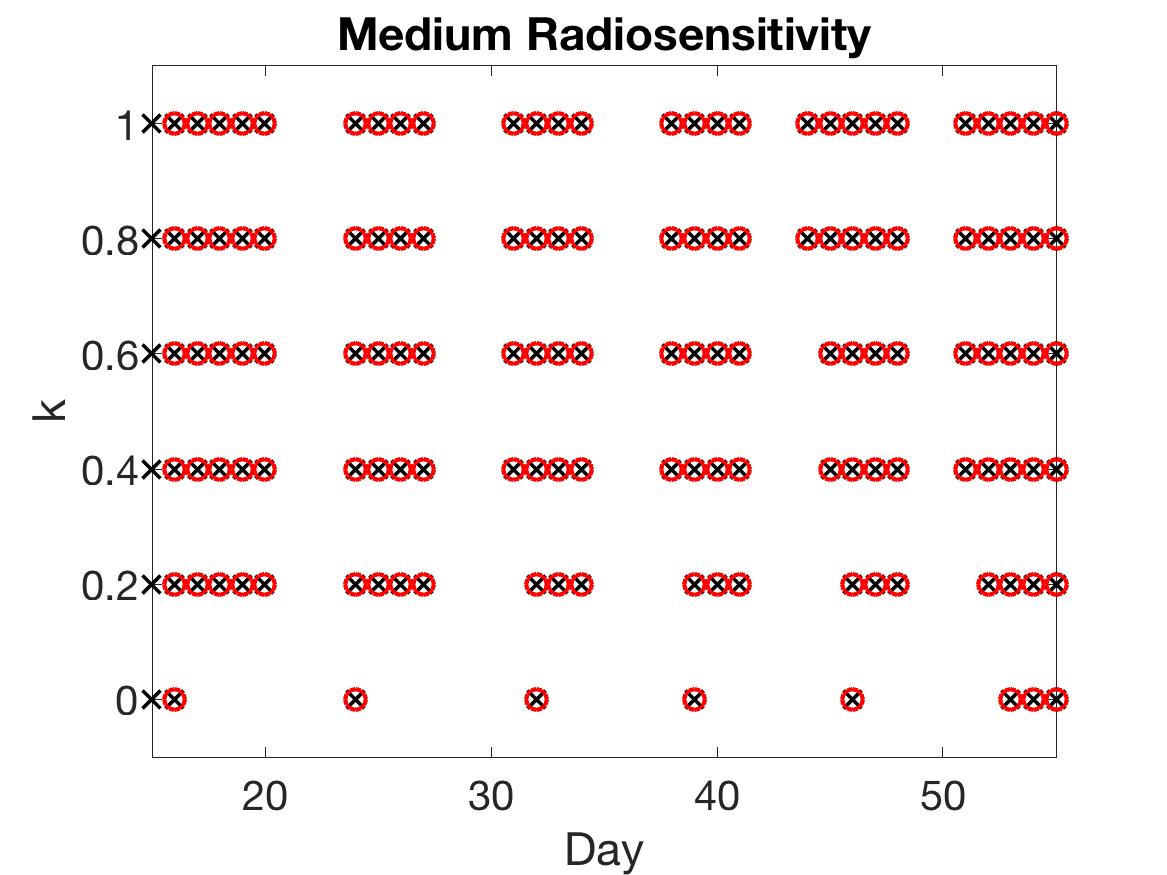}  \includegraphics[width=1.9in]{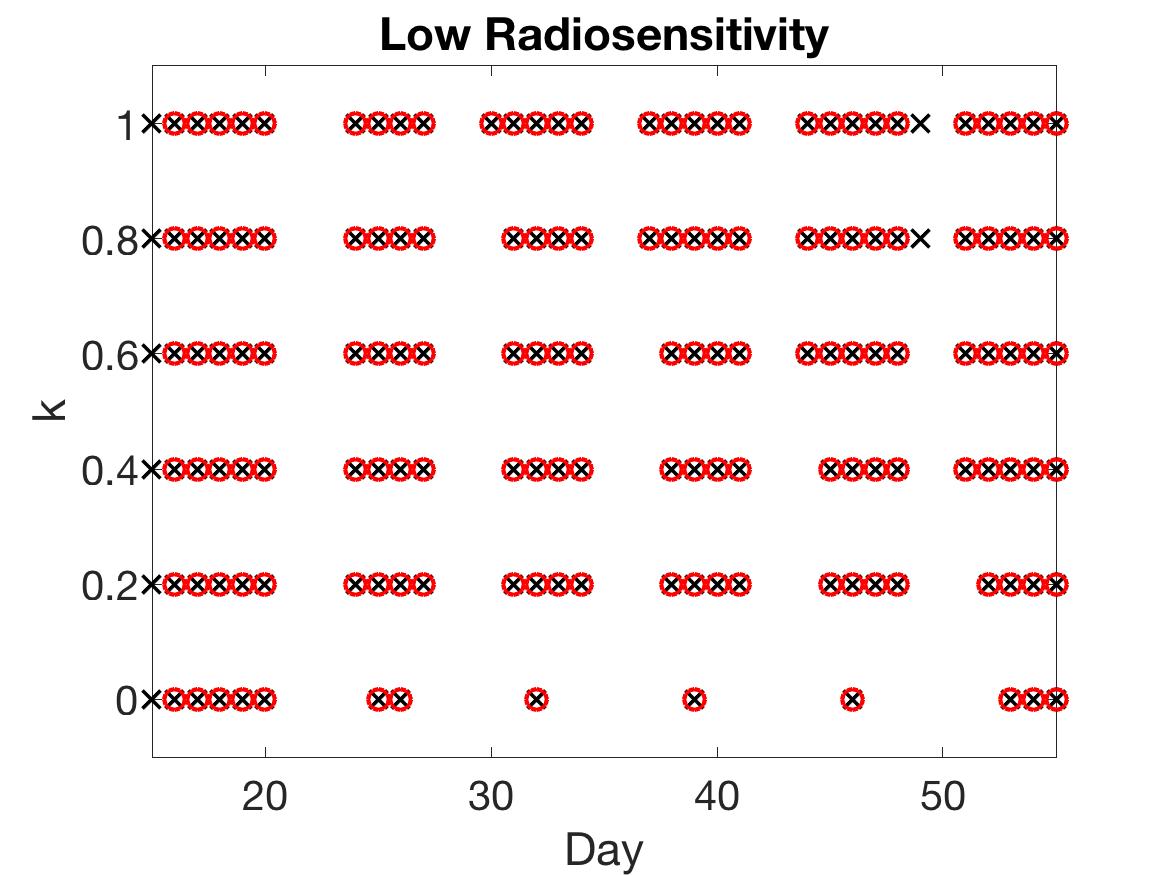} 
         }
    \caption{Two-compartment model with tumor volume ($\times$) automatically included when necrotic volume (\textcolor{red}{$\circ$}) is chosen. Choice of scan for different values of score function parameter $k = 0, 0.2, ..., 1$. 
    Larger values of $k$ result in choosing earlier time points in all three radiosensitivity levels.}
    \label{fig:ptschosen_twocompbothdata}
\end{figure}

In Figure \ref{fig:ptschosen_twocompbothdata}, we display the design conditions chosen for each of six different $k$ values.  In the high radiosensitivity case, there are numerous scenarios in which the MI score function selects tumor volume as the most informative metric, particularly for large values of $k$ when there is a larger penalty for skipping days.  In the medium and low radiosensitivity cases, however, the algorithm nearly always chooses necrotic proportion as the more informative metric.  Generally, these results agree with those found when tumor volume was not automatically incorporated in Section \ref{sec:twocompOAT}.

\begin{figure}[]
\centerline{  
	\includegraphics[width=1.9in]{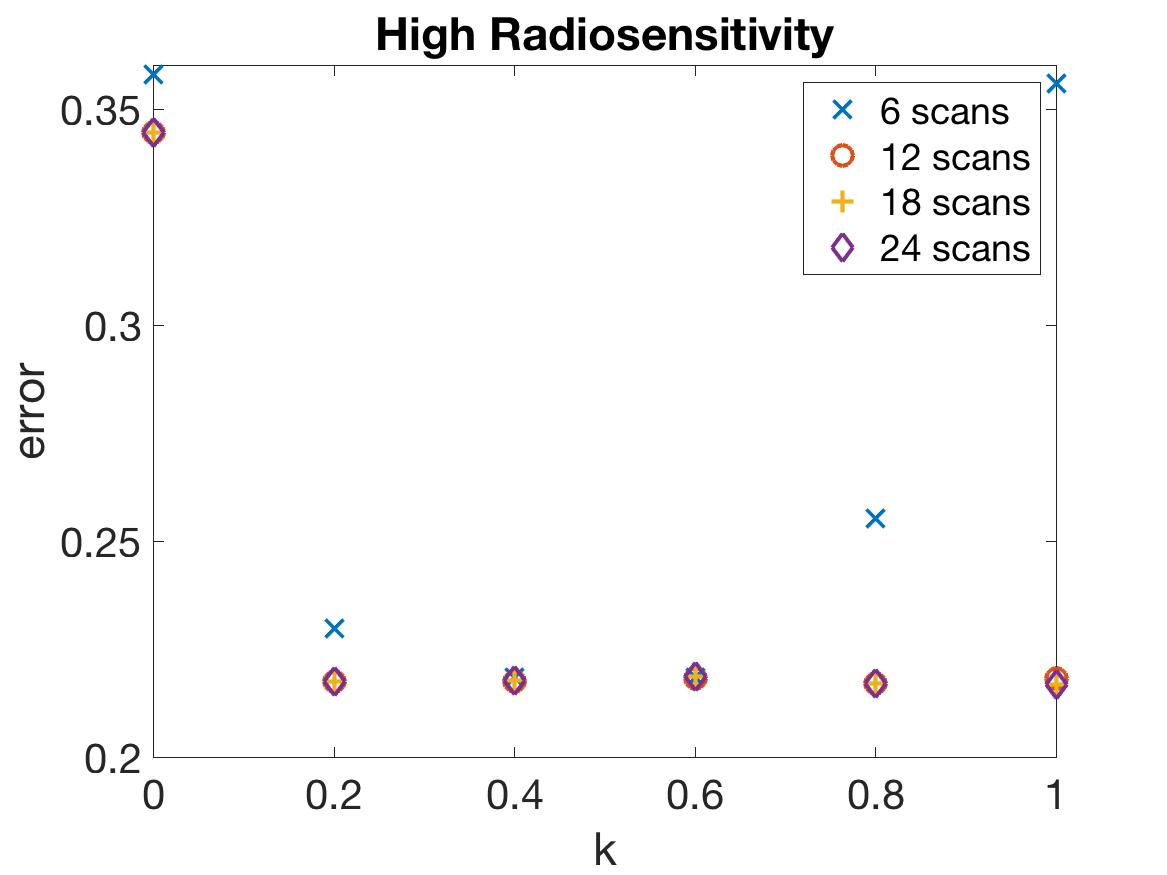}
	\includegraphics[width=1.9in]{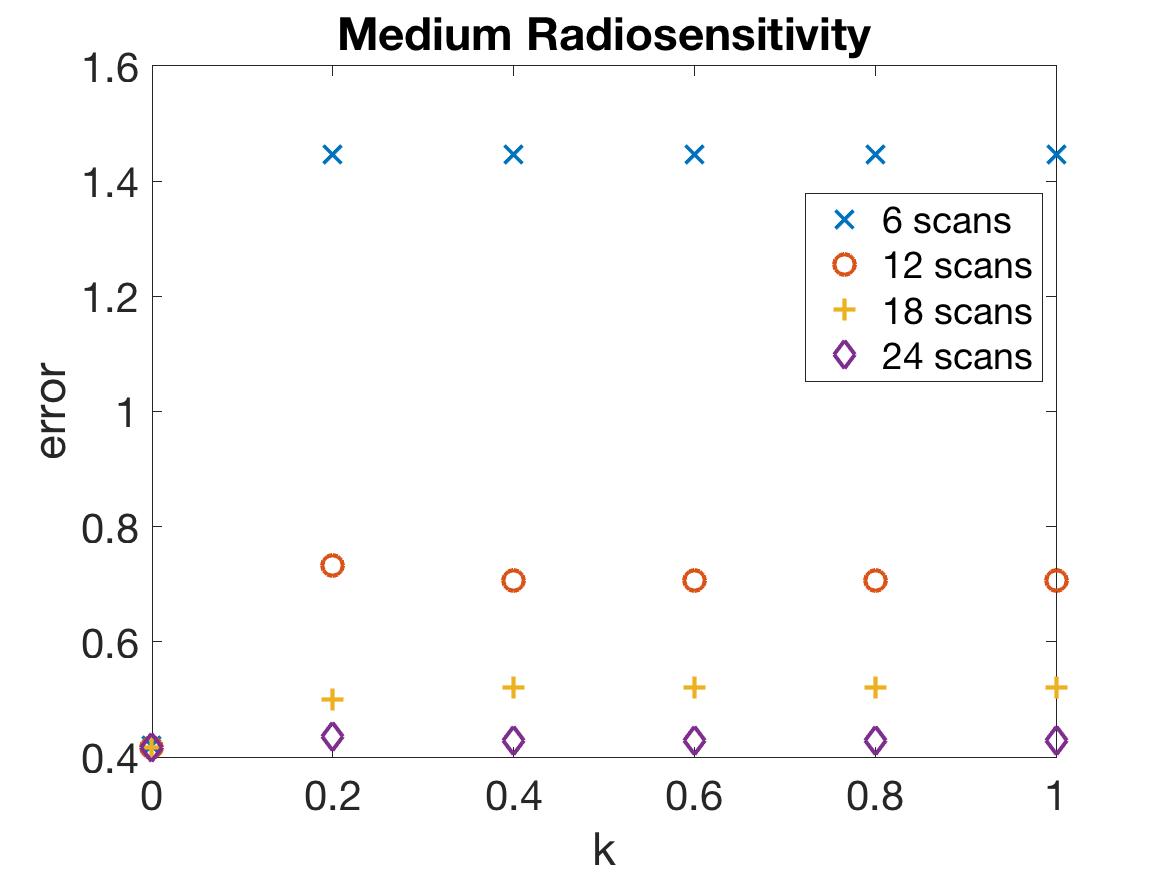}
 	\includegraphics[width=1.9in]{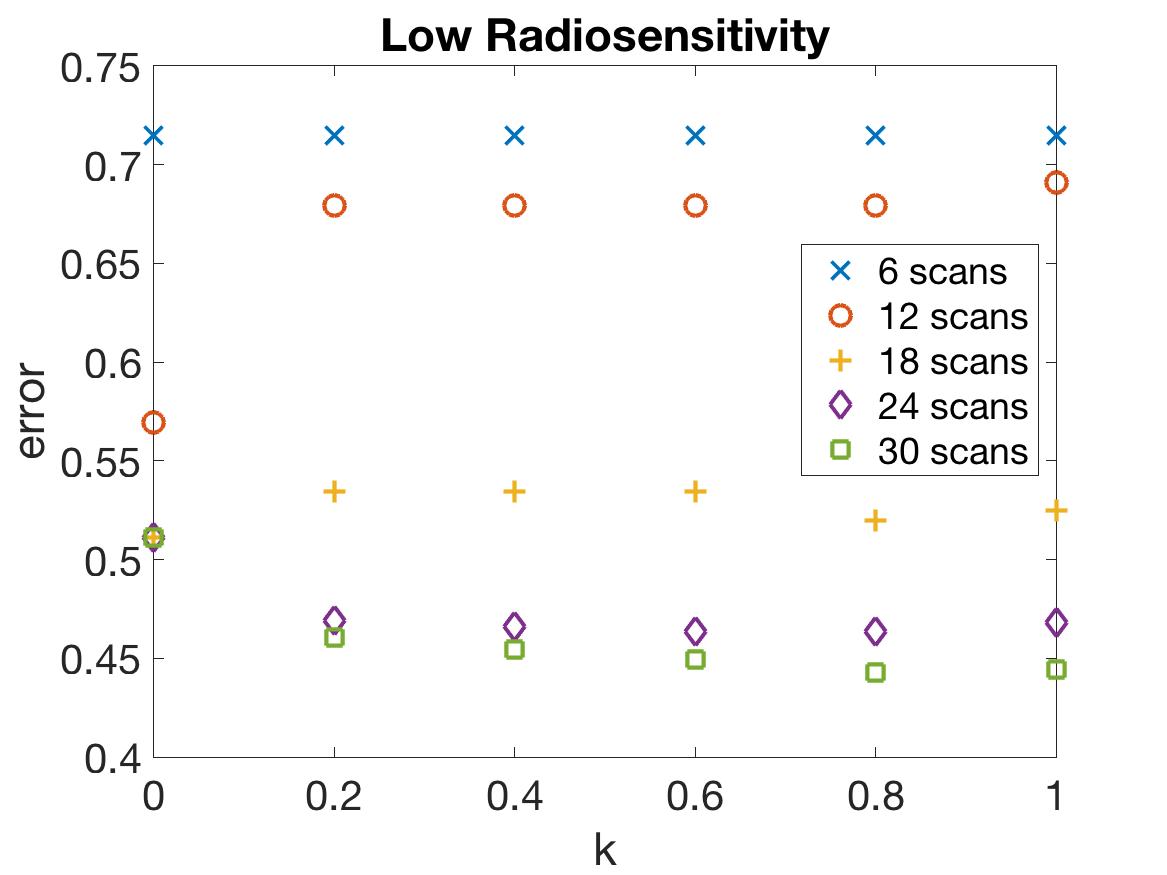}
	}	
\caption{Two-compartment model testing with tumor volume automatically included when necrotic is chosen. Plots of error \eqref{eq:error} vs $k$. In case of high radiosensitivity, using larger values of $k>0$ is more accurate for all scan budgets. For medium and low radiosensitivity, using $k=0$ gives the most accurate result when using 6 scans, however, larger values of $k$ yield more accurate predictions when larger scan budgets are available. } 
	\label{fig:errorVSk_twocompbothdata} 
\end{figure}

\begin{figure}[!h]
\centerline{ 
	\includegraphics[width=1.9in]{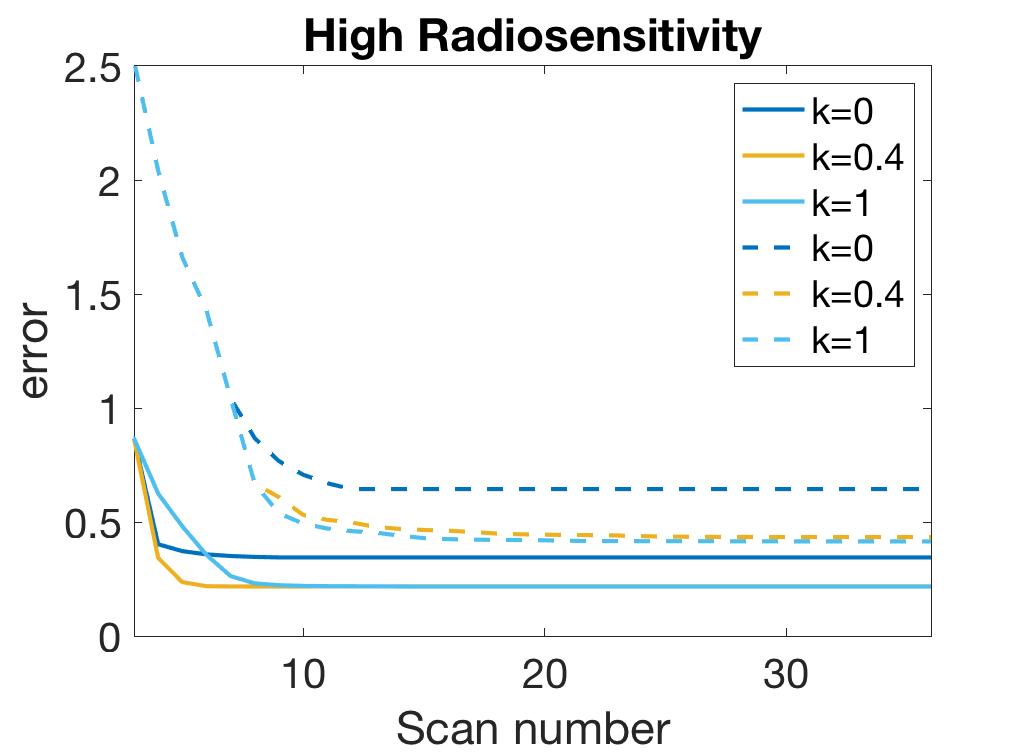}
	\includegraphics[width=1.9in]{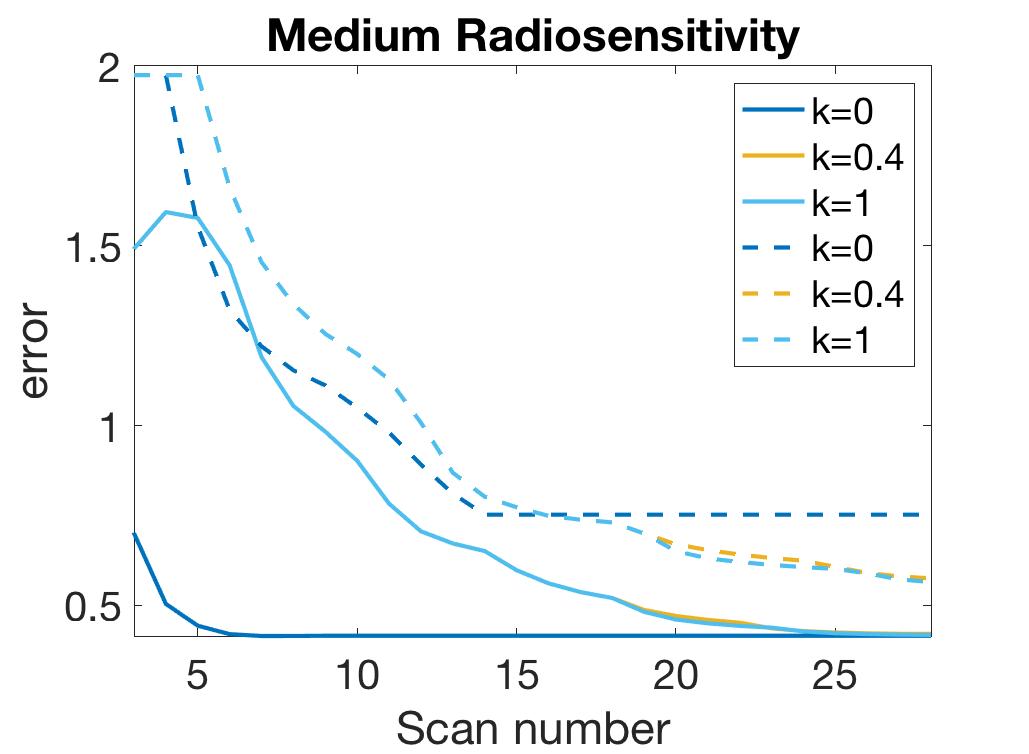}
	\includegraphics[width=1.9in]{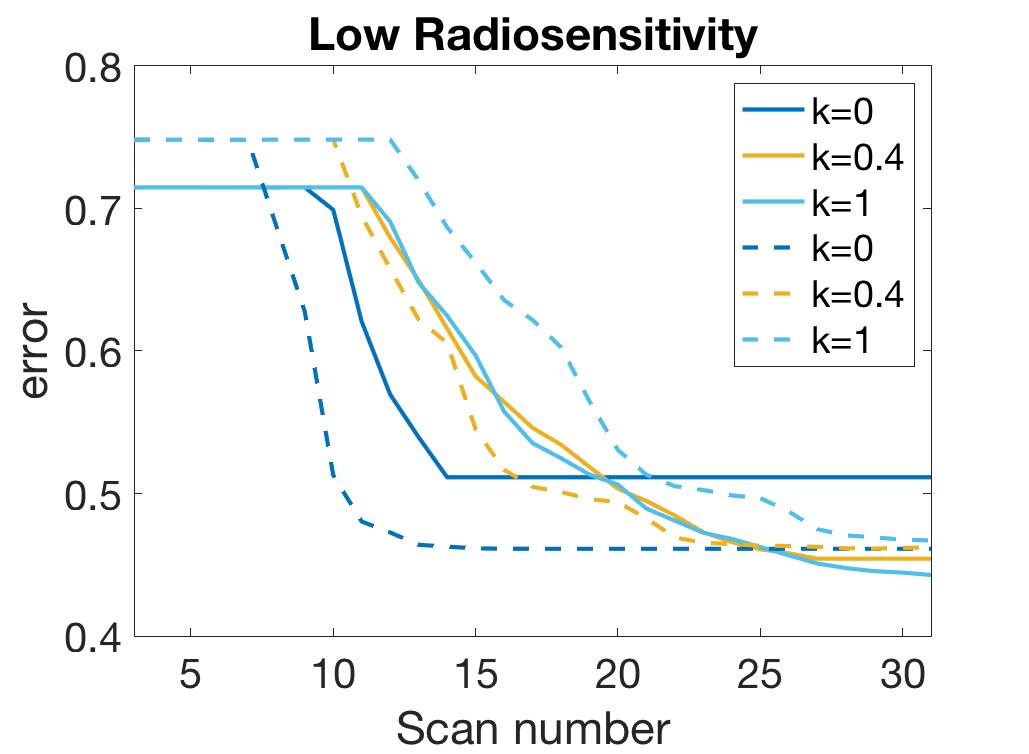}
	}
\centerline{  
	\includegraphics[width=1.9in]{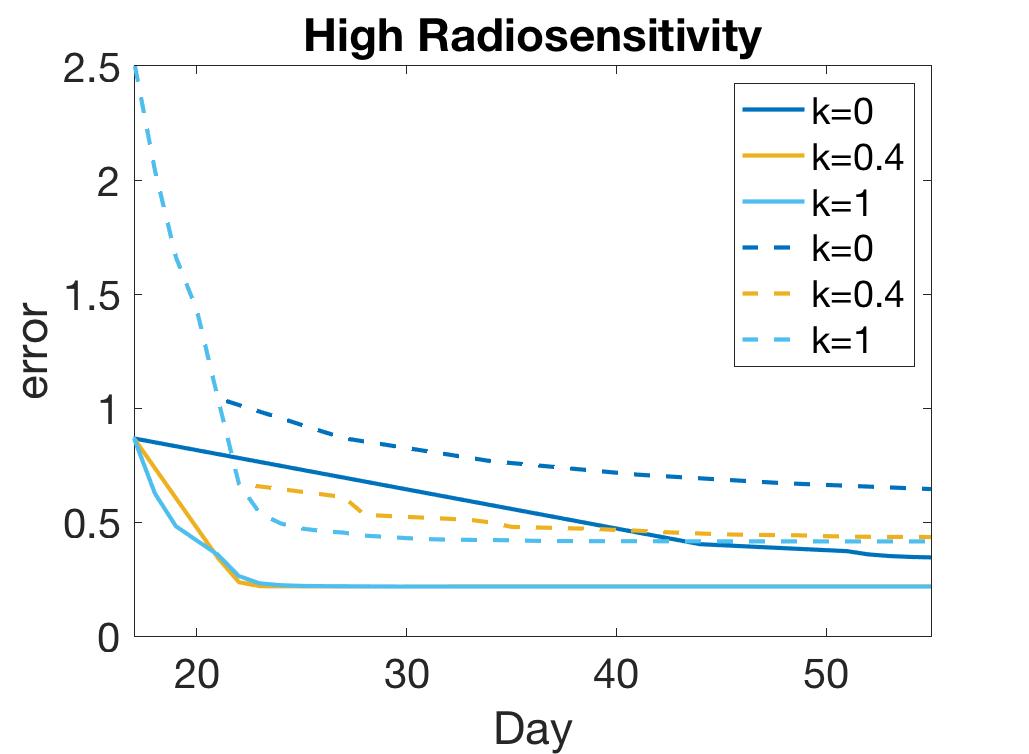}
	\includegraphics[width=1.9in]{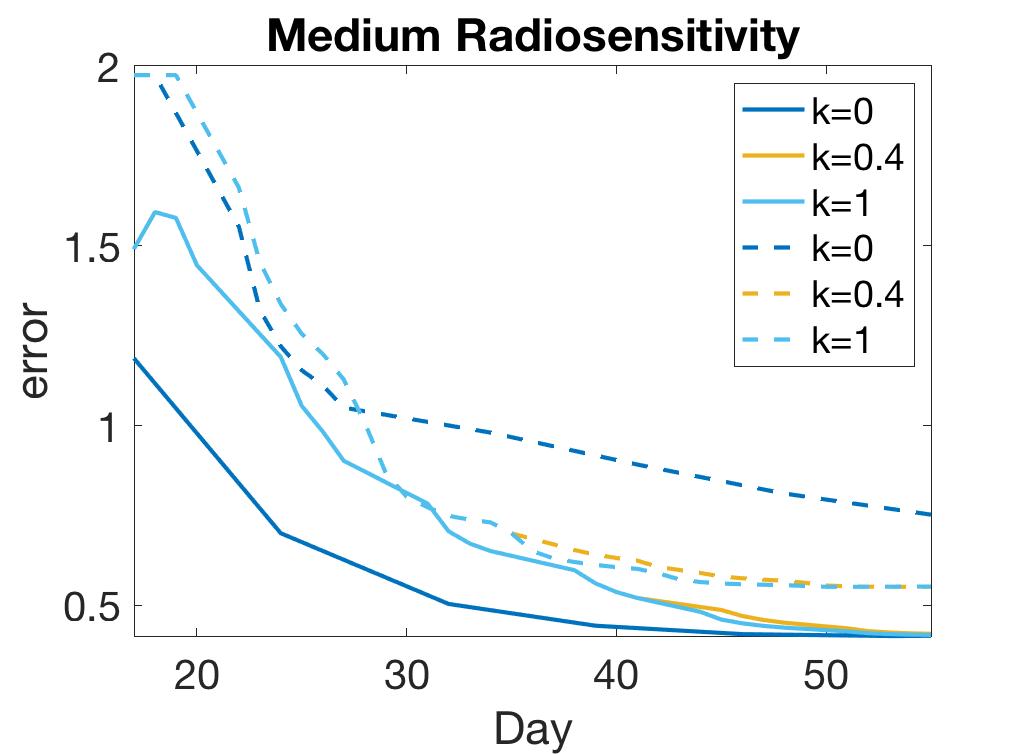}
	\includegraphics[width=1.9in]{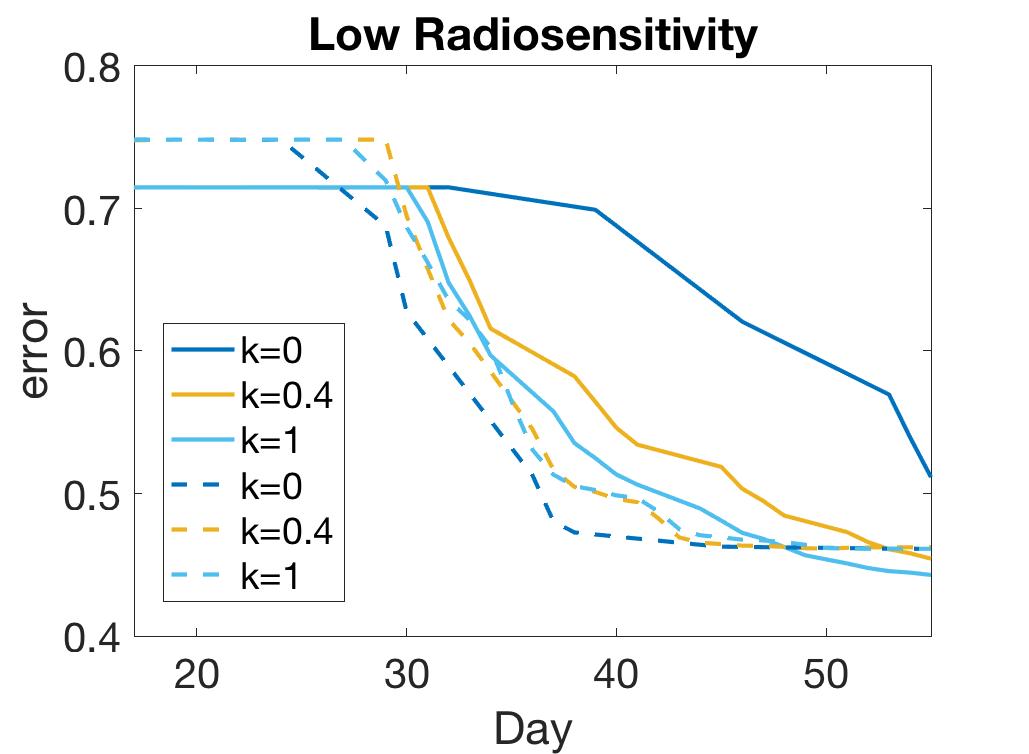}
	}	
\centerline{  
	\includegraphics[width=7.0cm]{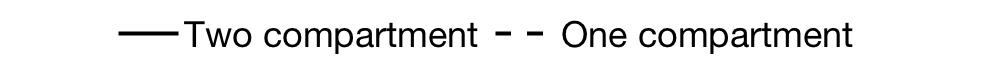}
	}	
\caption{Comparison of accuracy between the one-compartment model calibration \eqref{eqn:eqn1comp} and two-compartment model calibration \eqref{eqn:twocomp}. The shown result compares the error \eqref{eq:error} with respect to scan number (top) and time in days (bottom). Using the same number of scans, the two compartment model show more accurate results in high and medium radiosensitivity, especially using our recommended $k$ value. In case of low radiosensitivity, the one-compartment model with $k=0$ shows better accuracy for scan budgets around 10 to 20. 
} 
	\label{fig:errorVSscan_time} 
\end{figure}

\begin{figure}[!h]
\centerline{ \footnotesize \textsf{High Radiosensitivity} \hspace{4cm} \textsf{Medium Radiosensitivity} }
\centerline{ $k=0.0$ \hspace{2.5cm} $k=0.4$ \hspace{3cm} $k=0.0$ \hspace{2.5cm} $k=0.4$ }
\centerline{ 
	\includegraphics[width=1.5in]{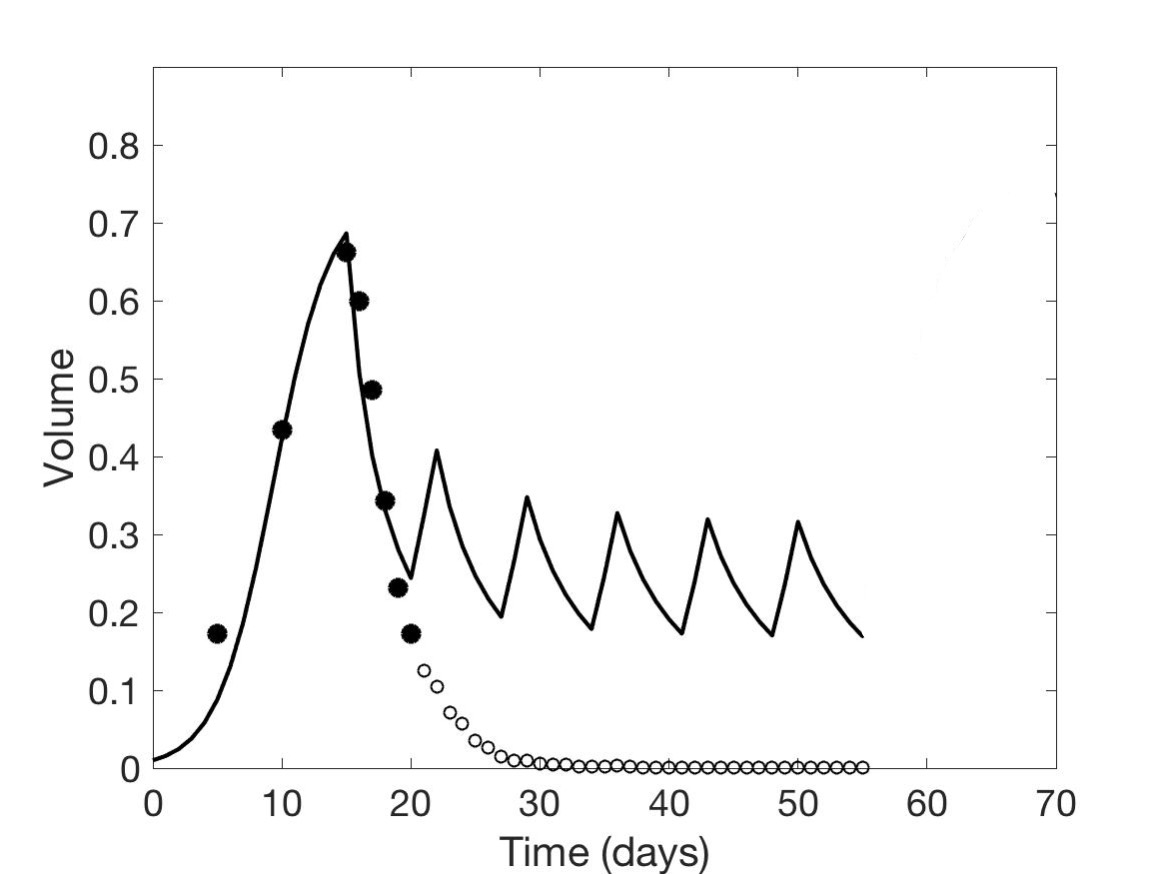}
	\includegraphics[width=1.5in]{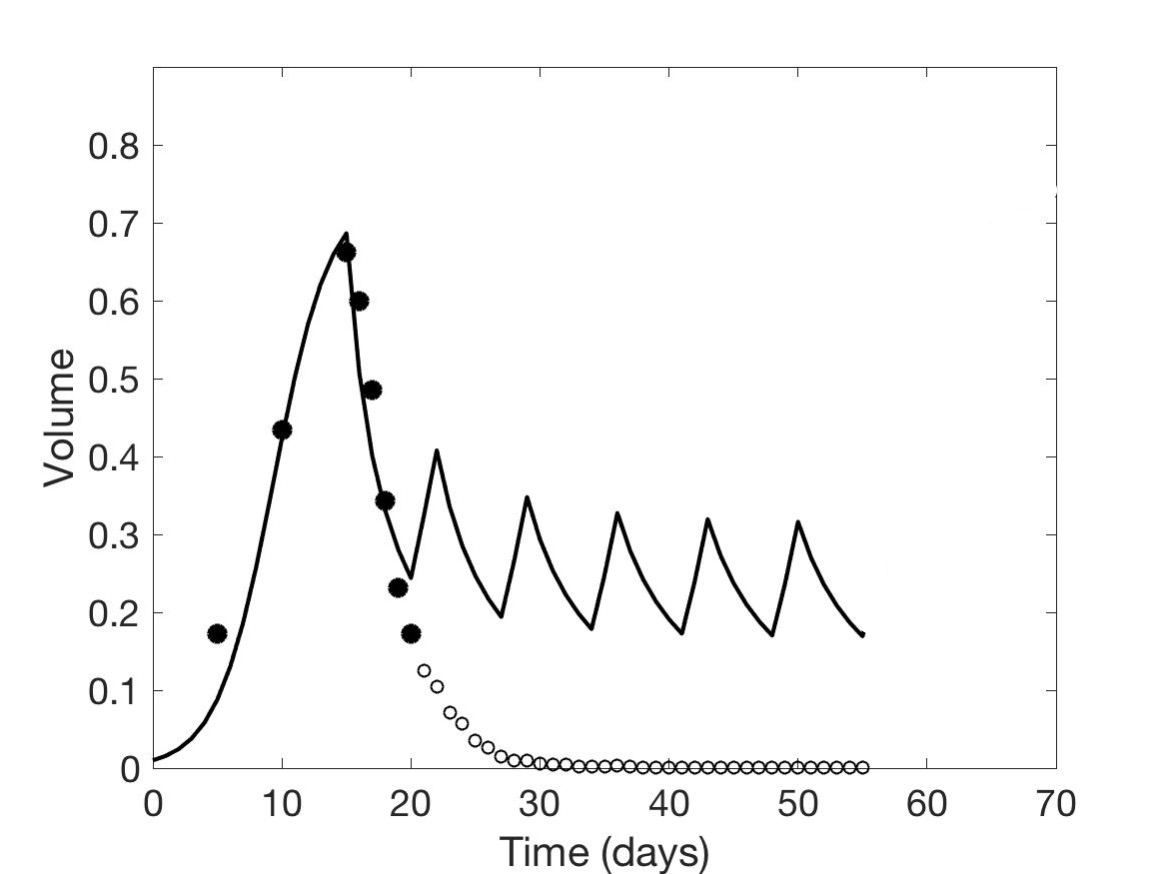}
	\includegraphics[width=1.5in]{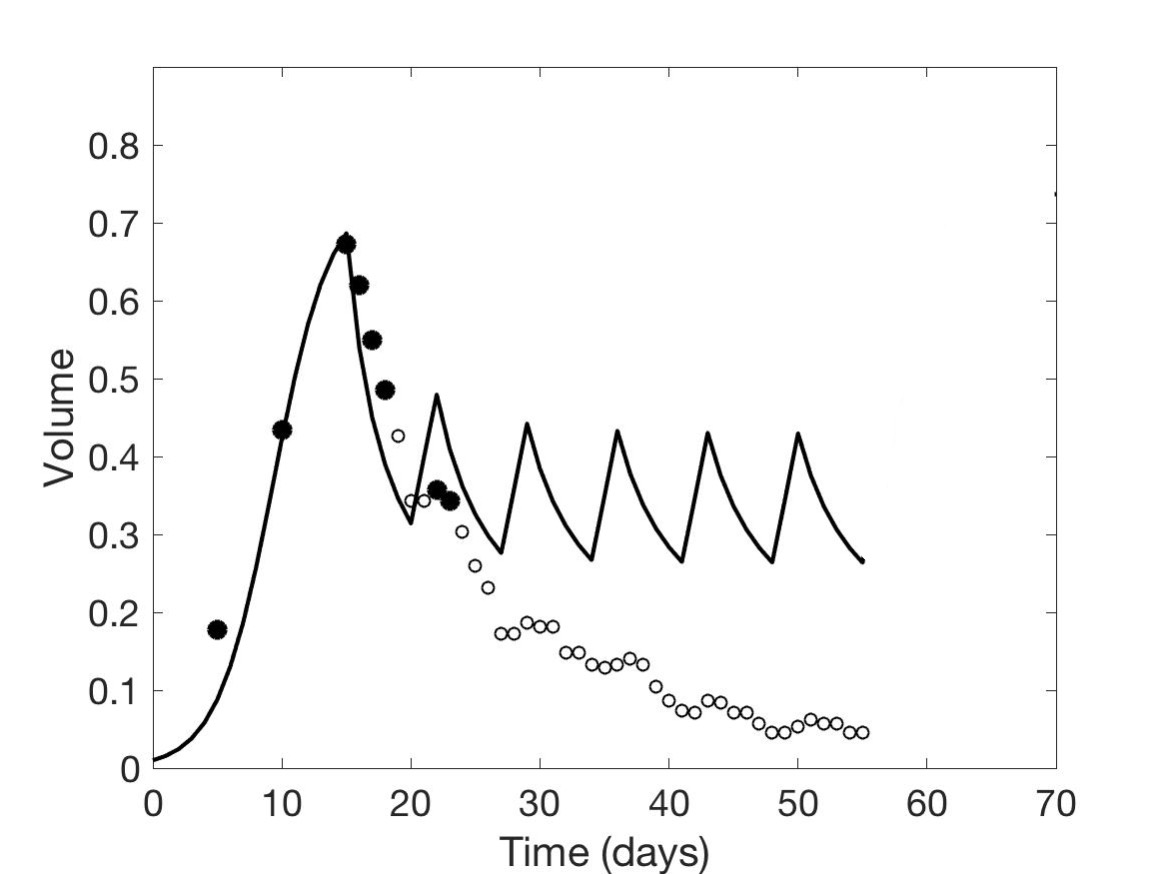}
	\includegraphics[width=1.5in]{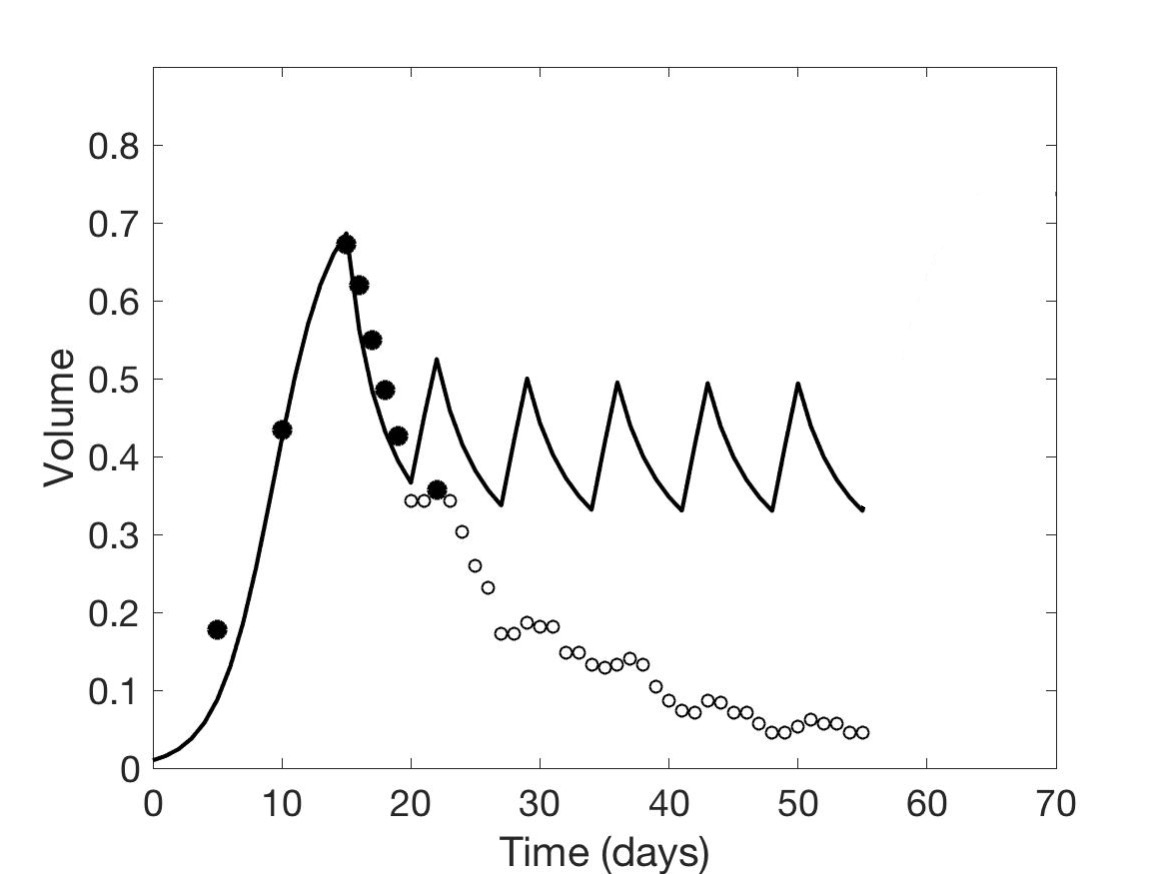}
	}
\centerline{ 
	\includegraphics[width=1.5in]{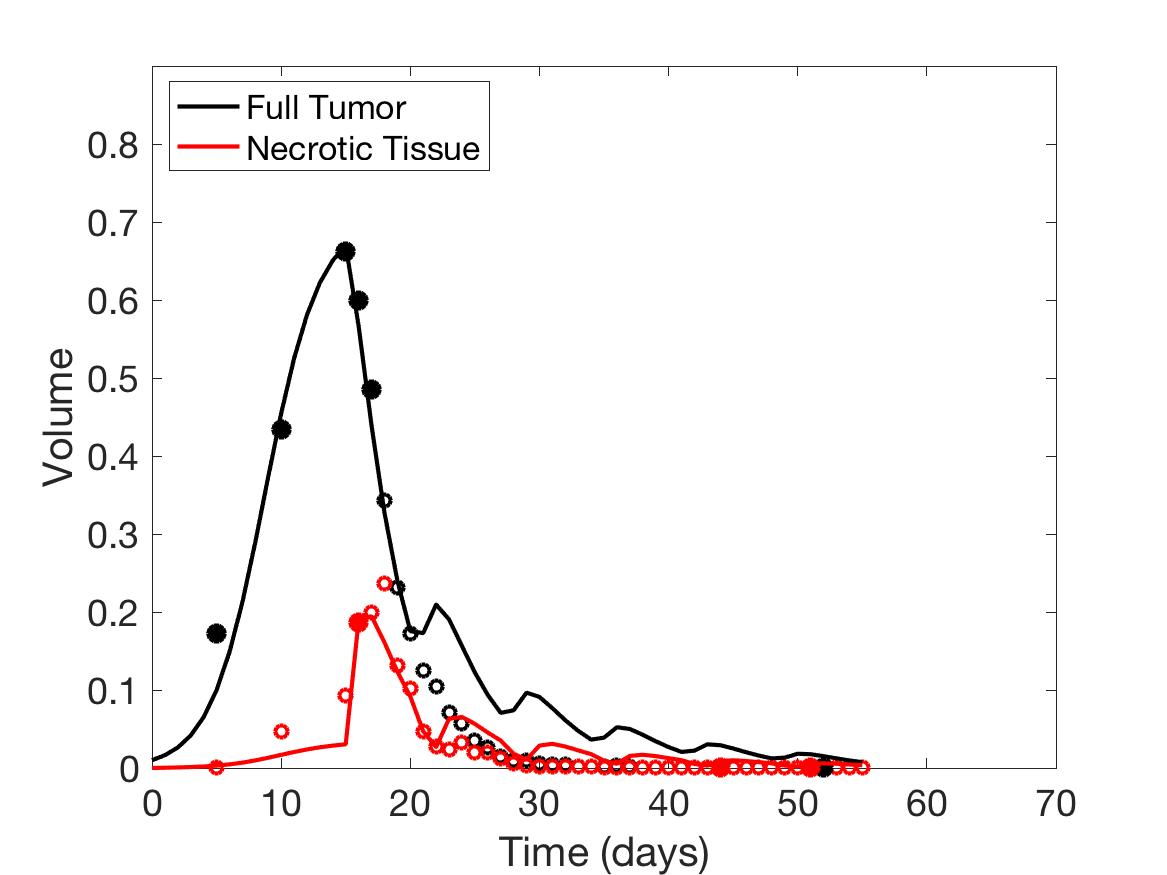}
	\includegraphics[width=1.5in]{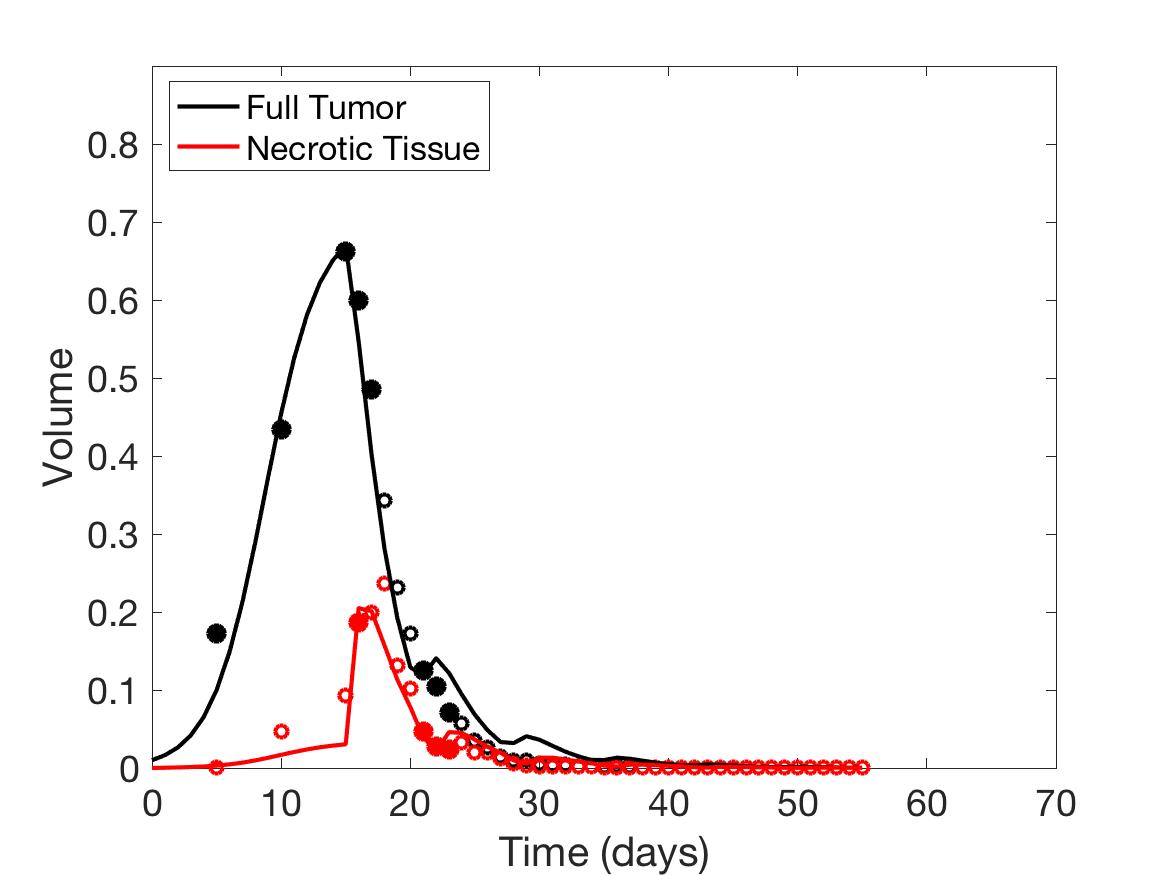}
	\includegraphics[width=1.5in]{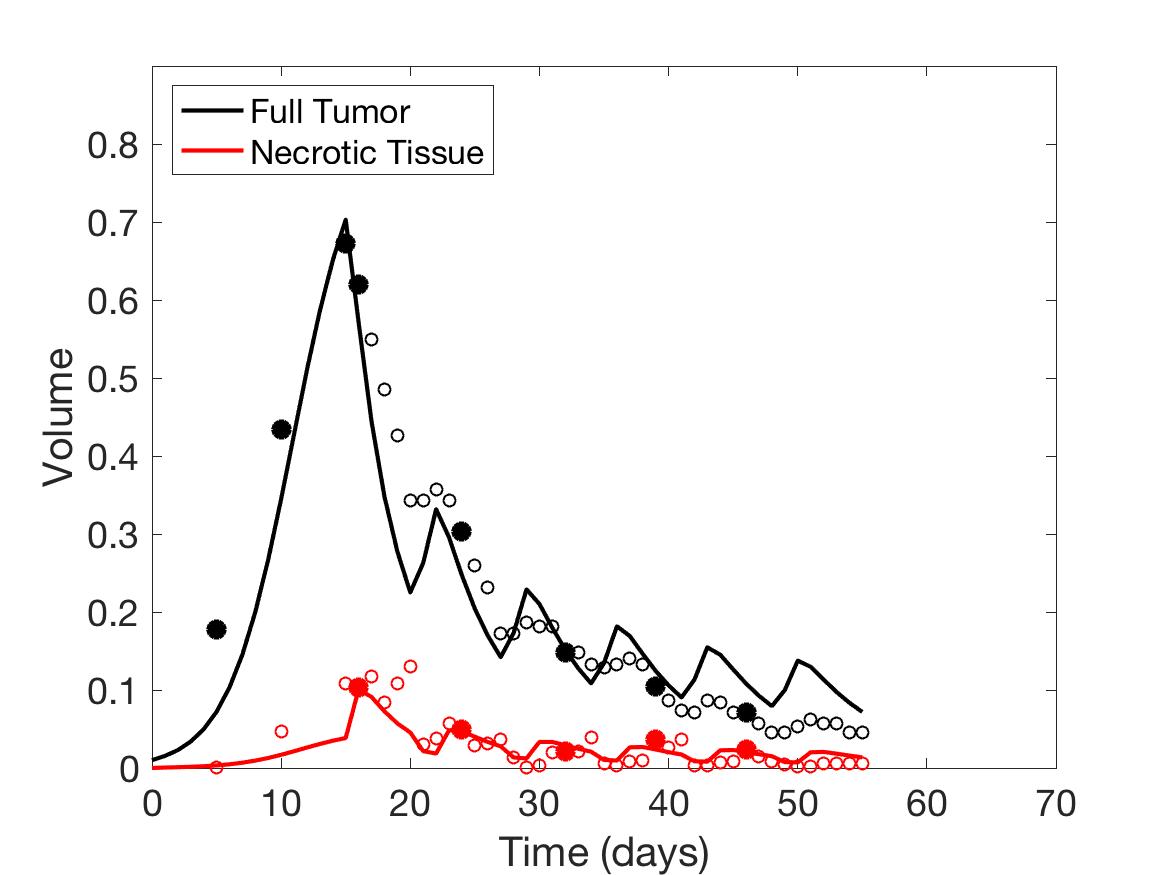}
	\includegraphics[width=1.5in]{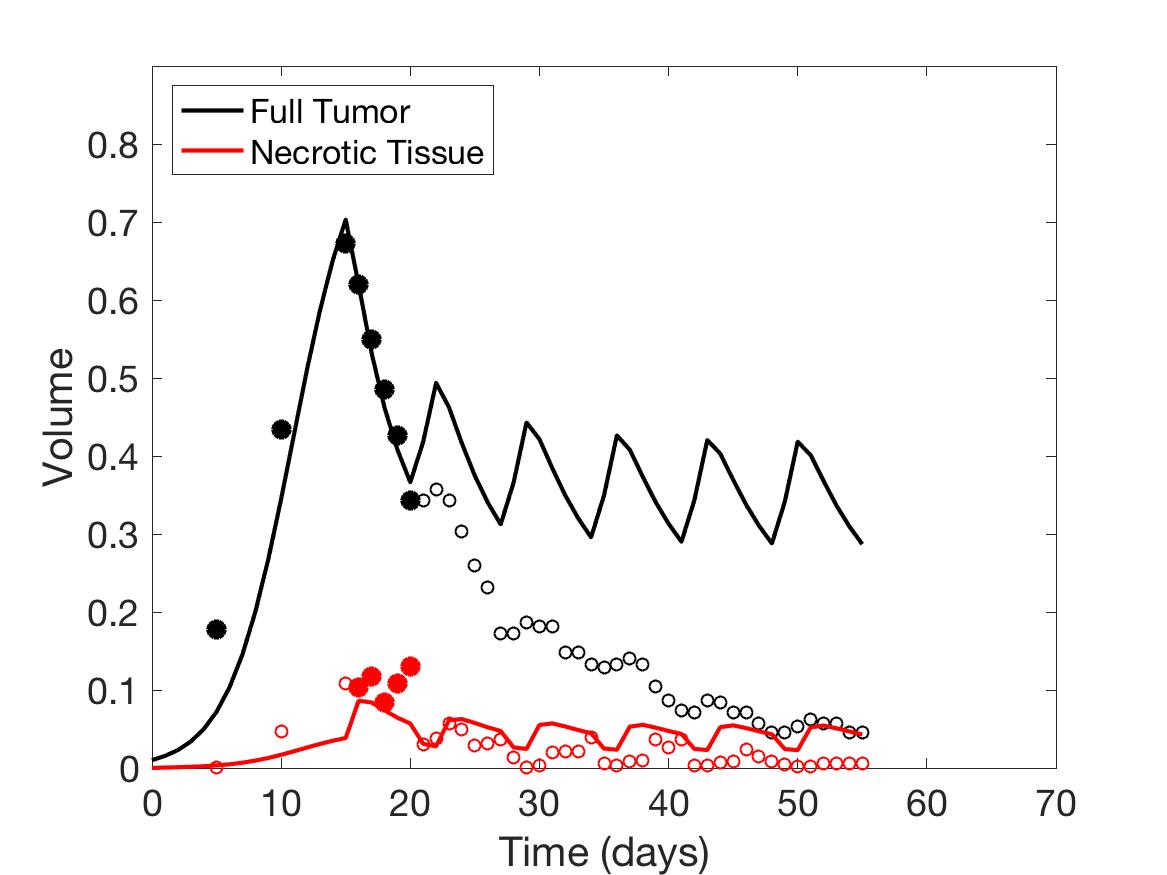}
	}
\caption{Comparison of data calibration between the one-compartment model \eqref{eqn:eqn1comp} (top) and two-compartment model \eqref{eqn:twocomp} (bottom) using a budget of 6 scans. 
The selected scan of tumor volume data ($\bullet$) and necrotic volume data (\textcolor{red}{$\bullet$}) among the potential data ($\circ$, \textcolor{red}{$\circ$}) and the fitted model results are plotted. 
The shown results show how collecting both tumor volume and necrotic fraction and using the two-compartment model calibration can improve the model prediction significantly from the one compartment model. In particular, using $k=0.4$ in high radiosensitivity and using $k=0$ in medium radiosensitivity show the best calibrated results.  
} 
	\label{fig:Modelfit_compare} 
\end{figure}

As in the previous two-compartment model analysis, the high radiosensitivity simulation favors larger score function parameter $k$ values in terms of prediction error, while the other two simulations favor smaller $k$ values when scan budget is limited---see Figure \ref{fig:errorVSk_twocompbothdata}. Moreover, we emphasize that the prediction error using the two-compartment model is smaller compared to using the one-compartment model when the same number of scans are available, although this is likely due to the additional information about the necrotic fraction. For example, we observe that the two-compartment model using as few as 6 scans in high radiosensitivity with $k=0.4$---or medium radiosensitivity with $k=0$---shows far more accurate results compared to using the same number of scans with the one-compartment model; see Figure \ref{fig:errorVSscan_time}, where we compare the error with respect to the number of scans and time in days, in addition to the calibrated data fit in Figure \ref{fig:Modelfit_compare}.

% \begin{figure}[]
% \centerline{  
%          \includegraphics[width=1.5in]{Fig/TotalMIvsDays_TwoCompBothData_Rad1.jpg}\includegraphics[width=1.5in]{Fig/TotalMIvsDays_TwoCompBothData_Rad3.jpg}   \includegraphics[width=1.5in]{Fig/TotalMIvsDays_TwoCompBothData_Rad9.jpg}}
%          \centerline{
%          \includegraphics[width=1.5in]{Fig/TotalMIvsPts_TwoCompBothData_Rad1.jpg}\includegraphics[width=1.5in]{Fig/TotalMIvsPts_TwoCompBothData_Rad3.jpg}   \includegraphics[width=1.5in]{Fig/TotalMIvsPts_TwoCompBothData_Rad9.jpg}\\
%     }
%     \caption{THESE ARE UP TO DATE. Two-compartment model testing with tumor volume automatically included when necrotic is chosen. Total MI versus day and versus number of points.}
%     \label{fig:miplots_twocompbothdata}
% \end{figure}

\subsection{Summary of scan schedule recommendations} \label{sec:discussion}

In what follows, we summarize the major observations from our simulation results in a clinically relevant manner. However, we note that these observations are based on very specific scenarios; results may differ when this methodology is applied to different tumor growth models, treatment regimens, or data collection protocols.

In our first scenario, which investigated this framework under the assumption that a clinician could collect tumor volume data once per week, our results suggest that the scan should be taken on the first day of treatment for the first three weeks. At this point, if roughly half or more of the pre-treatment tumor volume still remains, it would be beneficial to switch to day 6 (Saturday) for weekly data collection, in order to maximize the mutual information provided by each scan.

However, if clinicians are not restricted to taking one scan per week, as in Scenarios 2-4, we use our score function, $S_k(i,r)$---defined in Equation \eqref{eq:scorefxn_eqn}---to enforce a penalty for choosing later scans. Our results suggest that in the the high radiosensitivity case, large values of penalty parameter $k \approx 1$ provide the optimal scan schedule for any scan budget number, allowing for the inclusion of numerous scans early in the treatment schedule. In the case of low radiosensitivity, small values of $k=0$ give the optimal scan schedule, especially when the scan budget is small---the use of a small $k$ allows the algorithm to skip ahead and obtain data toward the end of the treatment protocol so that an overall data trend can be captured. Such a score function recommends data collection largely on the first and last days of the week, skipping most intermediate time points.  In the case of medium radiosensitivity, $k=0$ is optimal when the scan budget is less than 15, with larger $k$ values favored to improve accuracy as more scans are available.

In general, the optimal choice of $k$ is highly dependent upon the shape of the data; data with a rapidly changing gradient (i.e., high radiosensitivity) will require different handling of the scanning schedule than data that is nearly constant, such as our low radiosensitivity scenario. This suggests that we might benefit from updating $k$ as we learn about a patient's sensitivity to radiotherapy, thus gaining information about the trend of their data. Though this will require an abundance of future work to investigate in depth, we give the following examples of the type of suggestion that might be made to clinicians based on our preliminary results:

\begin{quote}
\textit{If only total tumor volume is measured, then for a small scan budget, our results recommend to start with a score function with small $k$, within $0 \leq k \leq 0.3$. Then, if the patient is highly responsive to radiotherapy, increase $k \geq 0.5$, or if the patient is less responsive to radiotherapy, reduce to $k=0$. When the budget of total scans is high (for example, more than 15), we suggest using large $k$ in the range of $k > 0.3$, and then to increase $k$ further if the patient is highly responsive to radiotherapy. } \end{quote}

\begin{quote}
\textit{If both the tumor volume and necrotic volume can be measured, then for a small scan budget, our results suggest to start with a score function with parameter $k\approx 0.2$. Similarly to the one-compartment case, we might further recommend increasing $k$ for highly responsive patients, or reducing $k$ to $k=0$ for less responsive patients. For a scan budget at or above 15 scans, we suggest using a non-zero $k$, for instance, $k\geq 0.2$, in all scenarios.  }

\end{quote}

\section{Conclusion} \label{sec:future}

In summary, we have illustrated a Bayesian information-theoretic framework for determining the optimal sampling of time-series data in order to accurately calibrate low-fidelity models for use in clinical decision-making. We applied this methodology to one- and two-compartment ODE models of tumor response to radiotherapy, calibrated using synthetic data generated from a cellular automaton model, simulating tumors with varying radiosensitivity levels. 

We first enforced a budget of exactly one tumor volume scan per week, and used the mutual information between the low-fidelity model parameters and high-fidelity data to determine the optimal day within each week for data collection. Our results suggested that clinicians with a weekly scan budget should start by measuring the tumor volume on the first day of the weekly treatment regime. If the tumor does not respond well after three weeks of treatment, then scans should be collected on the sixth day of the treatment schedule for the final three weeks to garner additional information about the magnitude of tumor reduction over the course of one dosage cycle; otherwise, scans should continue to be collected on day 1.

In order to relax the scan schedule restrictions to allow for $n$ scans collected at any time throughout the course of treatment, we developed and applied an algorithm relying on a score function, which rewards the user for choosing a data point that yields a large mutual information between the low-fidelity model parameters and high-fidelity data, and enforces a penalty with weight $k$ for skipping forward to later time points. We tested this algorithm on both the one- and two-compartment models; for the latter, we first incorporated a single metric at each step, and then repeated the analysis in a more practical setting by providing the total tumor volume whenever the necrotic volume metric was chosen. We assessed the predictive power of the model resulting from the calibration algorithm using the error between all high-fidelity data points and the low-fidelity model approximation. 

In all cases, we observed the algorithm favoring a larger penalty parameter, $k$, for tumors with a high level of radiosensitivity and any scan budget size; for cases such as this, the accuracy of the model benefits from the inclusion of data at numerous early time points. For tumors with low and medium levels of radiosensitivity and a small scan budget, the calibration was most accurate when using small $k$ values, allowing for samples interspersed throughout the full treatment schedule and observance of the overall data trend. However, in these lower radiosensitivity cases, larger $k$ values became more favorable as the scan budget increased, allowing for the inclusion of more data. from early in the treatment regimen. When calibrating the two-compartment model, the algorithm largely preferred necrotic data to tumor volume, with some exceptions in the high radiosensitivity case. Overall, we observed a smaller prediction error when using the two-compartment model than when using the one-compartment model, due to the additional information provided by the necrotic fraction.

Although this work used only fixed values of the penalization parameter $k$, as future work we propose the development of an adaptive framework to alter $k$ throughout the course of the treatment regimen as we gain information about a patient's sensitivity to treatment.  Additionally, we plan to further investigate robustness of the optimal measurement protocol on different potential patients, based not only on radiosensitivity, but also in terms of other characteristics---including tumor aggressiveness and micro-environment---and study the impact of measurement errors on optimal measurement protocol. 
The results in this work are highly dependent upon the models, treatment regimen, and data collection protocols chosen. We plan to expand this investigation to demonstrate the effectiveness of this framework on models that can incorporate two less closely-related metrics---such as tumor volume and immune cell count---as well as models containing other treatment types, such as chemotherapy, immunotherapy, or some combination of these modalities with radiotherapy. We plan to use such extensions of our scanning framework to help inform decision-making about adjusting therapy types and schedules during tumor treatment at the clinical level.

%%%%%%%%%%%%%%%%%%%%%%%%%%%%%%%%%%%%%%%%%%
\vspace{6pt} 
\section{Acknowledgments}
We would like to thank Dr. Helen Byrne of the University of Oxford for providing helpful feedback. 

%%%%%%%%%%%%%%%%%%%%%%%%%%%%%%%%%%%%%%%%%%
%\authorcontributions{
%Conceptualization, Allison Lewis; Data curation, Kathleen Storey; Formal analysis, Allison Lewis, Heyrim Cho and Kathleen Storey; Investigation, Allison Lewis, Heyrim Cho and Kathleen Storey; Methodology, Allison Lewis, Heyrim Cho and Kathleen Storey; Software, Allison Lewis, Heyrim Cho and Kathleen Storey; Validation, Allison Lewis, Heyrim Cho and Kathleen Storey; Visualization, Heyrim Cho and Kathleen Storey; Writing – original draft, Allison Lewis, Heyrim Cho and Kathleen Storey; Writing – review \& editing, Allison Lewis, Heyrim Cho and Kathleen Storey.}

%%%%%%%%%%%%%%%%%%%%%%%%%%%%%%%%%%%%%%%%%%
%\funding{This research received no external funding.} 

%%%%%%%%%%%%%%%%%%%%%%%%%%%%%%%%%%%%%%%%%%
%\acknowledgments{We would like to thank Dr. Helen Byrne of the University of Oxford for providing helpful feedback. }

%%%%%%%%%%%%%%%%%%%%%%%%%%%%%%%%%%%%%%%%%%
%\conflictsofinterest{The authors declare no conflict of interest.} 

%%%%%%%%%%%%%%%%%%%%%%%%%%%%%%%%%%%%%%%%%%
%% optional
\section{Abbreviations}
The following abbreviations are used in this manuscript:\\

\noindent 
\begin{tabular}{@{}ll}
ODE & Ordinary Differential Equation\\
L-Q & Linear-Quadratic\\
RT & Radiotherapy\\
CA & Cellular Automaton\\ 
$k$NN & $k$th-Nearest-Neighbor\\
DRAM & Delayed Rejection Adaptive Metropolis\\
\end{tabular}

%%%%%%%%%%%%%%%%%%%%%%%%%%%%%%%%%%%%%%%%%%
%% optional
%\appendixtitles{yes} %Leave argument "no" if all appendix headings stay EMPTY (then no dot is printed after "Appendix A"). If the appendix sections contain a heading then change the argument to "yes".
\appendix

\section{Parameter Values} \label{app:params}

\begin{table}[!htb]
\begin{center}
\begin{tabular}[c]{|c|c|c|c|}
        \hline
        \textbf{Parameter} &\textbf{Description} & \textbf{Value} & \textbf{Units} \\

\hline 
 $l$ & Cell size & 0.0018 & cm \\
\hline
 $L$ & Domain length & 0.36 & cm \\
\hline
 $\bar{\tau}_{cycle}$& Mean (standard deviation) cell cycle time   & 18.3 (1.4) & h \\
\hline 
 $c_{\infty}$& Background O$_2$ concentration & $2.8\times 10^{-7}$ & mol cm$^{-3}$  \\
\hline 
 $D$ & O$_2$ diffusion constant & $1.8\times 10^{-5}$ & cm$^2$s$^{-1}$\\
\hline 
 $c_Q$ & O$_2$ concentration threshold for proliferating cells & 1.82$\times10^{-7}$ & mol cm$^{-3}$ \\
\hline 
$c_N$ & O$_2$ concentration threshold for quiescent cells & 1.68$\times10^{-7}$ & mol cm$^{-3}$ \\
\hline 
 $\kappa_P$ & O$_2$ consumption rate of proliferating cells & 1.0$\times10^{-8}$ &mol cm$^{-3}$s$^{-1}$\\
\hline 
 $\kappa_Q$ & O$_2$ consumption rate of quiescent cells & 5.0$\times10^{-9}$ & mol cm$^{-3}$s$^{-1}$  \\
\hline 
 $p_{NR}$ & Rate of lysis of necrotic cells & 0.015  & hr$^{-1}$ \\
\hline 
\end{tabular}
\caption{A summary of the parameters used in the CA Model and their default values. Parameter values are estimated using experimental data from the prostate cancer cell line, PC3, in \cite{Kannan2019}.}
\label{table:CA_pars}
\end{center}
\end{table}

\begin{table}[!htb]
    \centering
    \begin{tabular}{|c||p{0.8cm}|p{1cm}|p{0.8cm}||p{0.8cm}|p{1cm}|p{0.8cm}|}
    \hline
    & \multicolumn{3}{c||}{\textbf{One-Compartment Model}} & \multicolumn{3}{c|}{\textbf{Two-Compartment Model}} \\
    \hline
    \textbf{Parameter} & High &Medium & Low & High & Medium & Low \\
    \hline
$A$ & 0.4584 & 0.4584 & 0.4584  & & & \\
$B$ & 0.6213 & 0.6213 & 0.6213  & & & \\
$\lambda$ & & & &0.5193 &0.4845 &0.6236   \\
$k$ & & & &0.7201 &0.9529 &0.9031   \\
$\eta$ & & & &0.0366 &0.0828 &0.1762   \\
$\zeta$ & & & &0.7270 &1.3476 &1.6638 \\
\hline
    \end{tabular}
\caption{A summary of the fixed pre-treatment parameter values used for low-fidelity model simulations.   Throughout the investigation, we assume that pre-treatment parameters have been identified prior to the administration of RT, and leave the work of ensuring pre-treatment parameter identifiability during calibration to a future investigation.}
\label{table:ODE_pars}
\end{table}

\section{$k$NN estimate of mutual information}
\label{app:kNN_est}

Our definition of mutual information in Equation \eqref{eq:mi_eqn} requires an integral that often cannot be evaluated directly and may be prohibitively expensive, even using many numerical methods. As such, we utilize the $k$NN ($k$th-Nearest-Neighbor) estimate of mutual information to make our procedure computationally feasible.  This estimate (and a proof of convergence) was fully developed in \cite{Kraskov}---below, we summarize the major details. 

To begin, at each calibration step, we estimate the posterior distribution $p(\theta | D_{n-1})$ using the Delayed Rejection Adaptive Metropolis  (DRAM) algorithm for model calibration \cite{Haario}. This variation on a traditional Metropolis-Hastings algorithm includes an additional two steps; an adaptation step that allows for periodic updating of the covariance matrix as information is learned about parameter dependencies, and a delayed rejection step, whereby the algorithm avoids stagnating for long periods of time by replacing rejected parameter set candidates with an alternate candidate chosen from a narrowed proposal distribution. We refer the interested reader to \cite{Haario, Smith} for additional details about the DRAM algorithm, and to \cite{Lewis} for more information on how this calibration procedure fits into our mutual information framework.

To estimate the mutual information between the low-fidelity model parameters and a specified design condition $\xi_n$, we draw $M$ samples from the prior distribution $p(\theta | D_{n-1})$ computed via our Metropolis algorithm, and append low-fidelity model estimates to each parameter set drawn---this results in a chain $X = (\theta, d_n(\xi_n))$ of size $M\times (p+1)$. We satisfy the $k$NN algorithm assumption of independent parameter samples by systematically thinning our dependent Metropolis chains, selecting every 10th sample in the posterior distribution to include in our set of $M$ samples. 

\begin{figure}[!h]
\centering \includegraphics[width=2.2in]{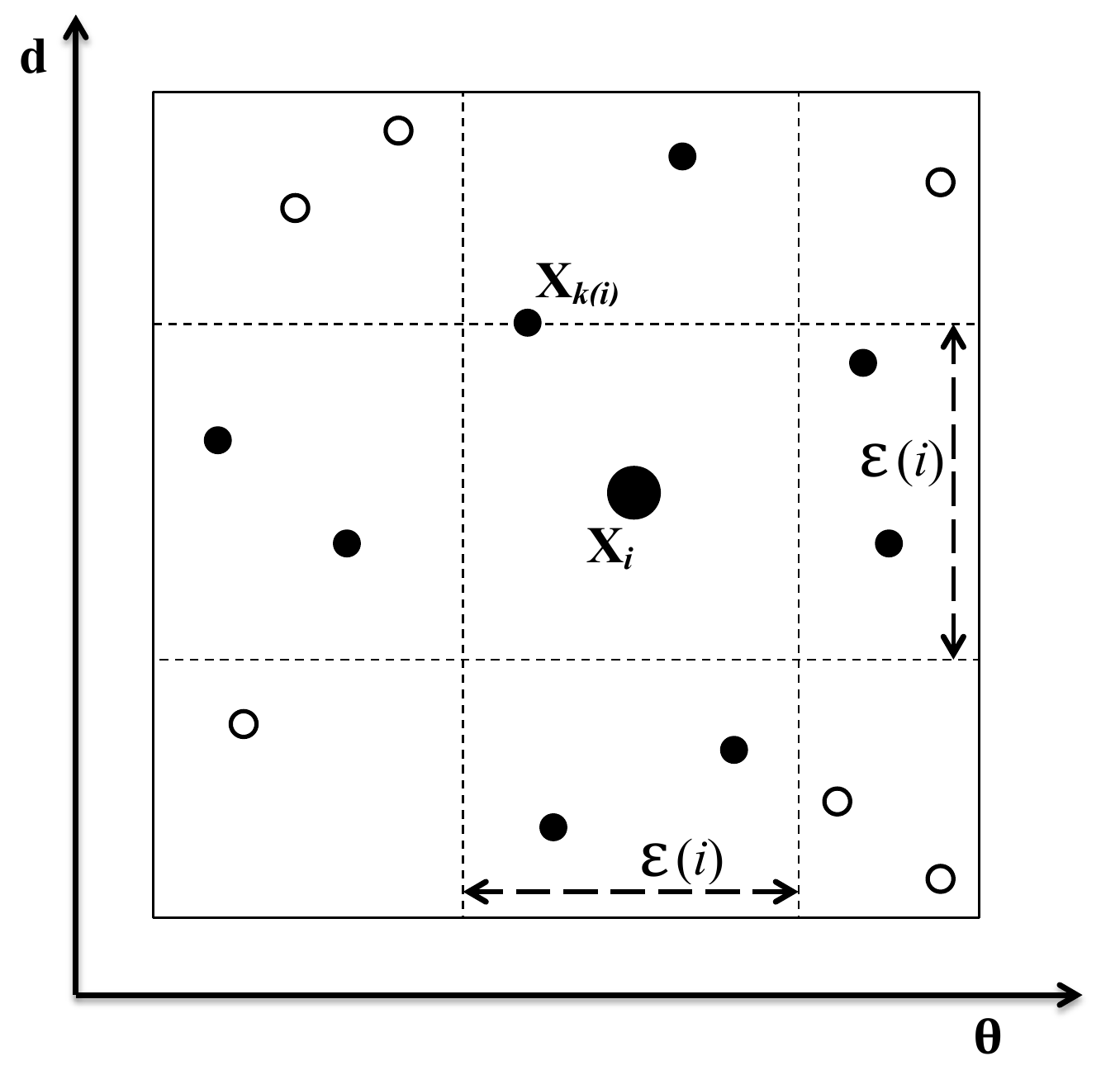}
\caption{Calculation of $\epsilon(i)$, $n_{\theta}(i)$, and $n_d(i)$ for the case $k=1$ from \cite{Kraskov}. Here we have $n_{\theta}(i) = 3$ and $n_d(i) = 4$.  Note that the $k^{th}$-nearest neighbor is not included in the determination of $n_{\theta}$ and $n_d$.}
\label{mi_calculation}
\end{figure}

Following the procedure laid out in \cite{Kraskov}, for each chain element $X_i$, we compute the distance $\epsilon(i)/2 = || X_i-X_{k(i)} ||_{\infty}$, where $X_{k(i)}$ represents the $k^{th}$-nearest neighbor to $X_i$ in the chain $\{X_i\}_{i=1}^M$ (determined using the supremum norm). We determine the number of points in each marginal subspace $n_{\theta}$ and $n_d$ that lie within $\epsilon(i)/2$ of the projected point; see Figure \ref{mi_calculation} for an example of this computation. The mutual information can approximated by
\begin{eqnarray}
I(\theta; d_n | D_{n-1},\xi_n) \approx \psi(k) - \frac{1}{M}\left[\sum_{i=1}^M \psi(n_{\theta}(i)+1)+\sum_{i=1}^M \psi(n_d(i)+1)\right] + \psi(M),\nonumber
%\label{knnEqn}
\end{eqnarray}
where $\psi(\cdot)$ is the digamma function.  

For this study, we estimate mutual information using $k=6$, as suggested in \cite{Terejanu}, though we note that one could calculate the mutual information for a vector of different $k$ values and choose the value that yields the maximum mutual information.

\bibliography{OTMC_reference}
\bibliographystyle{unsrt}

%%%%%%%%%%%%%%%%%%%%%%%%%%%%%%%%%%%%%%%%%%
%% optional
% \sampleavailability{Samples of the compounds ...... are available from the authors.}

%% for journal Sci
%\reviewreports{\\
%Reviewer 1 comments and authors’ response\\
%Reviewer 2 comments and authors’ response\\
%Reviewer 3 comments and authors’ response
%}

%%%%%%%%%%%%%%%%%%%%%%%%%%%%%%%%%%%%%%%%%%
\end{document}